\documentclass[amsmath,amssymb,aps,bm,prd,nofootinbib,floatfix,superscriptaddress,twocolumn]{revtex4-2}

\usepackage{graphicx}
\usepackage{dcolumn}
\usepackage{bm}

\usepackage{amsmath,graphicx,epsfig}
\usepackage{epstopdf}
\usepackage{euscript}
\usepackage{amsfonts}
\usepackage{amssymb}
\usepackage{float}
\usepackage[colorlinks = true,
            linkcolor = blue,
            urlcolor  = blue,
            citecolor = blue,
            anchorcolor = blue]{hyperref}

\def\prn#1{{\left(#1\right)}}

\def\sbrk#1{{\left[#1\right]}}

\def\bra#1{{\langle#1|}}

\def\cg(#1,#2)(#3,#4)(#5,#6){\bra{#1,#2,#3,#4}#5,#6\rangle}

\def\ts#1{{_{\mbox{\scriptsize #1}}}}

\def\threej(#1,#2)(#3,#4)(#5,#6){\begin{pmatrix}#1&#3&#5\\#2&#4&#6\end{pmatrix}}
\def\sixj(#1,#2,#3)(#4,#5,#6){\begin{Bmatrix}#1&#2&#3\\#4&#5&#6\end{Bmatrix}}
\def\ninej(#1,#2,#3)(#4,#5,#6)(#7,#8,#9){\begin{Bmatrix}#1&#2&#3\\#4&#5&#6\\#7&#8&#9\end{Bmatrix}}

\renewcommand{\eqref}[2][]{Eq{#1}.~(\ref{eq:#2})}

\newcommand{\figref}[2][]{Fig{#1}.~(\ref{fig:#2})}
\newcommand{\secref}[2][]{Sec{#1}.~\ref{sec:#2}}

\newcommand{\LL}{\mathcal{L}}
\newcommand{\PHI}{\bm{\hat\phi}}
\newcommand{\THETA}{\bm{\hat\theta}}

\def\mr{\mathrm}

\def\bs{\boldsymbol}
\def\mc{\mathcal}

\usepackage[arrows]{ezedits}
\defineEdit{DK}{\color[rgb]{1.0,0.0,0.0}}{\color[rgb]{1.0,0.0,0.0}}
\defineEdit{TS}{\color[rgb]{0.63,0.12,0.94}}{\color[rgb]{0.63,0.12,0.94}}
\defineEdit{IAS}{\color[rgb]{0,0.7,0.5}}{\color[rgb]{0,0.7,0.5}}
\defineEdit{AK}{\color[rgb]{1.0,0.0,0.0}}{\color[rgb]{1.0,0.0,0.0}}
\defineEdit{SK}{\color[rgb]{0.0,0.0,1.0}}{\color[rgb]{0.0,0.0,1.0}}
\defineEdit{AAA}{\color[rgb]{0.8,0.0,0.8}}{\color[rgb]{0.8,0.0,0.8}}
\defineEdit{JES}{\color[rgb]{0,0.8,0.8}}{\color[rgb]{0.,0.8,0.8}}
\defineEdit{IBM}{\color[rgb]{0.3,0.1,0.7}}{\color[rgb]{0.3,0.1,0.7}}
\defineEdit{MF}{\color[RGB]{250, 128, 114}}{\color[RGB]{250, 128, 114}}

\begin{document}

\title{A Hunt for Magnetic Signatures of Hidden-Photon and Axion Dark Matter in the Wilderness}

\author{Ibrahim A. Sulai}
\email{ibrahim.sulai@bucknell.edu}
\affiliation{Department of Physics \& Astronomy, Bucknell University, Lewisburg, Pennsylvania 17837, USA}

\author{Saarik Kalia}
\email{kalias@umn.edu}
\affiliation{School of Physics \& Astronomy, University of Minnesota, Minneapolis, MN 55455, USA}

\author{Ariel Arza}
\affiliation{Tsung-Dao Lee Institute (TDLI), Shanghai Jiao Tong University, Shanghai 200240, China}

\author{Itay M. Bloch}
\affiliation{Berkeley Center for Theoretical Physics, University of California, Berkeley, CA 94720, USA}
\affiliation{Theory Group, Lawrence Berkeley National Laboratory, Berkeley, CA 94720, USA}

\author{Eduardo Castro Mu\~{n}oz}
\affiliation{Department of Physics \& Astronomy, Oberlin College, Oberlin, Ohio 44074, USA}

\author{Christopher Fabian}
\affiliation{Department of Physics \& Astronomy, Bucknell University, Lewisburg, Pennsylvania 17837, USA}

\author{Michael A.~Fedderke}
\affiliation{The William H.~Miller III Department of Physics and Astronomy, The Johns Hopkins University, Baltimore, MD 21218, USA}

\author{Madison Forseth}
\affiliation{Department of Physics, California State University -- East Bay, Hayward, California 94542-3084, USA}

\author{Brian Garthwaite}
\affiliation{Department of Physics \& Astronomy, Bucknell University, Lewisburg, Pennsylvania 17837, USA}

\author{Peter W.~Graham}
\affiliation{Stanford Institute for Theoretical Physics, Department of Physics, Stanford University, Stanford, CA 94305, USA}
\affiliation{Kavli Institute for Particle Astrophysics \& Cosmology, Stanford University, Stanford, CA 94305, USA}

\author{Will Griffith}
\affiliation{Department of Computing, Math and Physics, Messiah University, Mechanicsburg, PA 17055, USA}

\author{Erik Helgren}
\affiliation{Department of Physics, California State University -- East Bay, Hayward, California 94542-3084, USA}

\author{Katie Hermanson}
\affiliation{Department of Physics, California State University -- East Bay, Hayward, California 94542-3084, USA}

\author{Andres Interiano-Alvarado}
\affiliation{Department of Physics, California State University -- East Bay, Hayward, California 94542-3084, USA}

\author{Brittany Karki}
\affiliation{Department of Physics, California State University -- East Bay, Hayward, California 94542-3084, USA}

\author{Abaz Kryemadhi}
\affiliation{Department of Computing, Math and Physics, Messiah University, Mechanicsburg, PA 17055, USA}

\author{Andre Li}
\affiliation{Department of Physics, California State University -- East Bay, Hayward, California 94542-3084, USA}

\author{Ehsanullah Nikfar}
\affiliation{Department of Physics \& Astronomy, Oberlin College, Oberlin, Ohio 44074, USA}

\author{Jason E.\ Stalnaker}
\affiliation{Department of Physics \& Astronomy, Oberlin College, Oberlin, Ohio 44074, USA}

\author{Yicheng Wang}
\affiliation{Department of Physics \& Astronomy, Bucknell University, Lewisburg, Pennsylvania 17837, USA}

\author{Derek F. Jackson Kimball}
\email{derek.jacksonkimball@csueastbay.edu}
\affiliation{Department of Physics, California State University -- East Bay, Hayward, California 94542-3084, USA}

\date{\today}

\begin{abstract}
Earth can act as a transducer to convert ultralight bosonic dark matter (axions and hidden photons) into an oscillating magnetic field with a characteristic pattern across its surface. 
Here we describe the first results of a dedicated experiment, the Search for Non-Interacting Particles Experimental Hunt (SNIPE Hunt), that aims to detect such dark-matter-induced magnetic-field patterns by performing correlated measurements with a network of magnetometers in relatively quiet magnetic environments (in the wilderness far from human-generated magnetic noise). 
Our experiment constrains parameter space describing hidden-photon and axion dark matter with Compton frequencies in the 0.5-5.0 Hz range.
Limits on the kinetic-mixing parameter for hidden-photon dark matter represent the best experimental bounds to date in this frequency range.
\end{abstract}

\maketitle

\section{Introduction}

Understanding the nature of dark matter is of paramount importance to astrophysics, cosmology, and particle physics.
A well-motivated hypothesis is that the dark matter consists of ultralight bosons (masses $\ll$ 1~eV$/c^2$) such as hidden photons, axions, or axion-like particles (ALPs)~\cite{kimball2022search,graham2015experimental,Arias:2012az}.
If ultralight bosons are the dark matter, under reasonable assumptions\footnote{Here we assume models where the self-interactions among the bosons are sufficiently feeble that they do not collapse into large composite structures (such as boson stars \cite{braaten2019colloquium}). Therefore, the bosons can be treated as an ensemble of independent particles described by the standard halo model (SHM) of dark matter \cite{freese2013colloquium,pillepich2014distribution,evans2019refinement}.} the ensemble of virialized bosons constituting the dark matter halo has extremely large mode-occupation numbers and can be well described as a stochastic classical field~\cite{hui2017ultralight,foster2018revealing,lin2018self,centers2021stochastic,lisanti2021stochastic}.

Ultralight bosonic fields can couple to Standard Model particles through various ``portals''~\cite{graham2016dark,safronova2018search}, one of which is the interaction between the ultralight bosonic dark matter (UBDM) and the electromagnetic field.
Several ongoing laboratory experiments employ sensitive magnetometers located within controlled magnetic environments to search for electromagnetic signatures of UBDM; see, for example, Refs.~\cite{sikivie1983experimental,asztalos2010squid,braine2020extended,zhong2018results,backes2021quantum,salemi2021search,gramolin2021search,andrew2023axion,wagner2010search,chaudhuri2015radio,phipps2020exclusion}.
As noted in Refs.~\cite{Fedd21concept,Fed21search,Arza22}, the conceptual framework for UBDM-to-photon conversion upon which these aforementioned laboratory searches are based also applies to Earth as a whole.
For hidden-photon dark matter (HPDM), the non-conducting atmosphere sandwiched between the conductive Earth interior and the ionosphere acts as a transducer to convert the hidden photon field into a real magnetic field, just as laboratory-scale shields act as transducers in lumped-element or resonant-cavity experiments~\cite{wagner2010search,chaudhuri2015radio,phipps2020exclusion}.
For axion dark matter, Earth's geomagnetic field causes axion-to-photon conversion via the inverse Primakoff effect~\cite{primakoff1951photo,raffelt1988bounds}, playing the role of the applied magnetic field in laboratory-scale axion haloscope experiments~\cite{sikivie1983experimental,asztalos2010squid,braine2020extended,zhong2018results,backes2021quantum,salemi2021search,gramolin2021search,andrew2023axion}.
Thus, unshielded magnetometers can be used to search for ambient oscillating magnetic fields generated by UBDM.

In this paper we describe initial results of the ``Search for Non-Interacting Particles Experimental Hunt'' (SNIPE Hunt~\cite{SNIPEHuntStory}): a campaign to search for axion\footnote{We use the term ``axion'' as a generic descriptor of both QCD axions (that solve the strong-CP problem) and axion-like particles (ALPs).} and hidden-photon dark matter using magnetometers located in the ``wilderness'' (away from the high levels of magnetic noise associated with urban environments \cite{Bowen2019Geo,Dumont2022JAP}).
This work extends to higher axion/hidden-photon Compton frequencies (covering the range from 0.5-5 Hz) than earlier analyses of archival data from the SuperMAG network of magnetometers \cite{SuperMAGwebsite,Gjerloev:2009wsd,Gjerloev:2012sdg} published in Refs.~\cite{Fed21search,Arza22}.
In this frequency range, the dominant magnetic field noise sources are anthropogenic~\cite{Constable}, so we anticipate that the sensitivity to UBDM can be drastically enhanced by measuring in a remote location.

The rest of this paper is structured as follows.
Section~\ref{sec:signal} reviews the model developed in Refs.~\cite{Fedd21concept,Arza22} to predict the global magnetic field patterns induced by hidden-photon and axion dark matter and used to interpret our data.
In Sec.~\ref{sec:expt-details}, we discuss the experimental setup for the magnetometers that measured the magnetic fields at three different locations in July 2022 as well as the time and frequency characteristics of the acquired data.
In Sec.~\ref{sec:data-analysis}, the data analysis procedure is described, which is closely based on that presented in Refs.~\cite{Fed21search,Arza22}.
Section~\ref{sec:data-analysis} is subdivided into one subsection on the hidden-photon dark-matter analysis and another on the axion dark-matter analysis; in both cases no evidence of a dark-matter-induced magnetic signal was discovered, so each subsection concludes by summarizing the constraints obtained on relevant parameters.
In Sec.~\ref{sec:future}, we summarize the next steps for the SNIPE Hunt research program, namely developing and carrying out an experiment for higher Compton frequencies with more sensitive magnetometers.
Finally, in our conclusion we summarize results and compare them to other experiments and observational limits.

\section{Dark-Matter Signal}
\label{sec:signal}

First, we review relevant features of the theory motivating our hidden-photon dark-matter search.
The hidden photon is associated with an additional $U(1)$ symmetry, beyond that corresponding to electromagnetism, which is a common feature of beyond-the-Standard-Model theories, such as string theory \cite{Cve96}.
In our case, we are interested in hidden photons that kinetically mix with ordinary photons \cite{Hol86}.
This allows hidden and ordinary photons to interconvert via a phenomenon akin to neutrino mixing \cite{graham2014parametrically}; i.e., the mass (propagation) and interaction eigenstates are misaligned.
Hidden photons possess a non-zero mass $m_{A'}$ and can be generated in the early universe (see, for example, Refs.~\cite{Graham:2015rva,ahmed2020gravitational,kolb2021completely,Adshead:2023qiw}), which means that they have the right characteristics to be wave-like dark matter \cite{nelson2011dark}.
A useful way to understand the impact of the existence of hidden-photon dark matter on electrodynamics is to write the Lagrangian describing real and hidden photons in the ``interaction'' basis \cite{chaudhuri2015radio,Fedd21concept}:\footnote{Throughout, we use natural units where $\hbar=c=1$.}
\begin{equation}
\begin{split}
\mc{L} &\supset  -\frac{1}{4} \sbrk{ F_{\mu\nu}F^{\mu\nu} + \prn{F'}_{\mu\nu}\prn{F'}^{\mu\nu} } \\ &+ \frac{1}{2} m_{A'}^2 \prn{A'}_\mu\prn{A'}^\mu  + \varepsilon m_{A'}^2 \prn{A'}^\mu A_\mu - J_{\mr{EM}}^\mu A_\mu\:,
\label{eq:hidden-photon-Lagrangian-interaction-basis}
\end{split}
\end{equation}
where only terms up to first order in the kinetic mixing parameter $\varepsilon \ll 1$ are retained.
In \eqref{hidden-photon-Lagrangian-interaction-basis}, $F_{\mu\nu}$ is the field-strength tensor for the ``interacting'' mode of the electromagnetic field that couples to charges, $\prn{F'}_{\mu\nu}$ is the field-strength tensor for the ``sterile'' mode that does not interact with charges, $A_\mu$ is the four-potential for the interacting mode, $\prn{A'}_\mu$ is the four-potential for the sterile mode, and $J_{\mr{EM}}^\mu$ is the electromagnetic four-current density.
In our case of interest, the hidden-photon dark-matter field in the vicinity of Earth is a coherently oscillating vector field with random polarization:\footnote{In this work, we assume that both the hidden-photon phase and its polarization state randomize on the coherence timescale. It is also possible, depending on the production mechanism and subsequent structure-formation processing, that the hidden-photon polarization state could be fixed in inertial space; see, e.g., the discussions in Refs.~\cite{Paola_Arias:2012jcap,Caputo:2021eaa}. We do not explicitly consider this case in this work; a closely related, but different, analysis would need to be undertaken. However, absent accidental geometrical cancellations that are made unlikely by virtue of the length of the data-taking period compared to Earth's sidereal rotational period and the widely separated geographical locations of the magnetic-field stations on which we report, limits in that case are expected to be of the same order of magnitude as those we obtain.}
\begin{align}
\bs{A}'(\bs{r},t) \approx \frac{ \sqrt{2\rho_{\mr{DM}}} }{m_{A'}} e^{-i m_{A'} t} \sum_{i=1}^3\xi_i\prn{\bs{r},t}\bs{\hat n}_ie^{i \phi_i(\bs{r},t)}~,
\end{align}
where $\bs{A}'$ is the sterile vector potential, $\rho_{\mr{DM}} \approx 0.3~{\rm{GeV/cm^3}}$ is the local dark-matter density \cite{read2014local}, $\bs{\hat n}_i$ are a set of orthonormal unit vectors, $\xi_i(\bs{r},t)$ are slowly varying $\mathcal O(1)$ amplitudes, and $\phi_i(\bs{r},t)$ are slowly varying random phases.
Both the amplitudes $\xi_i\prn{\bs{r},t}$ and phases $\phi_i(\bs{r},t)$ of the hidden-photon dark-matter field change stochastically on length scales given by the dark-matter coherence length,
\begin{align}
\ell\ts{coh} \approx \frac{2\pi}{m_{A'} v\ts{DM}}~,
\label{eq:coherence-length}
\end{align}
and time scales given by the coherence time of the field,
\begin{align}
\tau\ts{coh} \approx \frac{\ell\ts{coh}}{v\ts{DM}} \approx \frac{2\pi}{m_{A'} v_{\mr{DM}}^2}~,
\label{eq:coherence-time}
\end{align}
where $v\ts{DM} \sim 10^{-3}$ is the characteristic dispersion (virial) velocity of the dark matter in the vicinity of Earth \cite{bland2016galaxy,evans2019refinement}.
Note that the timelike component of the four-potential $\prn{A'}^\mu$ is suppressed relative to the spacelike component (the vector potential $\bs{A}'$) by $\sim v\ts{DM} \sim 10^{-3}$.
From inspection of \eqref{hidden-photon-Lagrangian-interaction-basis}, it can be seen that the physical effects due to the hidden-photon dark-matter field $\prn{A'}^\mu$ are to leading order the same as those generated by an effective current density
\begin{align}
\bs{J}_{A'} = -\varepsilon m_{A'}^2 \bs{A}'~.
\label{eq:hidden-photon-induced-current}
\end{align}
Inside a good conductor, the interacting mode vanishes, $F_{\mu\nu} = 0$ and $A_\mu = 0$, whereas the sterile mode can propagate into a conducting region with essentially no perturbation.
Outside a conducting region, the effective current density due to the sterile mode acts to generate a non-zero interacting mode.
These effects, where Earth's conducting interior and the conducting ionosphere provide relevant boundary conditions, give rise to the oscillating magnetic-field pattern we seek to measure in our experiment, as described in detail in Ref.~\cite{Fedd21concept}.

The second theoretical scenario we consider is the hypothesis that the dark matter consists primarily of axions~\cite{Peccei:1977hh,Weinberg:1977ma,Wilczek:1977pj,preskill1983cosmology,Abbott:1982af,Dine:1982ah}.
Axions are pseudoscalar particles arising from spontaneous symmetry breaking at a high energy scale associated, for example, with grand unified theories (GUTs) or even the Planck scale \cite{graham2018stochastic}.
Combined with explicit symmetry breaking at lower energy scales, such pseudoscalar particles acquire small masses ($\ll 1~{\rm{eV}}$) and couplings to Standard Model particles and fields \cite{graham2015experimental}.
Like hidden photons, axions are ubiquitous features of beyond-the-Standard-Model theories \cite{preskill1983cosmology,svrcek2006axions,arvanitaki2010string,graham2015cosmological,alexander2023piaxion}, and have all the requisite characteristics to be the dark matter \cite{kimball2022search,graham2015experimental,Arias:2012az}.
The focus of our experiment is the axion-to-photon coupling which is described by the Lagrangian:
\begin{align}
\mc{L} \supset -\frac{1}{4} F_{\mu\nu}F^{\mu\nu} + \frac{1}{2}\prn{\partial_\mu a}^2 - \frac{1}{2}m_a^2 a^2 + \frac{1}{4}g_{a\gamma}aF_{\mu\nu}\tilde{F}^{\mu\nu}~,
\label{eq:axion-Lagrangian}
\end{align}
where $a$ is the axion field, $m_a$ is the axion mass, $g_{a\gamma}$ parameterizes the axion--photon coupling, and $\tilde{F}^{\mu\nu}$ is the dual field-strength tensor.
The last term appearing in \eqref{axion-Lagrangian} describes the interaction between the axion and electromagnetic fields:
\begin{align}
\frac{1}{4}g_{a\gamma}aF_{\mu\nu}\tilde{F}^{\mu\nu} = -g_{a\gamma}a \bs{E}\cdot\bs{B}~,
\label{eq:axion-photon-coupling-term}
\end{align}
where $\bs{E}$ and $\bs{B}$ are the electric and magnetic fields.
In the non-relativistic limit, the leading-order correction to Maxwell's equations arising from the existence of the axion--photon coupling described by \eqref{axion-photon-coupling-term} appears in the Amp\`{e}re--Maxwell Law:
\begin{align}
\bs{\nabla}\times\bs{B} - \partial_t\bs{E} = \bs{J} - g_{a\gamma} \prn{\partial_t a} \bs{B}~.
\label{eq:Ampere-Maxwell-Law-with-axion}
\end{align}
It follows that the physical effects of the axion--photon coupling in the presence of a magnetic field $\bs{B}$, as in the case of hidden photons [\eqref{hidden-photon-induced-current}], manifest as an effective current:
\begin{align}
\bs{J}_a = - g_{a\gamma} \prn{\partial_t a} \bs{B} = ig_{a\gamma}m_a a\prn{\bs{r},t} \bs{B}~,
\label{eq:axion-induced-current}
\end{align}
where
\begin{align}
a\prn{\bs{r},t} = a_0\prn{\bs{r},t} e^{-i m_a t}
\end{align}
is the axion field with a stochastically (slowly) varying amplitude $\left| a_0 \right| \sim \sqrt{2 \rho\ts{DM}} / m_a$, with coherence length $\ell\ts{coh}$ and coherence time $\tau\ts{coh}$ analogous to those for hidden photons described by Eqs.~(\ref{eq:coherence-length}) and (\ref{eq:coherence-time}), with the replacement $m_{A'} \rightarrow m_a$.
The interaction of an axion dark-matter field with the geomagnetic field of Earth thus generates an oscillating magnetic-field pattern, which is discussed in detail in Ref.~\cite{Arza22}. 

In this work, we aim to analyze the first dedicated measurements of the SNIPE Hunt experiment in the frequency range 0.5--5 Hz. The lower frequency bound of 0.5~Hz for our analysis was chosen for practical reasons: $1/f$ noise begins to reduce our sensitivity below $\approx 0.5~{\rm Hz}$ and there is ongoing analysis of SuperMAG data covering frequencies up to $\approx 1~{\rm Hz}$ that is expected to surpass the sensitivity of this experiment. For the upper bound of 5 Hz, we are limited by the well-studied Schumann resonances of the Earth-ionosphere cavity~\cite{sentman2017schumann,Rodriguez-Camacho}. We cannot make a robust prediction for frequencies corresponding to the Schumann resonances because of finite conductivity effects and inhomogeneities in the ionosphere refractive index \cite{sentman2017schumann}. Indeed, the first Schumann resonance occurs at a frequency around 7.8 Hz with time-dependent fluctuations of the order of 0.5 Hz. Most importantly, its width is about 2 Hz, which makes $f \leq 5~\text{Hz}$ a region where the dark-matter-induced magnetic-field pattern can be reliably derived (see \secref{model_uncertainty} for further discussion). The analyses carried out in Refs.~\cite{Fedd21concept,Arza22} considered a quasi-static limit valid only when the UBDM Compton wavelengths are much larger than Earth's radius $R$: $\lambda_{A'} \approx 1/m_{A'} \gg R$ and $\lambda_a \approx 1/m_a \gg R$.
This sets an upper limit on the hidden-photon mass $m_{A'}$ and axion mass $m_a$ of $\sim 3 \times 10^{-14}~{\rm eV}$ and, correspondingly, for their Compton frequencies: $f_{A'}$ and $f_a$ must be $\ll 7~{\rm Hz}$. As we are working at frequencies up to 5 Hz, the formulas used in Refs.~\cite{Fedd21concept,Arza22} are only marginally correct, and therefore more robust formulas are needed here. 

In the following we calculate a more general signal for dark-matter masses close to $\sim1/R$. We write the magnetic and electric fields in terms of vector spherical harmonics (VSH; see Appendix D of~\cite{Fedd21concept}) $\bs{Y}_{\ell m}$, $\bs{\Psi}_{\ell m}$, $\bs{\Phi}_{\ell m}$ as

\begin{multline}
\bs{B}(\bs{x},t)  = e^{-i \omega t}\sum_{\ell,m}\big(B^{(r)}_{\ell m}(r)\bs{Y}_{\ell m} +B^{(1)}_{\ell m}(r)\bs{\Psi}_{\ell m}\\
+B^{(2)}_{\ell m}(r)\bs{\Phi}_{\ell m} \big)
\end{multline}

\begin{multline}
\bs{E}(\bs{x},t) = e^{-i \omega t}\sum_{\ell,m}\big(E^{(r)}_{\ell m}(r)\bs{Y}_{\ell m}+E^{(1)}_{\ell m}(r)\bs{\Psi}_{\ell m}\\
+E^{(2)}_{\ell m}(r)\bs{\Phi}_{\ell m}\big),
\end{multline}
where $\omega$ is the oscillation angular frequency of the dark-matter effective current. For the dark-matter effective current $\bs{J}$ which stands for both hidden photons and axion-like particles, we use the fact that it satisfies $\bs{\nabla}\times\bs{J}=0$ to write
\begin{align}
\bs{J}(\bs{x},t) &= e^{-i \omega t}\sum_{\ell,m}\left(J^{(r)}_{\ell m}(r)\bs{Y}_{\ell m}+J^{(1)}_{\ell m}(r)\bs{\Psi}_{\ell m}\right).
\end{align}
Inserting the above ansatz into Maxwell's equations, we get
\begin{align}
\left({1\over r^2}{d\over dr}\left(r^2{d\over dr}\right)+\omega^2-{\ell(\ell+1)\over r^2}\right)
\left(
\begin{array}{cc}
B^{(2)}_{\ell m} \\
E^{(2)}_{\ell m}
\end{array}
\right)=0~,
\end{align}
and the other components are determined by
\begin{align}
E^{(r)}_{\ell m}&={1\over i\omega}\left({\ell(\ell+1)\over r}B^{(2)}_{\ell m}+J^{(r)}_{\ell m}\right)
\\
E^{(1)}_{\ell m}&={1\over i\omega}\left({1\over r}{d\over dr}\left(r B^{(2)}_{\ell m}\right)+J^{(1)}_{\ell m}\right)
\\
B^{(r)}_{\ell m}&=-{1\over i\omega}{\ell(\ell+1)\over r}E^{(2)}_{\ell m}
\\
B^{(1)}_{\ell m}&=-{1\over i\omega}{1\over r}{d\over dr}\left(r E^{(2)}_{\ell m}\right).
\end{align}

This system is solved with boundary conditions such that $E^{(1)}_{\ell m}$ and $E^{(2)}_{\ell m}$ vanish at both Earth's surface $r=R$ and ionosphere $r=R+h$, where $h$ is the ionosphere height. Because we work in the regime $\omega h \ll 1$, the boundary condition for $E^{(2)}_{\ell m}$ implies immediately that it is zero everywhere; it follows that $B^{(r)}_{\ell m}$ and $B^{(1)}_{\ell m}$ also vanish identically.

Writing $B^{(2)}_{\ell m}=u_{\ell m}/r$, in the limit in which $h\ll R$ we find
\begin{align}
u_{\ell m}''-\lambda_\ell^2u_{\ell m}=0,
\end{align}
where $\lambda_\ell^2=\ell(\ell+1)/R^2-\omega^2$. We write the solution for $u_{\ell m}$ as $u_{\ell m}=\alpha_{\ell m}\cosh(\lambda_\ell(r-R))+\beta_{\ell m}\sinh(\lambda_\ell(r-R))$. Notice that the magnetic field signal at Earth's surface ($r=R$) is simply given by 
\begin{align}
\bs{B}=\sum_{\ell,m}{\alpha_{\ell m}\over R}\bs{\Phi}_{\ell m}.
\end{align}
From the boundary condition $u_{\ell m}'=-rJ^{(1)}_{\ell m}$ at $r=R$ and $r=R+h$, we find at zeroth order in $h/R$
\begin{align}
\alpha_{\ell m}=-{J^{(1)}_{\ell m}(R)+R J^{(1)}_{\ell m}{}^{\prime}(R)\over\lambda_\ell^2}.
\end{align}

\subsection{Hidden-Photon Signal}

In terms of vector spherical harmonics, the hidden-photon effective current, given in \eqref{hidden-photon-induced-current}, is written as

\begin{align}
\bs{J}_{A'}=-\sqrt{4\pi\over3}\varepsilon m_{A'}^2\sum_{m=-1}^1A'_m(\bs{Y}_{1 m}+\bs{\Psi}_{1 m})e^{-i \omega_mt} \enspace. \label{eq:hidden-photon-current}
\end{align}
Here $\omega_m=m_{A'}-2\pi f_dm$, where $f_d$ is the frequency associated to the sidereal day,\footnote{The appearance of $f_d$ here is due to the rotation of Earth.  While the direction of the hidden photon is fixed in the inertial celestial frame, our measurements are performed by magnetometers which are fixed to the rotating Earth.  Transforming the hidden-photon amplitude from the inertial to co-rotating frame, introduces an additional time dependence related to Earth's rotational frequency.} and the hidden-photon amplitudes $A'_m$ (for polarizations $m=0,\pm1$) appearing in \eqref{hidden-photon-current} are normalized via
\begin{equation}
    \frac{1}{2}m_{A'}^2 \langle |A'|^2 \rangle = \rho_{\mr{DM}},
\label{eq:normalization}
\end{equation}
where $\rho_{\mr{DM}}=0.3\,\mathrm{GeV/cm}^3$ is the local dark-matter density. Extracting $J^{(1)}_{1m}$ from \eqref{hidden-photon-current}, we find

\begin{align}
\bs{B}_{A'}=\sqrt{4\pi\over3}{\varepsilon\, m_{A'}^2R\over2-m_{A'}^2R^2}\sum_{m=-1}^1A'_m\bs{\Phi}_{1m}e^{-i\omega_mt}.
\label{eq:Bhpdm}
\end{align}

\subsection{Axion Signal}

For axion dark matter, the orientation of the effective current is determined by Earth's dc magnetic field [see \eqref{axion-induced-current}].  As in Ref.~\cite{Arza22}, we utilize the IGRF-13 model~\cite{IGRF}, which parameterizes Earth's magnetic field $\bm B_\oplus$ in terms of a scalar potential $V_0$, such that $\bm B_\oplus=-\nabla V_0$, where $V_0$ is expanded as
\begin{equation}
    V_0=\sum_{\ell=1}^\infty\sum_{m=0}^\ell\frac{R^{\ell+2}}{r^{\ell+1}}(g_{\ell m}\cos(m\phi)+h_{\ell m}\sin(m\phi))P_\ell^m(\cos\theta),
\end{equation}
where $P_\ell^m$ are the Schmidt-normalized associated Legendre polynomials.  The Gauss coefficients $g_{\ell m}$ and $h_{\ell m}$ are specified by the IGRF model at five-year intervals (see Tab.~2 of Ref.~\cite{IGRF}).  The last of these coefficients correspond to the year 2020, with time derivatives provided for their subsequent evolution.  In this work, we extrapolate the 2020 values (up to $\ell=4$) forward to July 23, 2022 using these time derivatives, and adopt the conventions $g_{\ell,-m}=(-1)^mg_{\ell m}$ to $h_{\ell,-m}=(-1)^{m+1}h_{\ell m}$ to extend to negative $m$.

Once Earth's dc field has been parametrized in this way, the effective current that axion dark matter of mass $m_a$ and axion--photon coupling $g_{a\gamma}$ generates can be written as~\cite{Arza22}
\begin{equation}\label{eq:axion-current}
 \bs{J}_a = ig_{a\gamma}a_0m_a\sum_{\ell,m}C_{\ell m}\left(R\over r\right)^{\ell+2}((\ell+1)\bs{Y}_{\ell m}-\bs{\Psi}_{\ell m})e^{-im_at},
\end{equation}
where $a_0$ is the (complex) axion amplitude, normalized by $\frac12m_a^2\langle|a_0|^2\rangle=\rho_{\mr{DM}}$, and
\begin{equation}
    C_{\ell m}=(-1)^m\sqrt{\frac{4\pi(2-\delta_{m0})}{2\ell+1}}\frac{g_{\ell m}-ih_{\ell m}}2.
\end{equation}
Now, by identifying $J^{(1)}(r)$ in \eqref{axion-current}, the magnetic-field signal from axion dark matter is found to be
\begin{align}
\bs{B}_a=-ig_{a\gamma}a_0m_aR\sum_{\ell,m}{(\ell+1)C_{\ell m}\over\ell(\ell+1)-m_a^2R^2}\bs{\Phi}_{\ell m}e^{-im_at}.
\label{eq:Baxion}
\end{align}

\section{Experimental Details}
\label{sec:expt-details}

\begin{figure}[h!]
\centering
\includegraphics[width = \linewidth]{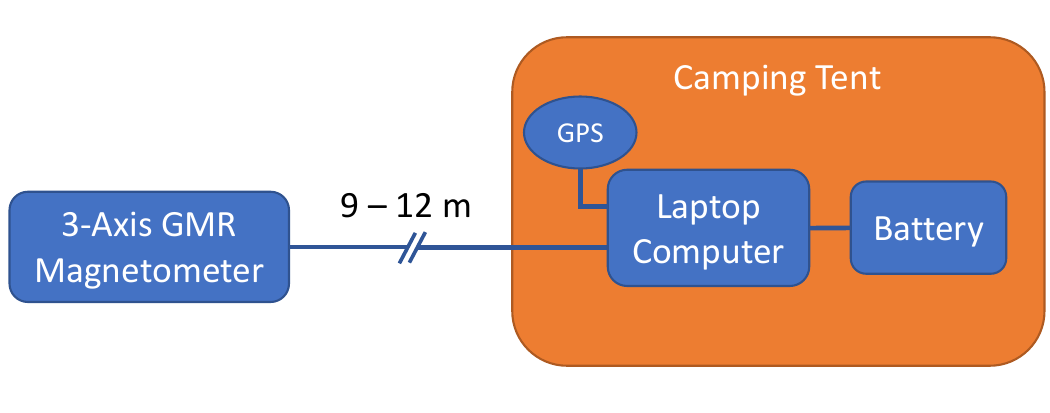}
\caption{Block diagram of SNIPE station setup. A three-axis GMR magnetometer was connected via USB to a laptop located 9--12 m from the sensor. The data were recorded with a laptop and time stamped using the laptop computer time, which was steered to GPS time using a GPS timing receiver. The laptop was powered with battery power banks that were swapped out every 6--10 hours.} \label{fig blockDiagram}
\end{figure}

From 21 July 2022 to 24 July 2022, we conducted the first coordinated SNIPE Hunt science run. Measurements were made with battery-operated  magnetometers located at three sites which were chosen to have minimal magnetic-field interference from power lines, traffic, and other anthropic sources. A block diagram of the experimental setup at an individual station is shown in Fig.~\ref{fig blockDiagram}. The magnetometers were Vector Magnetoresistive (VMR) sensors manufactured by Twinleaf LLC. 
The VMRs use three mutually perpendicular giant magnetoresitive (GMR) field sensors to measure all three components of the magnetic field. 
The sensitivity of the GMR sensors is specified to be 300 $\mathrm{pT}/\sqrt{\mathrm{Hz}}$ over a frequency range of 0.1--100~Hz. 
Prior to deploying the sensors in the field, we verified the calibration of the magnetometers with well-known external oscillating fields applied to the sensors within a magnetically shielded environment. An accurate determination of the oscillating magnetic fields used for calibration was independently attained by observing and measuring magneto-optical resonances in alkali-metal vapor magnetometers \cite{budker2002nonlinear,gawlik2006nonlinear,kimball2009magnetometric}.

\begin{figure*}[h!]
\includegraphics[width = 0.8\textwidth]{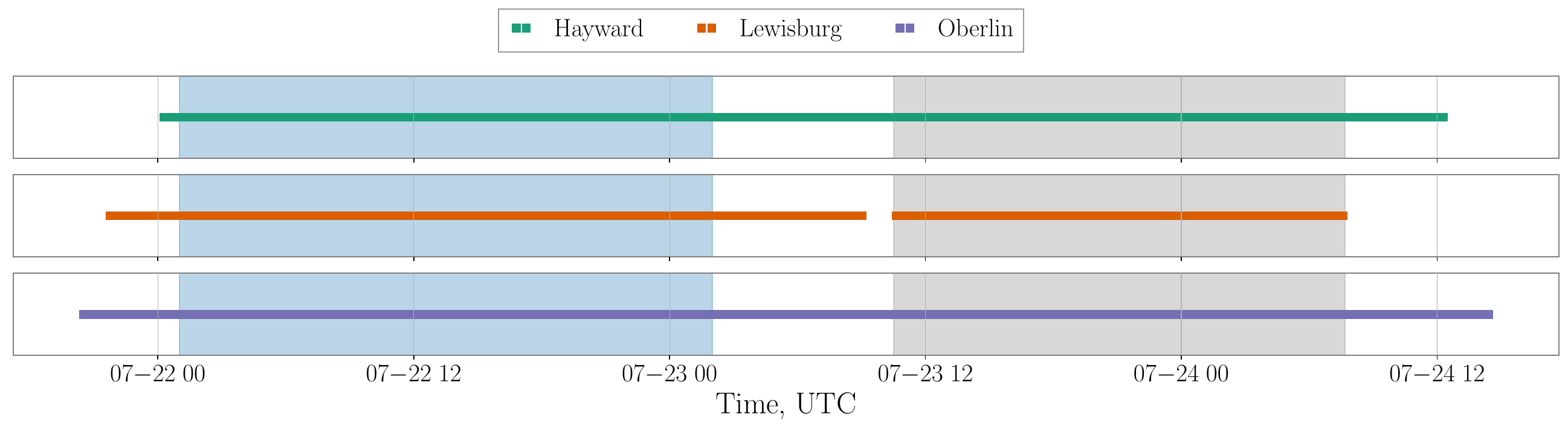}
\caption{Activity for the 2022 SNIPE science run. The horizontal bars indicate when the Hayward, Lewisburg, and Oberlin stations were operational. Two subsets of the data were analyzed independently: Scan-1 covering the interval shown as the light blue shaded region on the left, and Scan-2, the grey shaded region on the right.
}\label{fig:dutyCycle}

\end{figure*}

In addition to the magnetic field sensors, the VMR also has a three-axis gyroscope, a three-axis accelerometer, a barometer, and a thermometer. 
The measurements from all of these sensors were recorded during the course of the science run on a laptop computer which also provided power to the VMR via a USB connection. 
The sample rate for the data acquisition was set to 160 samples/s. In order to limit the influence of magnetic noise from the laptop on the VMR, the laptop was located in a camping tent 9--12~m from the sensor, depending on the station. 
The laptops were powered by 50~${\rm{A \cdot hr}}$ powerbanks, which were swapped with fully charged powerbanks every 6--10 hours and recharged using a solar generator. 
Fig.~\ref{fig:dutyCycle} shows the operation times for the three stations.

The data were time stamped using the computer clocks, which were steered to GPS time using a receiver antenna and synchronization software. To account for the software lag present in the timing calibration, the timing offset correction was set prior to the science run using a time server from the National Institute for Standards and Technology. The accuracy of the timing was tested in the laboratory by applying magnetic-field signals that were triggered by an external GPS receiver before and after the science run. Based on these tests, we estimate the accuracy of the timing to be $\lesssim 100$ ms. 

\begin{figure}[h!]
\centering
\includegraphics[width = 0.5\textwidth]{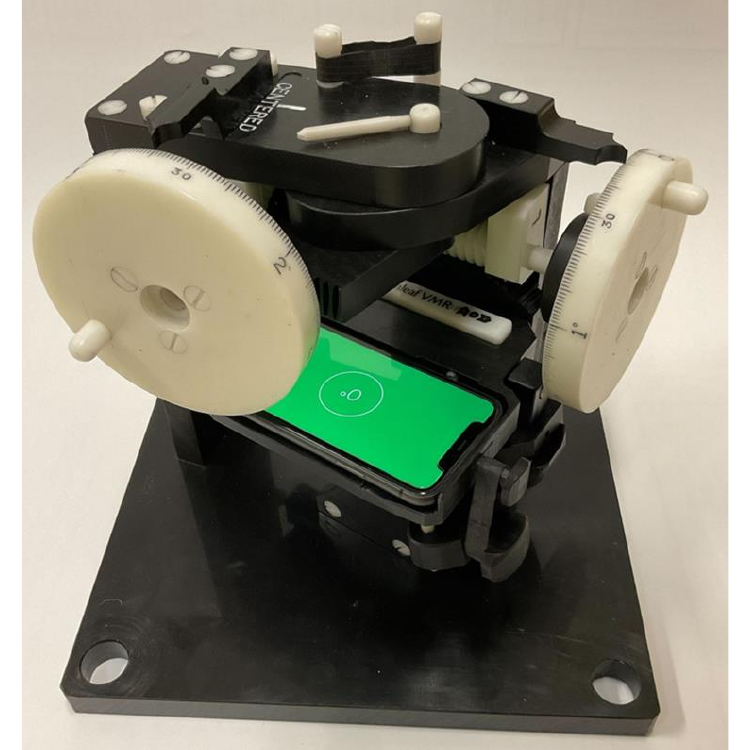}
\caption{Mount for the detector. The pitch, roll, and yaw can be adjusted. A smart phone fits onto the table that holds the sensor for alignment. The phone is removed during data collection. The mount was attached to the ground using heavy-duty plastic tent screws.} \label{fig mount}
\end{figure}

The location of the three stations is shown in Table \ref{tab stations}. The magnetometers were aligned so that the $y$ axis of the magnetometers was vertical, relative to local gravity, and the $z$ axis of the detectors was pointing to true north 
as determined by smart-phone compasses. We estimate the pointing accuracy of the detectors to be $\lesssim 1^\circ$. An example of one of the mounts used for the alignment of the magnetometers is shown in Fig.~\ref{fig mount}. The sensors and mounts were covered with a plastic container that was secured to the ground to guard against rain.

\begin{table*}[ht]
\centering
{\renewcommand{\arraystretch}{1.1}
\begin{tabular}{l | c| ccc}
\hline
\hline
Station & Location & Latitude & Longitude & Elevation  \\
& & (deg) & (deg) & (meters)  \\
\hline
Hayward & Auburn State Recreation Area &  39.1017 & -120.924  & 355.0  \\
Lewisburg & Penn Roosevelt State Park & 40.7404 & -77.7113  & 692.2 \\
Oberlin & Findley State Park &  41.1303 & -82.2069  & 277.4 \\
\hline
\hline
\end{tabular}}
\caption{Locations of sensors used in the 2022 SNIPE Hunt. The stations are referred to by the location of the home institution for the groups in charge of each station.} \label{tab stations}
\end{table*}

\begin{figure*}[h!]
\centering
\includegraphics[width = 0.8\textwidth]{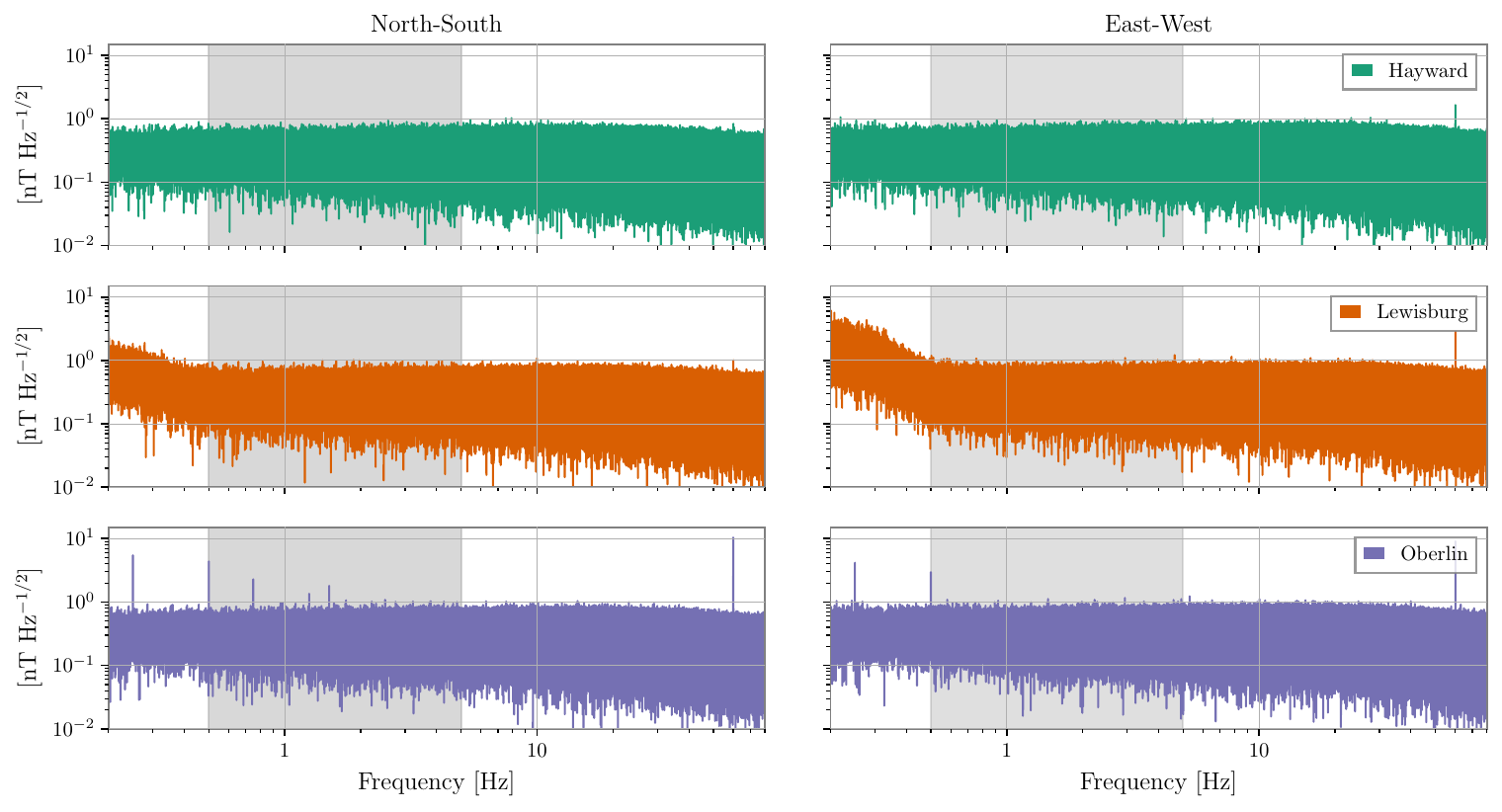}
\caption{Amplitude spectral densities of the North-South and East-West components of the magnetic field measurements from the three measurement sites. The shaded band 0.5--5.0 Hz shows the range of frequencies probed in this work. In this band, the noise floor is limited by the instrumental sensitivity of $\sim 300\,\mathrm{pT}/\sqrt{\mathrm{Hz}}$.}

\label{fig:Psds}
\end{figure*}


\subsection{Noise Characteristics}

For the three sites, we show in Fig.~\ref{fig:Psds} the amplitude spectral density for the East-West and North-South components of the magnetic field -- the components relevant for this search. A couple of features are evident. The Hayward station had noticeably smaller power-line noise at 60 Hz than the Lewisburg and Oberlin stations. The Lewisburg station had a significant $1/f$ pedestal in the 0.1 to 0.5 Hz band that was absent in the other two stations. Also, the  Oberlin station had narrow peaks at 0.25, 0.5, and 0.75 Hz suggesting a common origin as harmonics of some fundamental frequency. As the local magnetic environments are distinct, this difference in noise profile between the stations is expected even though we have not identified the origins of the particular features noted above. However, for the three stations, the amplitude spectral density in most of the band of interest is flat and corresponds to approximately $300\,\mathrm{pT/\sqrt{Hz}}$, the noise floor of the sensors. 

\begin{figure*}[ht]
\centering
\includegraphics[width = 0.8\textwidth]{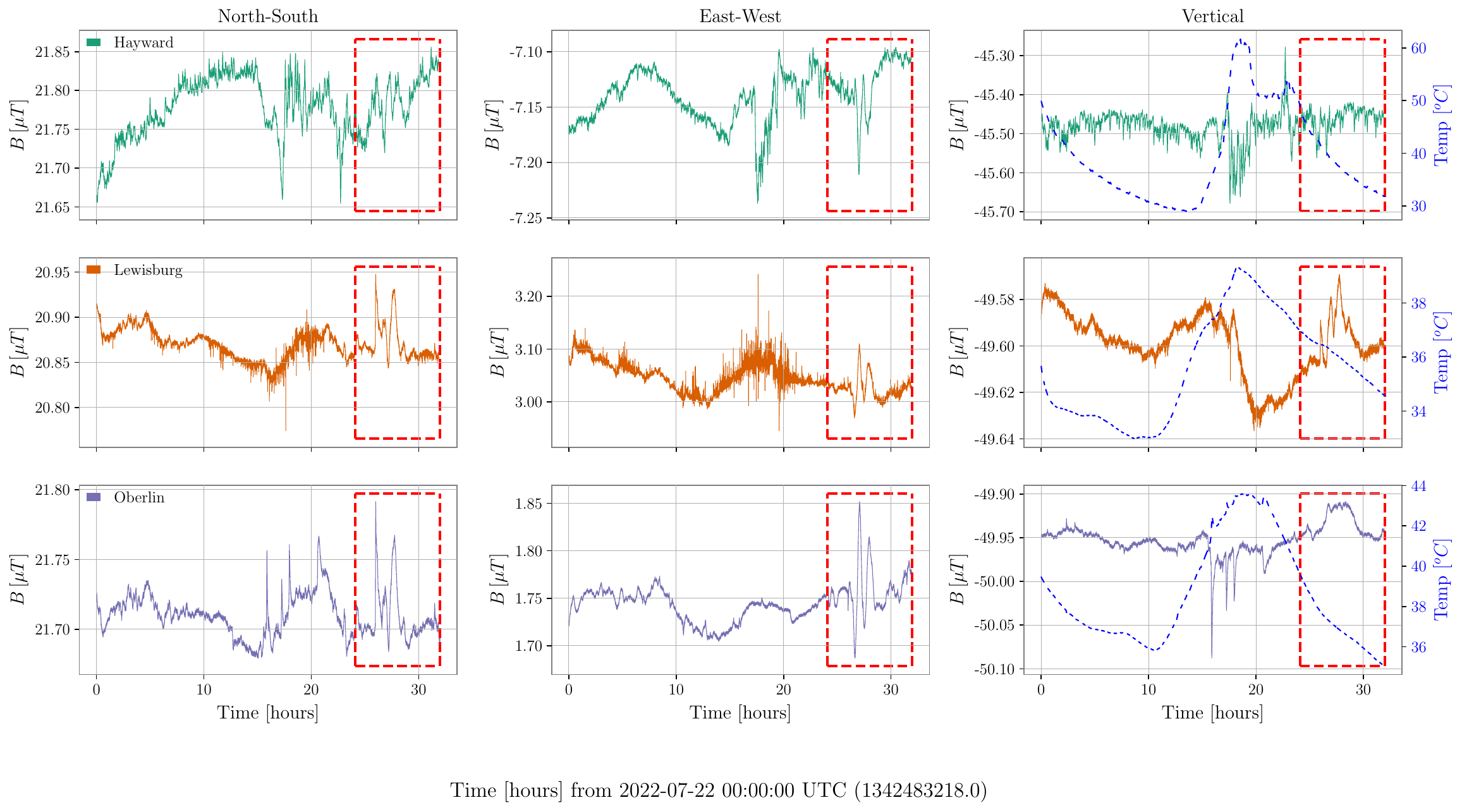}
\caption{Time series of magnetic fields made at the Hayward, Lewisburg, and Oberlin measurement stations. The North-South, East-West, and Vertical (normal to Earth's surface) directions are shown. Scan-1 begins at time $t=0$, and covers the first 24 hours of the data shown. The red dashed boxes correspond to the occurrence of a geomagnetic storm. During that time, we noticed correlated low-frequency oscillations in all three stations. Data from this period were not included in Scan-1, as discussed in the main text. The blue dashed line shows the sensor temperature measured at the different locations.}
\label{Fig:TimeSeries}
\end{figure*}

In Fig.~\ref{Fig:TimeSeries}, we plot time series of the sensor temperature (shown as the blue dashed lines on the right), and of the temperature-corrected measurements of the magnetic field covering the first $\sim 30$ hours of the observing run. The rows correspond to the different sites, and the columns to the North-South, East-West, and Vertical components of the field. We apply the temperature correction purely for plotting purposes, as we noticed a temperature-dependent drift in the sensor calibration at dc of up to 10 percent in the case of the Hayward station and about 2 percent for the other two stations. However, in the analysis band -- 0.5 to 5.0 Hz -- we do not make any temperature correction. Instead, as we discuss in Sec.~\ref{sec:errorbudget}, we assign an uncertainty on the  quoted HPDM and axion limits due to temperature drifts.


Between hours $\sim 13$ and 20 of the time series, we observe increased fluctuations in the North and East components of the Lewisburg data -- fluctuations which were not present in the other stations. This interval coincides with an overnight thunderstorm, during which mechanical agitation of the sensor or lightning occurring nearby may have led to the fluctuations. However, in the temporal window between hours $\sim 25$ and $32$ (shown enclosed in the red dashed boxes of Fig.~\ref{Fig:TimeSeries}), we notice features which are clearly correlated across all three stations, and which we believe are due to a geomagnetic storm associated with the eruption of sunspot AR3060. This produced a C5-class solar flare and a coronal mass ejection directed toward Earth  \cite{USGS,SpaceDotCom}. The storm led to the modulation of Earth's magnetic field which we detected. Including data from this window in the analysis presented below led to noticeable non-gaussianities in the test statistic used for setting limits on the HPDM and axion parameters. For this reason, we excluded the time interval containing the geomagnetic storm in the analysis and instead separate the data into two independently analyzed measurement periods: Scan-1 and Scan-2. These time periods are shown as shaded regions in Fig.~\ref{fig:dutyCycle}. 


\section{Data Analysis}
\label{sec:data-analysis}

In this section, we outline how the SNIPE Hunt data is analyzed to search for both a hidden-photon dark-matter (HPDM) and axion dark-matter signal.


\subsection{Hidden-Photon Analysis}
\label{sec:HPDM}


We begin with the HPDM signal.  Our analysis follows a similar (but simplified) methodology to that described in Ref.~\cite{Fed21search}.  In this search, our data consist of six time series, corresponding to the south-directed and east-directed magnetic field components measured at each of the three SNIPE Hunt measurement locations: $B_\theta(\Omega_1,t_j)$, $B_\phi(\Omega_1,t_j)$, $B_\theta(\Omega_2,t_j)$, $B_\phi(\Omega_2,t_j)$, $B_\theta(\Omega_3,t_j)$, and $B_\phi(\Omega_3,t_j)$.\footnote{Here $\Omega_i=(\theta_i,\phi_i)$ denotes the geographic coordinates of each station.  Note that while $\phi_i$ is exactly the longitude of each station, the latitude of each station is given by $\frac\pi2-\theta_i$.  Likewise, $\PHI$ points east, while $\THETA$ points south.}  We model these time series as being given by (the real part of) the signal in \eqref{Bhpdm} plus Gaussian white noise.  Our goal is then to extract a bound on $\varepsilon$.  As the exact amplitudes $A'_m$ are unknown, we utilize a Bayesian framework and treat these as nuisance parameters.  We also take a Gaussian distribution for them,\footnote{$A'$ can be written as a sum of several independent plane wave solutions of different velocities $v_n\sim\mathcal O(v_\mathrm{DM})$.  These have corresponding frequencies $f_n\sim f_{A'}\left(1+\mathcal O(v_\mathrm{DM}^2)\right)$.  On timescales longer than $\tau_\mathrm{coh}\sim1/(f_{A'}v_\mathrm{DM}^2)$, the value of $A'$ is thus a sum of many contributions with random phases. By the central limit theorem, it is thus distributed as a Gaussian variable.}~normalized by \eqref{normalization}.

The signal in \eqref{Bhpdm} indicates that all relevant information is contained at the frequencies $f_{A'}$ and $f_{A'}\pm f_d$.  Thus we Fourier transform the six time series $B_\alpha(\Omega_i)$, and construct an 18-dimensional data vector\footnote{We use $\vec x$ to denote a vector $x$ with 18 components (or six components in \secref{axion}), and $\bm y$ to indicate a vector $y$ with three components.}~$\vec X$ which contains all information which may be relevant to setting a bound at $f_{A'}$.  Namely, $\vec X$ consists of the six values $\tilde B_\alpha\left(\Omega_i,f_{A'}-\hat f_d\right)$, followed by the six values $\tilde B_\alpha\left(\Omega_i,f_{A'}\right)$, followed by the six values $\tilde B_\alpha\left(\Omega_i,f_{A'}+\hat f_d\right)$.  In our analysis, we compute bounds only at discrete Fourier transform (DFT) frequencies $f_{A'}=n/T$ (where $T$ is the total duration of the time window in consideration).  Note that $f_d$ may not generically be a DFT frequency, and so we have instead used $\hat f_d$, which we define as the nearest DFT frequency to $f_d$.  With these choices, $\vec X$ can be computed via a fast Fourier transform (FFT).  (This allows us to compute $\vec X$ at all frequencies simultaneously, and perform the subsequent analysis for all frequencies in parallel.)  The first step of our analysis is to characterize the statistics of $\vec X$, namely its expectation and variance.

First, let us compute the expectation of $\vec X$.  As mentioned above, we model our measurements as being Gaussian noise on top of the signal in \eqref{Bhpdm}.  Since the expectation of the noise vanishes, the expectation of $\vec X$ simply comes from Fourier transforming \eqref{Bhpdm} and assembling its relevant components into a vector.  To remove the normalization from the amplitudes $A'_m$, let us define
\begin{equation}
c_m=\frac{\sqrt2\pi f_{A'}A'_m}{\sqrt{\rho_{\mr{DM}}}}.
\end{equation}
These now have $\sum_m\langle|c_m|^2\rangle=1$.  In the case $c_{\pm}=0$, (the real part of) \eqref{Bhpdm} takes the simple form
\begin{multline}
\bm B_0(\Omega,t)=-\frac{2\pi f_{A'}R}{2-(2\pi f_{A'}R)^2}\varepsilon\sqrt{2\rho_{\mr{DM}}}\sin\theta \times \\\text{Re}\left[c_0e^{-2\pi if_{A'}t}\right]\PHI,
\end{multline}
and the only nonzero components of $\langle\vec X\rangle$ are
\begin{widetext}
\begin{align}
\langle X_8\rangle_0&=\tilde B_{0,\phi}(\Omega_1,f_{A'})=-\frac{2\pi f_{A'}R}{2-(2\pi f_{A'}R)^2}c_0^*\varepsilon T\sqrt{\frac{\rho_{\mr{DM}}}2}\sin\theta_1\equiv c_0^*\varepsilon\mu_{0,8}\\
\langle X_{10}\rangle_0&=\tilde B_{0,\phi}(\Omega_2,f_{A'})=-\frac{2\pi f_{A'}R}{2-(2\pi f_{A'}R)^2}c_0^*\varepsilon T\sqrt{\frac{\rho_{\mr{DM}}}2}\sin\theta_2\equiv c_0^*\varepsilon\mu_{0,10}\\
\langle X_{12}\rangle_0&=\tilde B_{0,\phi}(\Omega_3,f_{A'})=-\frac{2\pi f_{A'}R}{2-(2\pi f_{A'}R)^2}c_0^*\varepsilon T\sqrt{\frac{\rho_{\mr{DM}}}2}\sin\theta_3\equiv c_0^*\varepsilon\mu_{0,12}.
\end{align}
On the other hand, if $c_0=c_-=0$, then the signal becomes
\begin{equation}
\bm B_+(\Omega,t)=\frac{2\pi f_{A'}R}{2-(2\pi f_{A'}R)^2}\varepsilon\sqrt{\rho_{\mr{DM}}}\cdot\text{Re}\left[c_+\left(i\THETA-\cos\theta\PHI\right)e^{-2\pi i(f_{A'}-f_d)t+i\phi}\right],
\end{equation}
and so the expectation of $\vec X$ is
\begin{equation}\label{eq:plus_exp}
\langle\vec X\rangle_+\approx-\frac{\pi f_{A'}R}{2-(2\pi f_{A'}R)^2}c_+^*\varepsilon\Delta t\sqrt{\rho_{\mr{DM}}}\begin{pmatrix}ie^{-i\phi_1}Q(f_d-\hat f_d)\\\cos\theta_1e^{-i\phi_1}Q(f_d-\hat f_d)\\ie^{-i\phi_2}Q(f_d-\hat f_d)\\\cos\theta_2e^{-i\phi_2}Q(f_d-\hat f_d)\\ie^{-i\phi_3}Q(f_d-\hat f_d)\\\cos\theta_3e^{-i\phi_3}Q(f_d-\hat f_d)\\ie^{-i\phi_1}Q(f_d)\\\cos\theta_1e^{-i\phi_1}Q(f_d)\\ie^{-i\phi_2}Q(f_d)\\\cos\theta_2e^{-i\phi_2}Q(f_d)\\ie^{-i\phi_3}Q(f_d)\\\cos\theta_3e^{-i\phi_3}Q(f_d)\\ie^{-i\phi_1}Q(f_d+\hat f_d)\\\cos\theta_1e^{-i\phi_1}Q(f_d+\hat f_d)\\ie^{-i\phi_2}Q(f_d+\hat f_d)\\\cos\theta_2e^{-i\phi_2}Q(f_d+\hat f_d)\\ie^{-i\phi_3}Q(f_d+\hat f_d)\\\cos\theta_3e^{-i\phi_3}Q(f_d+\hat f_d)\end{pmatrix}\equiv c_+^*\varepsilon\vec\mu_+,
\end{equation}
\end{widetext}
where
\begin{equation}
Q(f)=\frac{1-e^{-2\pi ifT}}{1-e^{-2\pi if\Delta t}},
\end{equation}
and $\Delta t=(1/160)\,\mathrm{s}$ is the time resolution.  Note that, in principle, \eqref{plus_exp} should have an additional term  proportional to $c_+$, which contains factors of $Q(2f_{A'}-f_d-\hat f_d)$, $Q(2f_{A'}-f_d)$, and $Q(2f_{A'}-f_d+\hat f_d)$.  Since $f_d\ll f_{A'}$ and $Q(f)\sim1/f$, these will all be significantly smaller than the $Q$ factors appearing in \eqref{plus_exp}.  Thus we are safe to neglect this additional term.  Similarly, $\langle\vec X\rangle_-\equiv c_-^*\varepsilon\vec\mu_-$ can be computed (for the case when $c_0=c_+=0$).  Then generically, the full expectation of $\vec X$ is
\begin{equation}\label{eq:Sigma}
\langle\vec X\rangle=\varepsilon(c_+^*\vec\mu_++c_0^*\vec\mu_0+c_-^*\vec\mu_-).
\end{equation}

Now that we have computed the expectation of $\vec X$, let us consider its variance.  In this analysis, we consider the frequency range $0.5\,\text{Hz}\leq f_{A'}\leq5\,\text{Hz}$, over which the noise is roughly frequency independent [see \figref{Psds}]. Therefore, we may consider each instance of $\vec X$ for different frequencies as independent realizations of the noise, and use these to estimate the noise.  In particular, we can compute the covariance matrix for $\vec X$ as
\begin{equation}\label{eq:Sigmaij}
\Sigma_{ij}\equiv\langle X_iX_j^*\rangle=\frac1{N}\sum_{k=1}^NX_i(f_k)X_j(f_k)^*,
\end{equation}
where $f_k$ indexes the DFT frequencies between $0.5\,\text{Hz}$ and $5\,\text{Hz}$ (for $k=1,\ldots,N\sim 10^5$).\footnote{Note that since the first six elements, the middle six elements, and the final six elements of $\vec X$ correspond to different frequencies, then covariances between elements from these different groups should vanish, i.e. $\Sigma$ should be block diagonal.  Moreover, the three diagonal blocks should be identical, since they correspond to the same averages in \eqref{Sigmaij} (only with the frequency $f_k$ shifted by $\hat f_d$).  Thus it suffices to only compute $\Sigma_{ij}$ for $7\leq i,j\leq12$.}

Now that we understand the statistics of $\vec X$, we can write down its likelihood
\begin{widetext}
\begin{equation}\label{eq:likelihood}
-\ln\LL\left(\varepsilon,\bm c|\vec X\right)=\left(\vec X-\varepsilon\sum_mc_m^*\vec\mu_m\right)^\dagger\Sigma^{-1}\left(\vec X-\varepsilon\sum_mc_m^*\vec\mu_m\right).
\end{equation}
\end{widetext}
From this likelihood, the computation of the bound on $\varepsilon$ proceeds as in Sec.~V\,D of Ref.~\cite{Fed21search}, but we reproduce it here for completeness. Let us write $\Sigma=LL^\dagger$ and then define
\begin{align}\label{eq:Yhpdm}
\vec Y&=L^{-1}\vec X, \\
\vec\nu_m&=L^{-1}\vec\mu_m.\label{eq:nuhpdm}
\end{align}
If we let $N$ be the $18\times3$ matrix whose columns are $\vec\nu_m$, then \eqref{likelihood} becomes
\begin{equation}\label{eq:Ylikelihood}
-\ln\LL\left(\varepsilon,\bm c|\vec Y\right)=\left|\vec Y-\varepsilon N\bm c^*\right|^2.
\end{equation}
Now if we perform a singular value decomposition $N=USV^\dagger$ (where $U$ is a $18\times3$ matrix with orthonormal columns, $S$ is a $3\times3$ diagonal matrix, and $V$ is a $3\times3$ unitary matrix) and further define
\begin{align}
\bm d&=V^\dagger\bm c^*, \\
\bm Z&=U^\dagger\vec Y,
\end{align}
then the likelihood in \eqref{Ylikelihood} can be reduced to
\begin{equation}\label{eq:Zlikelihood}
-\ln\LL\left(\varepsilon,\bm d|\bm Z\right)=\left|\bm Z-\varepsilon S\bm d\right|^2.
\end{equation}

\begin{figure*}[h!]
\centering
\includegraphics[width = 0.8\textwidth]{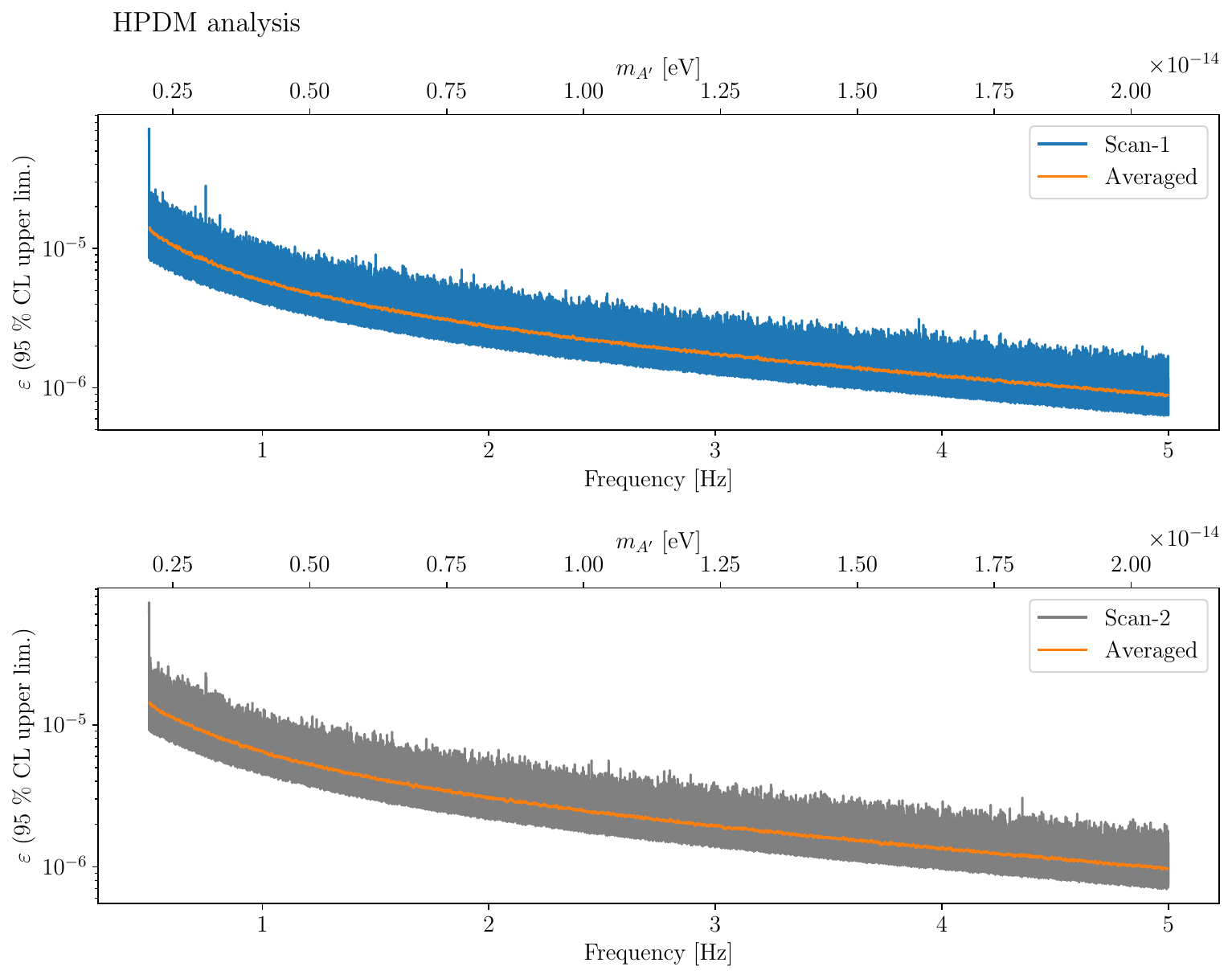}

\caption{95\% credible upper limit on $\varepsilon$, the HPDM kinetic-mixing parameter. The top figure shows the results for Scan-1, and the bottom figure shows the results for Scan-2. The orange traces on both plots are smoothed versions of the limits obtained by averaging over 100 adjacent frequency bins. 
}
\label{fig:HPDMbound}
\end{figure*}

As mentioned earlier, the polarization amplitudes $c_m$, and thus also the parameters $d_m$, are nuisance parameters over which we need to marginalize.  We take them to have a Gaussian likelihood
\begin{equation}
\LL(\bm d)=\exp(-3|\bm d|^2).
\end{equation}
Marginalizing over $\bm d$, the likelihood \eqref{Zlikelihood} reduces to
\begin{equation}\label{eq:marginalized}
\LL\left(\varepsilon|\bm Z\right)\propto\prod_m\frac1{3+\varepsilon^2s_m^2}\exp\left(-\frac{3|z_m|^2}{3+\varepsilon^2s_m^2}\right),
\end{equation}
where $z_m$ are the components of $\bm Z$ and $s_m$ are the diagonal entries of $S$ [see Appendix D\,1 of Ref.~\cite{Fed21search} for a derivation of \eqref{marginalized}].  In order to turn this into a posterior on $\varepsilon$, we must assume some prior.  We take a Jeffreys prior
\begin{equation}
p(\varepsilon)\propto\sqrt{\sum_m\frac{4\varepsilon^2s_m^4}{(3+\varepsilon^2s_m^2)^2}};
\end{equation}
again see Appendix D\,1 of Ref.~\cite{Fed21search}.  The posterior for $\varepsilon$ is thus
\begin{multline}
p(\varepsilon|\bm Z)=\mathcal N\sqrt{\sum_m\frac{4\varepsilon^2s_m^4}{(3+\varepsilon^2s_m^2)^2}} \times \\ \prod_m\frac1{3+\varepsilon^2s_m^2}\exp\left(-\frac{3|z_m|^2}{3+\varepsilon^2s_m^2}\right),
\end{multline}
where $\mathcal N$ must be calculated to normalize the integral of $p(\varepsilon|\bm Z)$ to 1.  We then set a 95\% credible upper limit $\hat\varepsilon$ by solving
\begin{equation}\label{eq:epsbound}
\int_0^{\hat\varepsilon}d\varepsilon\,p(\varepsilon|\bm Z)=0.95.
\end{equation}
By performing this analysis at all DFT frequencies between 0.5\,Hz and 5\,Hz, we arrive at a bound over a range of HPDM masses. \figref{HPDMbound} shows the results of our analysis for both Scan-1 and Scan-2.

\begin{figure*}[h!]
\centering
\includegraphics[width = 0.8\textwidth]{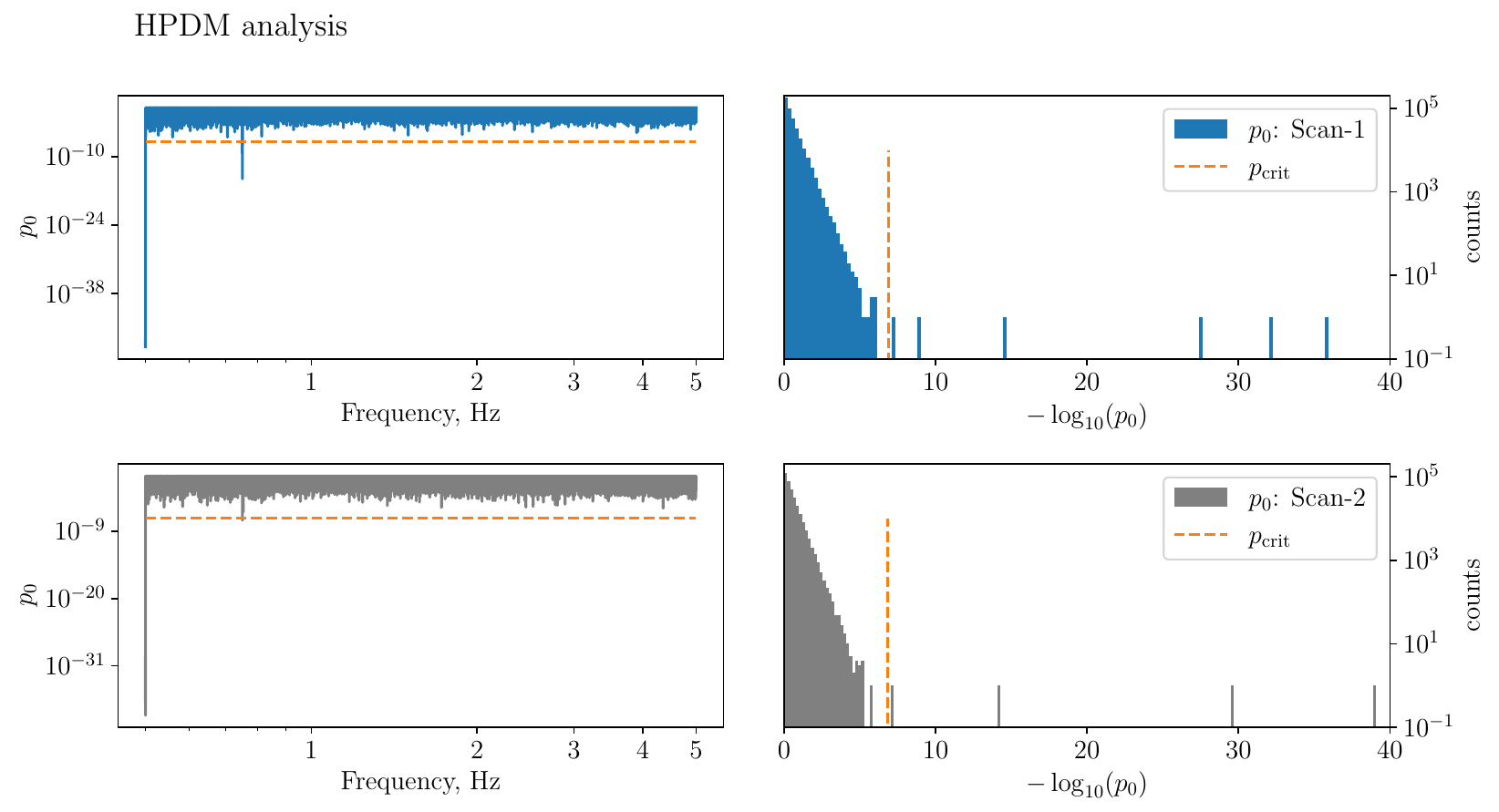}

\caption{The local $p_0$-values for each of the $N=414572$ frequency bins analyzed in the Scan-1, shown in the top (blue) figure, and each of the $N=340291$ bins searched in Scan-2, shown in the lower (grey) figure. The threshold value for declaring a dark-matter candidate at 95\% global confidence is shown by the dotted line (after accounting for the trials factor given by the multiplicity of frequencies searched; see Eq.~\ref{eq:pcrit}).  The left panels show $p_0$ as a function of frequency with candidates having $p$-values below the threshold. The right panels show histograms of $p_0$ for the two different scans and candidates as outliers to the right of the threshold.
}
\label{fig:PvalMain}
\end{figure*}

Following the methodology in Sec.~VI of Ref.~\cite{Fed21search}, we evaluate our data at each frequency for evidence of a significant dark-matter candidate.  From \eqref{Zlikelihood}, we see that under the null hypothesis of no dark matter signal ($\varepsilon=0$), the vector $\bm Z$ should be distributed as a multivariate Gaussian of mean zero.  Specifically, the statistic
\begin{equation}
Q=2\sum_m|z_m|^2
\end{equation}
should follow a $\chi^2$-distribution with six degrees of freedom.  We may therefore compute the corresponding local $p$-value
\begin{equation}
    p_0=1-F_{\chi^2(6)}(Q),
\end{equation}
where $F_{\chi^2(\nu)}$ denotes the cumulative distribution function for a $\chi^2$-distribution with $\nu$ degrees of freedom.  \figref{PvalMain} shows the local $p$-values at each frequency $f_{A'}$ for both Scan-1 and Scan-2.  We consider there to be evidence for a DM candidate at a given frequency (with 95\% global significance) if its local $p$-value is below the threshold $p_\text{crit}$ defined by
\begin{equation}
\label{eq:pcrit}
    (1-p_\text{crit})^N=0.95.
\end{equation}

\begin{figure*}[h!]
\centering
\includegraphics[width = 0.8\textwidth]{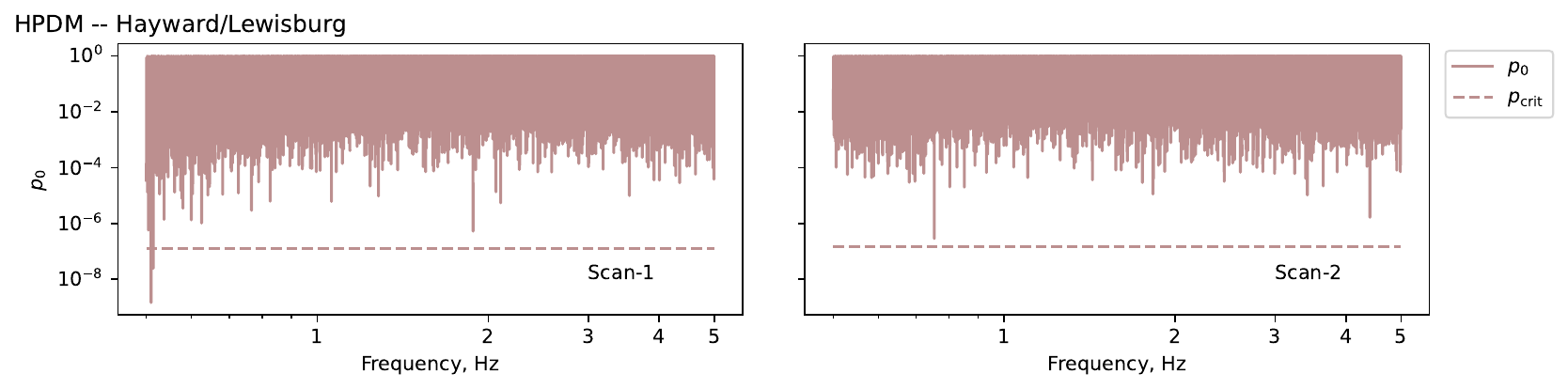}

\caption{The local $p_0$-values for each frequency bin when only data from the Hayward and Lewisburg stations are considered. No beyond-threshold candidates appear in common in \emph{both} Scan-1 and Scan-2. Also, the peaks at 0.50 and 0.75 Hz evident in \figref{PvalMain} are not present in this subset of stations. This indicates that those candidates were due to artefacts in the Oberlin data.
}
\label{fig:PvalHawayardLewisburg}
\end{figure*}

This threshold is shown as a dotted line in \figref{PvalMain}.  Scan-1 exhibits seven frequency bins which cross the threshold.  Four of these are clustered around 0.5\,Hz, while the other three are clustered around 0.75\,Hz. Scan-2, likewise, exhibits three candidate frequency bins clustered around 0.5\,Hz, and one at 0.75\,Hz.  We expect these candidates are associated with the narrow peaks observed in the Oberlin station data.  We have re-performed our analysis using only the Hayward and Lewisburg data, and find that these peaks do not cross the threshold for significance in either scan when restricting to these two stations [see \figref{PvalHawayardLewisburg}].  Since dark matter should be present in all locations at all times, this strongly suggests that these signal candidates do not correspond to dark matter.  Moreover, we note that the width of a dark-matter signal is given by $f_av_\mathrm{DM}^2$, where $v_\mathrm{DM}$ is the dark matter velocity dispersion.  Since the frequency bin size for our analysis is roughly $10^{-5}\,\mathrm{Hz}$ and each cluster spans multiple bins, these clusters represent signal candidates with widths of roughly $10^{-5}f_a$, corresponding to large velocity dispersions of $v_\mathrm{DM}\sim1000\,\mathrm{km/s}$ (which is far above the escape velocity of the Milky Way).  We therefore rule out these dark-matter candidates and conclude that our analysis finds no evidence for HPDM in the $0.5\,\text{Hz}\leq f_{A'}\leq5\,\text{Hz}$ range. 

We have verified our entire analysis by injecting artificial HPDM signals into our data set and ensuring that the analysis correctly identified them. For example, when we added a monochromatic signal of the form in \eqref{Bhpdm} with $\varepsilon=10^{-5}$ and $m_{A'}=10^{-14}\,\mathrm{eV}$ to the time series data from each station, and re-ran our analysis, we found the resulting limit only changed in the vicinity of $m_{A'}=10^{-14}\,\mathrm{eV}$, where it became $\hat\varepsilon\sim1.4\times10^{-5}$.  (Note that the limit is slightly weaker than the injected signal, as expected.)  Moreover, the candidate analysis correctly identified DM candidates near the injected masses with high significance.  We applied a similar verification process to the axion analysis described in the next section.

\subsection{Axion Analysis}
\label{sec:axion}
Now we move to the analysis for an axion dark-matter signal.  This analysis proceeds similarly to the HPDM analysis, but is slightly simpler.  As in the HPDM analysis, we construct a data vector $\vec X$ consisting of Fourier transforms of the measured magnetic field at each location.  Since the axion signal in \eqref{Baxion} contains no $f_d$ dependence, however, the only relevant information is contained at frequency $f_a$.  Therefore in this analysis, we only take $\vec X$ to be a six-dimensional vector, consisting of the measurements: $\tilde B_\theta(\Omega_1,f_a)$, $\tilde B_\phi(\Omega_1,f_a)$, $\tilde B_\theta(\Omega_2,f_a)$, $\tilde B_\phi(\Omega_2,f_a)$, $\tilde B_\theta(\Omega_3,f_a)$, and $\tilde B_\phi(\Omega_3,f_a)$.  The expectation of $\vec X$ is now given by
\begin{widetext}
\begin{equation}\label{eq:muaxion}
    \langle\vec X\rangle=ic^*g_{a\gamma}RT\sqrt{\frac{\rho_{\mr{DM}}}2}\sum_{\ell m}\frac{(\ell+1)C_{\ell m}}{\ell(\ell+1)-(2\pi f_aR)^2}
    \begin{pmatrix}\Phi^\theta_{\ell m}(\Omega_1)\\
    \Phi^\phi_{\ell m}(\Omega_1)\\
    \Phi^\theta_{\ell m}(\Omega_2)\\
    \Phi^\phi_{\ell m}(\Omega_2)\\
    \Phi^\theta_{\ell m}(\Omega_3)\\
    \Phi^\phi_{\ell m}(\Omega_3)\end{pmatrix}
    \equiv c^*g_{a\gamma}\vec\mu,
\end{equation}
\end{widetext}
where $\Phi^\theta_{\ell m}$ and $\Phi^\phi_{\ell m}$ denote the $\THETA$-component and $\PHI$-components of the VSH $\bm\Phi_{\ell m}$, and
\begin{equation}
    c=\frac{\sqrt2\pi f_aa_0}{\sqrt{\rho_{\mr{DM}}}}.
\end{equation}

\begin{figure*}[h!]
\centering
\includegraphics[width =0.8 \textwidth]{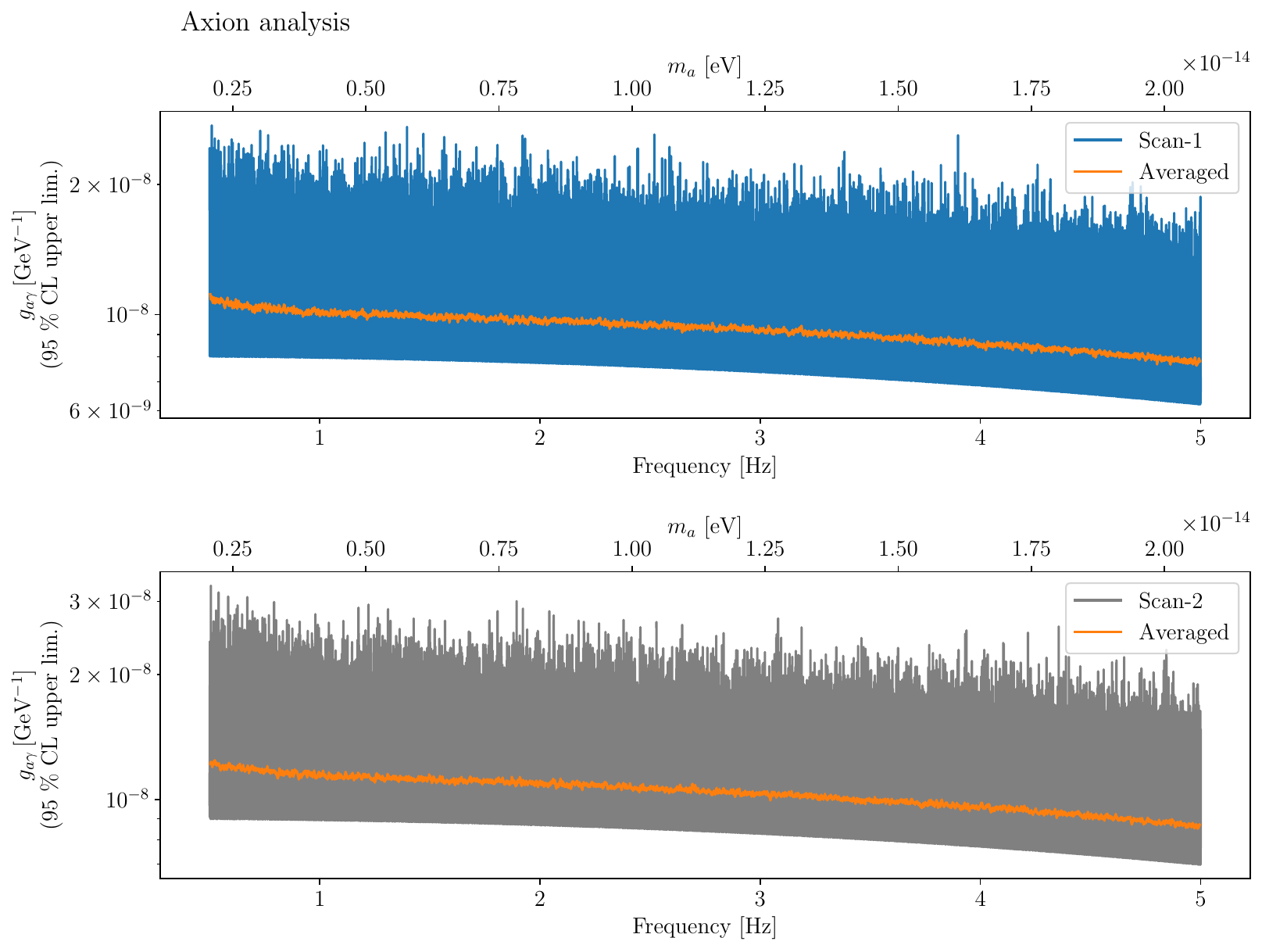}
\caption{95\% CL upper limit on $g_{a\gamma}$ for Scan-1 and Scan-2. The orange traces on both plots show smoothed versions of the limits obtained by averaging over 100 adjacent frequency
bins.
}
\label{fig:AxionBound}
\end{figure*}

The covariance matrix $\Sigma$ of $\vec X$ can again be determined by averaging over independent frequencies, as in \eqref{Sigma} [except that $\Sigma$ will now be a $6\times6$ matrix].  If we define $\vec Y$ and $\vec\nu$ as in \eqref[s]{Yhpdm} and (\ref{eq:nuhpdm}) [without the $m$ index], and further define
\begin{align}
    s&=|\vec\nu|,\\
    z&=\frac{\vec\nu^\dagger\vec Y}s,
\end{align}
we can write the likelihood function for the axion signal as
\begin{equation}\label{eq:axion_likelihood}
    -\ln\LL(g_{a\gamma},c|z)=\left|z-g_{a\gamma}c^*s\right|^2.
\end{equation}
Again marginalizing over $c$ (which we take to have a Gaussian distribution with $\langle|c|^2\rangle=1$), and utilizing a Jeffreys prior for $g_{a\gamma}$, we arrive at the posterior distribution
\begin{equation}\label{eq:axposterior}
    p(g_{a\gamma}|z)=\frac{|z|^2}{1-e^{-|z|^2}}\cdot \frac{2g_{a\gamma}s^2}{(1+g_{a\gamma}^2s^2)^2}\exp\left(-\frac{|z|^2}{1+g_{a\gamma}^2s^2}\right).
\end{equation}
Note that \eqref{axposterior} is properly normalized, which is possible because its integral over $g_{a\gamma}$ can be taken analytically.  The 95\% credible limit $\hat g_{a\gamma}$ can then be defined, as in \eqref{epsbound}.  In this case, we can solve for it analytically to find
\begin{equation}
    \hat g_{a\gamma}=\frac1s\sqrt{-\frac{|z|^2}{\log\left(0.95+0.05e^{-|z|^2}\right)}-1}.
\end{equation}
\figref{AxionBound} shows the resulting limit as a function of frequency, for both Scan-1 and Scan-2.  Note that the lower edge of the limit appears as a smooth curve.  This is due to the fact that $\hat g_{a\gamma}\rightarrow4.36/s$ in the limit $z\rightarrow0$.  Therefore, even when the measured data at a particular frequency becomes arbitrarily small (compared to the estimated noise level), the limit on $g_{a\gamma}$ asymptotes to a finite floor.\footnote{This floor exhibits a slight frequency dependence because of the $f_a$-dependence in \eqref{muaxion}.}

\begin{figure*}[h!]
\centering
\includegraphics[width = 0.8\textwidth]{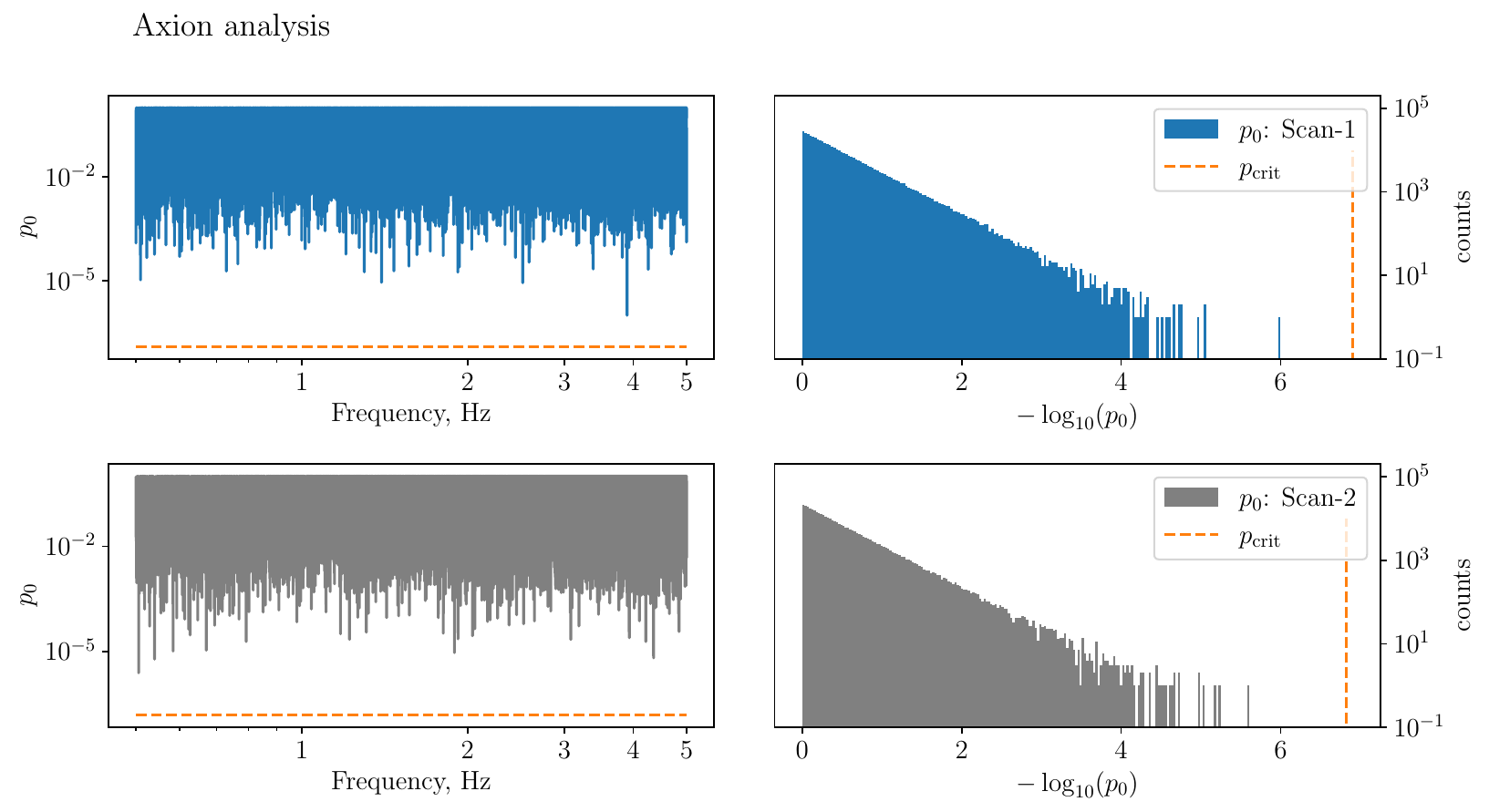}
\caption{The local $p_0$-values for each of the $N=414572$ frequency bins analyzed in Scan-1 (top), and each of the $N=340291$ frequency bins searched in Scan-2 (bottom). $p_\text{crit}$, the threshold value for declaring a candidate signal at $95 \%$ confidence is shown as the dotted line on each of the plots. The right panel shows a histogram of all the $p_0$-values for each scan. Signal candidates would appear as outliers to the right of the threshold.
} 
\label{fig:AxionPvalMain}
\end{figure*}

As in the HPDM case, we evaluate our data at each frequency in order to determine whether there is evidence for a significant DM signal.  We may compute the local $p$-value at a particular frequency under the null hypothesis ($g_{a\gamma}=0$) as
\begin{equation}
    p_0=1-F_{\chi^2(2)}(2|z|^2).
\end{equation}
(The $\chi^2$-distribution only has two degrees of freedom now, since the likelihood in \eqref{axion_likelihood} only has one $z$ variable.)  \figref{AxionPvalMain} shows these $p$-values as a function of frequency for both Scan-1 and Scan-2, along with the threshold value $p_\text{crit}$, as defined in \eqref{pcrit}.  Neither scan shows any significant signal candidates, and so we again conclude that our data contains no evidence for axion dark matter in the $0.5\,\text{Hz}\leq f_a\leq5\,\text{Hz}$ range.

\subsection{Error Budget}
\label{sec:errorbudget}
The results of this science run and analysis are summarized in Figs. \ref{fig:HPDMbound} and \ref{fig:AxionBound}. They show upper limits on $\varepsilon$, the HPDM kinetic mixing parameter, and on $g_{a\gamma}$, the axion--photon coupling constant, respectively. Below, we discuss the impact of uncertainties in the signal model and experimental conditions on the quoted limits.

\subsubsection{Signal model uncertainty}
\label{sec:model_uncertainty}
The signals in \eqref[s]{Bhpdm} and (\ref{eq:Baxion}) assume a simplified model of Earth and the ionosphere, where both are treated as spherical perfect conductors.  In Ref.~\cite{Fedd21concept}, it is argued that this model holds to a high degree of accuracy in the frequency range relevant to this work.  In particular, both Earth's crust~\cite{atlas} and the ionosphere~\cite{Takeda:1985hcf,GM118} achieve conductivities of at least $10^{-4}$\,S/m at certain depths/heights, which translate to skin depths of $\sim50\,$km for frequencies $f\sim1\,$Hz.  Given that the only relevant length scale appearing in \eqref[s]{Bhpdm} and (\ref{eq:Baxion}) is the radius of Earth $R\sim6000\,$km, finite-conductivity effects only modify the geometry of the system at the percent level.  In the absence of resonances, we conclude that the signal should also only be affected at the percent level.

Close examination of \eqref[s]{Bhpdm} and (\ref{eq:Baxion}), however, reveals that our model predicts resonances in the signal at $mR=\sqrt{\ell(\ell+1)}$ (for $\ell=1$ in the HPDM case, and $\ell\geq1$ in the axion case).  These are the Schumann resonances of the Earth-ionosphere cavity~\cite{sentman2017schumann,Rodriguez-Camacho}.  Our simplified spherical model predicts the first of these resonances to occur at $\sim 10$\,Hz, but the central frequency of this resonance has been measured to be $\sim 8$\,Hz~\cite{sentman2017schumann}, indicating that our spherical model does not accurately account for environmental effects on the Schumann resonances.  Moreover, since the signal nominally diverges at the Schumann resonances, small deviations in their central frequency can have a large impact on the predicted signal.  For this reason, we limit our analysis to $f\leq5$\,Hz, in order to remain below the measured Schumann resonances.

We note that the measured width of the Schumann resonances can, however, be quite large at certain times.  In the summer, during the day, the first Schumann resonance can reach widths as large as $\sim4$\,Hz~\cite{Rodriguez-Camacho}.  The upper end of our frequency range may therefore be mildly affected by the first Schumann resonance for certain portions of the runtime.  Such an effect would result in a slight \emph{enhancement} of the signal, beyond what our model predicted.  Therefore our exclusion limits are still conservative.  In principle, the effect of the Schumann resonances may, however, invalidate our signal-candidate rejection procedure.  This is because environmental effects could influence each station differently, meaning we cannot accurately characterize the spatial dependence of a true signal.  To this point, we simply note that our only signal candidates presented at the end of \secref{HPDM} were at $f\sim0.5,0.75\,\mathrm{Hz}$, and so are too low frequency to be affected by the Schumann resonances. We therefore conclude that both our exclusion analysis and our candidate rejection are robust to signal-model uncertainties. 

\subsubsection{Sensor orientation}
As discussed in section \ref{sec:expt-details}, we orient the magnetometers at each site such that the N-S, and E-W axes of each sensor lie in a horizontal plane with North indicating True (i.e., geographic) north, and the Normal (Up-Down) axis lies in the direction of the local force of gravity. We are able to achieve this orientation with repeatability $\lesssim 1^\circ$. By adjusting the orientation of the sensor in the analysis, we estimate that the impact of such an orientation error is to change the $\varepsilon$ and $g_{a\gamma}$ upper limits by $\lesssim 1 \%$.

\subsubsection{Calibration drift}

A temperature-dependent sensor calibration will lead to systematic errors in magnetic-field measurements.  As shown in Fig.~\ref{Fig:TimeSeries}, we observed that the temperature swing over the course of a day at the Hayward station was significantly greater than that in the Oberlin and Lewisburg stations.  In that period, we recorded changes in the dc magnetic-field readings that tracked the sensor temperature of up to $10 \%$ for the Hayward station, and less than $3\%$ for the Oberlin and Lewisburg stations. In the 0.5--5.0 Hz band, we estimate the impact of a possibly drifting calibration on the upper limits of $\varepsilon$ and $g_{a\gamma}$ by running analyses 
where we independently scaled the sensor readings by up to 10 percent for Hayward, and up to 3 percent for the other two stations. We then determined the resulting limits, concluding that a drifting calibration of the magnitude we observed would change the limits on $\varepsilon$ and $g_{a\gamma}$  by $\lesssim 3\%$.

\subsubsection{Timing synchronization}
As discussed in Sec.~\ref{sec:expt-details}, the magnetic-field measurements were digitized at 160 samples per second. An on-sensor real-time clock ensured sample-to-sample timing to better than 1 ppm and a GPS-referenced computer clock provided the absolute time reference for the time stamps. The absolute timing accuracy between sensors was limited to $\sim 100 \,\mathrm{ms}$ due to latencies in the steering of the  DAQ clock to GPS. This can be significantly improved. However, such an accuracy was adequate for an analysis covering the 0.5 to 5 Hz window. We estimate the systematic on the derived limits due to this error to be neglible.

\section{Future Directions}
\label{sec:future}

The current experiment is limited by the sensitivity of the magnetometers, rather than by the geomagnetic noise, and our model only accurately describes signals at frequencies below $\approx 5$~Hz. In the next generation of the experiment, we plan to use more sensitive magnetometers to reach the limit imposed by geomagnetic noise.
In addition, we propose to employ a novel experimental geometry to avoid model uncertainties in interpretation of our data.

At frequencies $\gtrsim 5$~Hz, the DM-induced magnetic field signal becomes sensitive to the details of Earth's atmosphere, which would require more careful modelling than that needed for the lower-frequency analysis presented in this paper.
In order to be sensitive to higher-mass ALPs and hidden photons, we are investigating the prospect of measuring spatial derivatives of the magnetic field. 
By measuring components of the magnetic field across multiple stations which are positioned $\lesssim 1~{\rm km}$ from one another, it is possible to compute the numerical derivatives of ${\bm B}$, and particularly components of $\nabla \times {\bm B}$. 
In the envisioned measurement scheme \cite{bloch2023curl}, we do not expect to have significant local electric currents, so the modified Amp\`ere--Maxwell law describing the sought-after effect of DM fields is 
\begin{equation}
\nabla \times {\bm B}-\partial_t{\bm E}={\bm J}_{\rm eff},
\end{equation}
where ${\bm J}_{\rm eff}$ encapsulates the effect of the dark matter [see Eqs.~(\ref{eq:hidden-photon-induced-current}) and (\ref{eq:axion-induced-current})]. 
Since ${\bm E}$ is negligible in directions tangent to the ground, a measurement of $\nabla\times\bm B$ in a tangent direction gives a direct measurement of the dark matter, which is insensitive to the atmospheric boundary conditions.
Moreover, we expect this scheme to reduce sensitivity to geomagnetic noise, as physical geomagnetic fields in the lower atmosphere should have $(\nabla\times\bm B)_\parallel=\bm J_\parallel=0$.
However, it is important to note that, unlike the low-frequency measurements whose signal is enhanced by the full radius of Earth, the effective enhancement here would only be the separation between stations. 
SNIPE Hunt is currently carrying out an investigation of the expected background and signal, while simultaneously taking steps to perform a search based on this new methodology. 





\section{Conclusions}

In this work, we reported on a search for axion and hidden-photon dark matter using a network of unshielded vector magnetoresistive (VMR) magnetometers located in relatively quiet magnetic environments, in wilderness areas far from anthropogenic magnetic noise.
The magnetic signal pattern targeted by our search could, in principle, be generated by the interaction of axion or hidden photon dark matter with Earth, which can act as a transducer to convert the dark matter into oscillating magnetic fields as described in Refs.~\cite{Fedd21concept,Fed21search,Arza22}.
Analysis of the data acquired over the course of approximately three days in July 2022 revealed no evidence of a persistent oscillating magnetic field matching the expected characteristics of a dark-matter-induced signal.
Consequently, we set upper limits on the kinetic-mixing parameter $\varepsilon$ for hidden-photon dark matter and on the axion--photon coupling constant $g_{a\gamma}$.

\begin{figure*}[h!]
\centering
\includegraphics[width = 0.8 \textwidth]{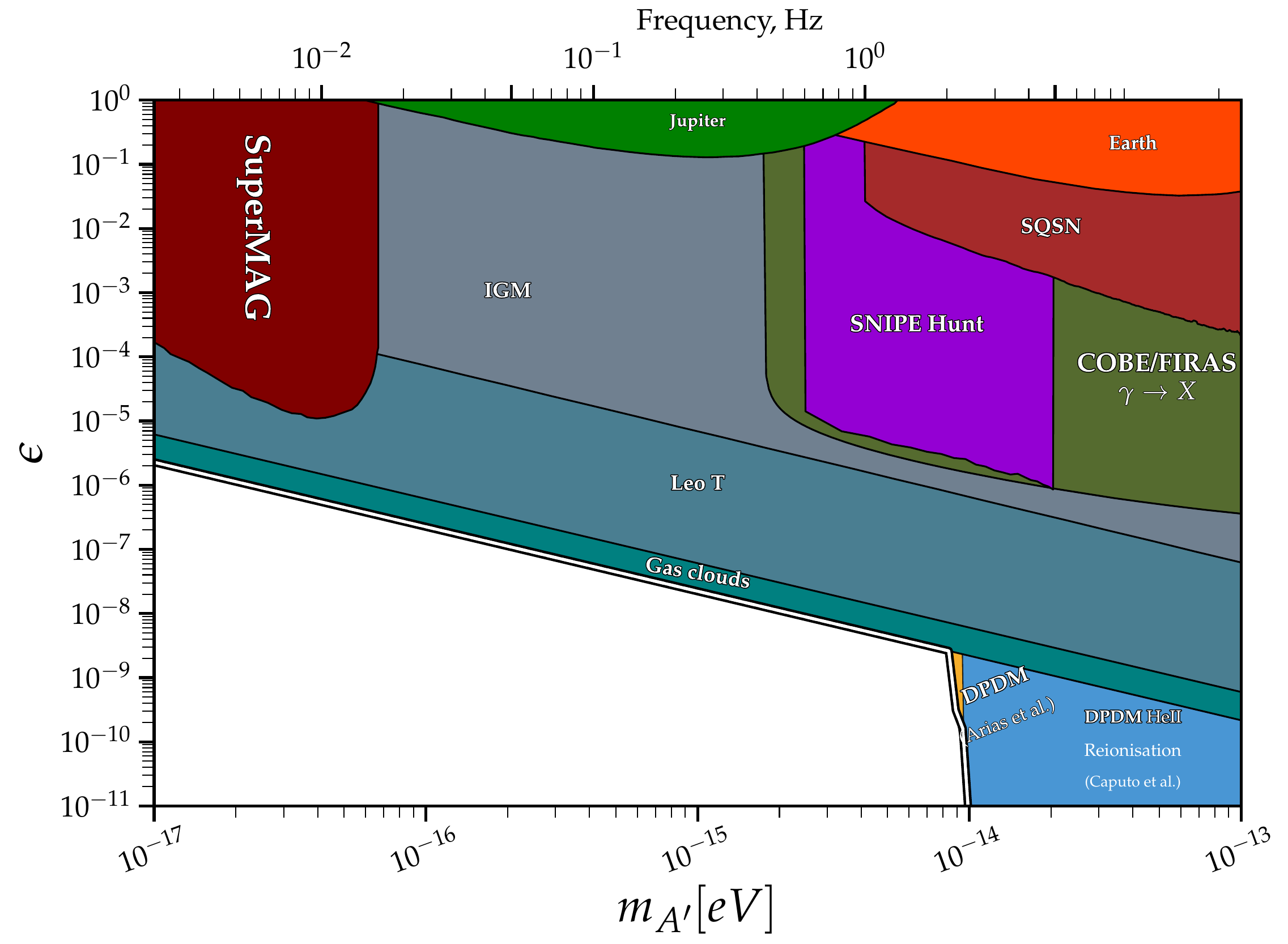}
\caption{Constraints on the hidden-photon kinetic-mixing parameter $\varepsilon$ as a function of hidden-photon mass $m_{A'}$. The plot was created based on Refs.~\cite{cohare_git} and~\cite{Caputo:2021eaa} and includes the SuperMAG limit~\cite{Fedd21concept,Fed21search} and the recent measurement using a network of magnetometers in meter-scale shielded rooms \cite{jiang2023search}, which we denote the ``Synchronized Quantum Sensor Network'' (SQSN). The results reported in Refs.~\cite{Fedd21concept,Fed21search,jiang2023search} are the only other laboratory measurements in this mass range. In addition to the laboratory constraints, the plot also shows various astrophysical bounds, including the geomagnetic limit obtained from satellite measurements of the Earth's magnetic field~\cite{DarkPhoton_Earth:1994}, the hidden-photon limits from magnetic-field measurements in Jupiter's magnetosphere~\cite{dph_Jupiter:2021}, limits from cold gas clouds at the Milky Way center~\cite{dark_photon_gas_cloud:2019prd}, heating of the ionized interstellar medium in the galaxy from hidden photons~\cite{dph_IGM_Dubovsky:2015jcap}, and the limit on heating/cooling due to DM in the Leo T dwarf galaxy~\cite{dph_LeoT:2021prd}. Cosmological bounds on hidden photons from COBE/FIRAS data estimated from potential hidden-photon interactions with plasmas in the universe are from Refs.~\cite{dph_cobe_firas_caputo:2021prl,dph_cobe_firas_witte:2020prg,Paola_Arias:2012jcap}.  Finally, the figure also displays cosmological/astrophysical bounds on hidden photons from He~II reionization~\cite{dph_HeII:2020prl}.
}
\label{fig:hidden-photon-parameter-exclusion-plot}
\end{figure*}

\begin{figure*}[h!]
\centering
\includegraphics[width = 0.8 \textwidth]{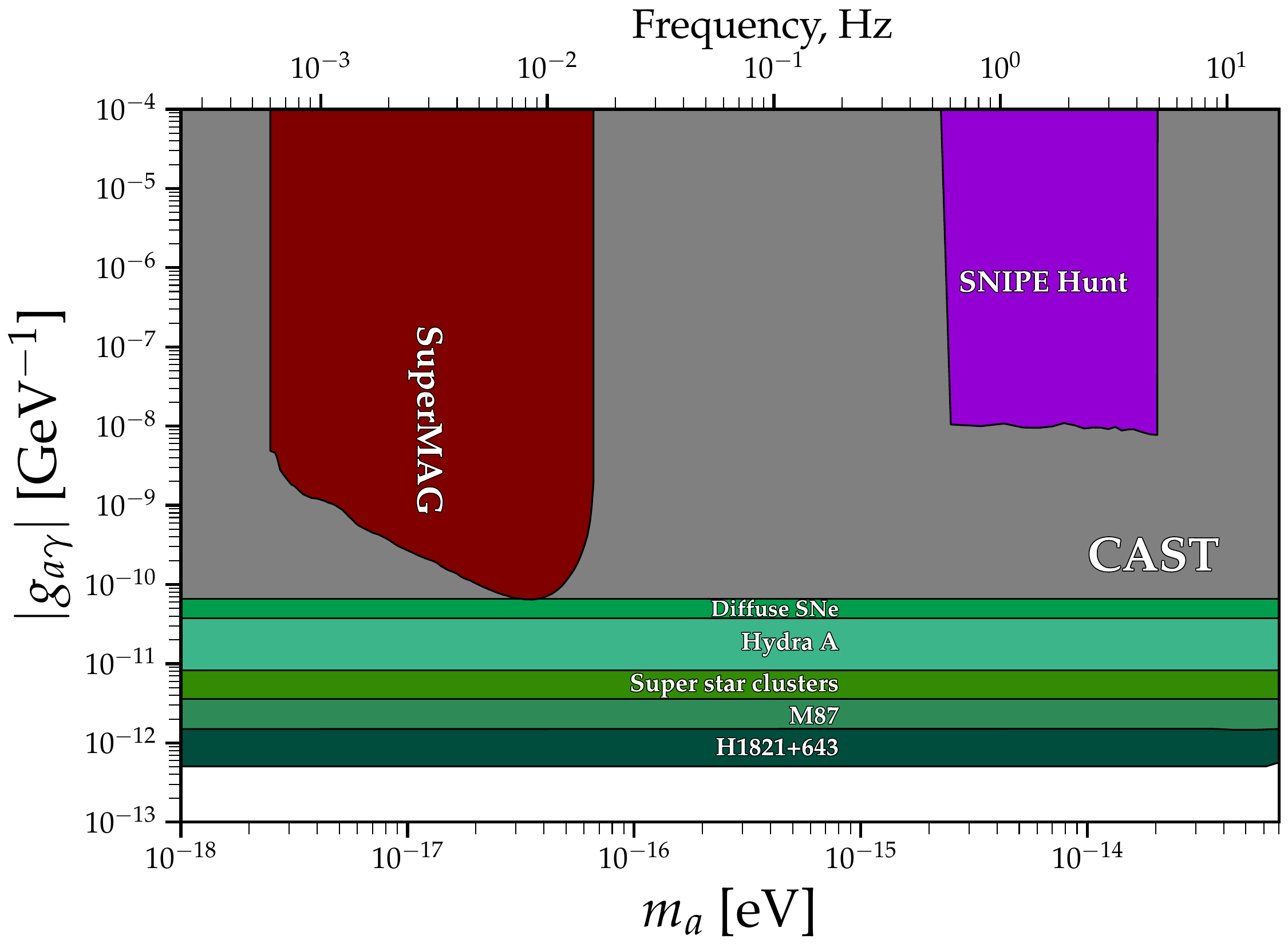}
\caption{Constraints on the axion--photon coupling constant parameter $g_{a\gamma}$ as a function of axion mass $m_a$. The plot was created based on Ref.~\cite{cohare_git}, and includes the relevant experimental bounds based on the SuperMAG analysis~\cite{Arza22} in maroon and the CAST result~\cite{CAST:2017} in grey. Additionally, the plot displays astrophysical limits on the axion--photon interaction, represented in various shades of green, including (Diffuse SNe)~\cite{SNe:2022prl}, (Hydra A)~\cite{hydra_a_Wouters_2013}, (Super star clusters)~\cite{Superstar_cluster:2020prl}, M87~\cite{M87_Marsh_2017}, and (H1821+643)~\cite{H1821_643:2021mnra}. }
\label{fig:axion-parameter-exclusion-plot}
\end{figure*}

Figure~\ref{fig:hidden-photon-parameter-exclusion-plot} displays constraints on $\varepsilon$ as a function of hidden-photon mass $m_{A'}$ obtained in our experiment as well as those from other experiments \cite{Fed21search,jiang2023search}, derived from planetary science \cite{DarkPhoton_Earth:1994,dph_Jupiter:2021}, and based on astrophysical observations \cite{dark_photon_gas_cloud:2019prd,dph_IGM_Dubovsky:2015jcap,dph_LeoT:2021prd,dph_cobe_firas_caputo:2021prl,dph_cobe_firas_witte:2020prg,Paola_Arias:2012jcap,dph_HeII:2020prl}.
We note that, in the studied frequency range, the results of the SNIPE Hunt experiment are the most stringent experimental bounds, and can be regarded as complementary to the more severe observational constraints.
Fig.~\ref{fig:axion-parameter-exclusion-plot} shows bounds on the axion--photon coupling constant parameter $g_{a\gamma}$ as a function of axion mass $m_a$.

We are actively pursuing further measurements based on this concept, but instead using induction-coil magnetometers \cite{votis2018new,poliakov2017range,hospodarsky2016spaced}. 
We anticipate an improvement in sensitivity to dark-matter-induced magnetic signals of several orders of magnitude.
Furthermore, as discussed in Sec.~\ref{sec:future}, we will use local multi-sensor arrays to measure the curl of the local magnetic field at the various sites and thereby extend the frequency range probed up to about a kHz.

\acknowledgments

The Oberlin group thank Michael Miller for his work on the construction of the sensor mount for the Oberlin station. This work was supported by the U.S.\ National Science Foundation under grants PHY-2110370, PHY-2110385, and PHY-2110388. S.K.~and M.A.F.~are supported by the U.S. Department of Energy, Office of Science, National Quantum Information Science Research
Centers, Superconducting Quantum Materials and Systems Center (SQMS) under contract number DE-AC02-07CH11359.
M.A.F.~is also supported by the Simons Investigator Award No.~827042. S.K.~and M.A.F.~thank the Aspen Center for Physics for hospitality during the final stages of this work, supported by NSF Grant No.~PHY-2210452.

\clearpage
\bibliography{SNIPERefs}

\begin{thebibliography}{91}%
\makeatletter
\providecommand \@ifxundefined [1]{%
 \@ifx{#1\undefined}
}%
\providecommand \@ifnum [1]{%
 \ifnum #1\expandafter \@firstoftwo
 \else \expandafter \@secondoftwo
 \fi
}%
\providecommand \@ifx [1]{%
 \ifx #1\expandafter \@firstoftwo
 \else \expandafter \@secondoftwo
 \fi
}%
\providecommand \natexlab [1]{#1}%
\providecommand \enquote  [1]{``#1''}%
\providecommand \bibnamefont  [1]{#1}%
\providecommand \bibfnamefont [1]{#1}%
\providecommand \citenamefont [1]{#1}%
\providecommand \href@noop [0]{\@secondoftwo}%
\providecommand \href [0]{\begingroup \@sanitize@url \@href}%
\providecommand \@href[1]{\@@startlink{#1}\@@href}%
\providecommand \@@href[1]{\endgroup#1\@@endlink}%
\providecommand \@sanitize@url [0]{\catcode `\\12\catcode `\$12\catcode
  `\&12\catcode `\#12\catcode `\^12\catcode `\_12\catcode `\%12\relax}%
\providecommand \@@startlink[1]{}%
\providecommand \@@endlink[0]{}%
\providecommand \url  [0]{\begingroup\@sanitize@url \@url }%
\providecommand \@url [1]{\endgroup\@href {#1}{\urlprefix }}%
\providecommand \urlprefix  [0]{URL }%
\providecommand \Eprint [0]{\href }%
\providecommand \doibase [0]{https://doi.org/}%
\providecommand \selectlanguage [0]{\@gobble}%
\providecommand \bibinfo  [0]{\@secondoftwo}%
\providecommand \bibfield  [0]{\@secondoftwo}%
\providecommand \translation [1]{[#1]}%
\providecommand \BibitemOpen [0]{}%
\providecommand \bibitemStop [0]{}%
\providecommand \bibitemNoStop [0]{.\EOS\space}%
\providecommand \EOS [0]{\spacefactor3000\relax}%
\providecommand \BibitemShut  [1]{\csname bibitem#1\endcsname}%
\let\auto@bib@innerbib\@empty
\bibitem [{\citenamefont {Jackson~Kimball}\ and\ \citenamefont {van
  Bibber}(2022)}]{kimball2022search}%
  \BibitemOpen
  \bibfield  {author} {\bibinfo {author} {\bibfnamefont {D.~F.}\ \bibnamefont
  {Jackson~Kimball}}\ and\ \bibinfo {author} {\bibfnamefont {K.}~\bibnamefont
  {van Bibber}},\ }\href {https://doi.org/10.1007/978-3-030-95852-7} {\emph
  {\bibinfo {title} {{The Search for Ultralight Bosonic Dark Matter}}}}\
  (\bibinfo  {publisher} {Springer},\ \bibinfo {year} {2022})\BibitemShut
  {NoStop}%
\bibitem [{\citenamefont {Graham}\ \emph
  {et~al.}(2015{\natexlab{a}})\citenamefont {Graham}, \citenamefont
  {Irastorza}, \citenamefont {Lamoreaux}, \citenamefont {Lindner},\ and\
  \citenamefont {van Bibber}}]{graham2015experimental}%
  \BibitemOpen
  \bibfield  {author} {\bibinfo {author} {\bibfnamefont {P.~W.}\ \bibnamefont
  {Graham}}, \bibinfo {author} {\bibfnamefont {I.~G.}\ \bibnamefont
  {Irastorza}}, \bibinfo {author} {\bibfnamefont {S.~K.}\ \bibnamefont
  {Lamoreaux}}, \bibinfo {author} {\bibfnamefont {A.}~\bibnamefont {Lindner}},\
  and\ \bibinfo {author} {\bibfnamefont {K.~A.}\ \bibnamefont {van Bibber}},\
  }\bibfield  {title} {\bibinfo {title} {Experimental searches for the axion
  and axion-like particles},\ }\href
  {https://doi.org/10.1146/annurev-nucl-102014-022120} {\bibfield  {journal}
  {\bibinfo  {journal} {Annu. Rev. Nucl. Part. Sci.}\ }\textbf {\bibinfo
  {volume} {65}},\ \bibinfo {pages} {485} (\bibinfo {year}
  {2015}{\natexlab{a}})}\BibitemShut {NoStop}%
\bibitem [{\citenamefont {Arias}\ \emph
  {et~al.}(2012{\natexlab{a}})\citenamefont {Arias}, \citenamefont {Cadamuro},
  \citenamefont {Goodsell}, \citenamefont {Jaeckel}, \citenamefont {Redondo},\
  and\ \citenamefont {Ringwald}}]{Arias:2012az}%
  \BibitemOpen
  \bibfield  {author} {\bibinfo {author} {\bibfnamefont {P.}~\bibnamefont
  {Arias}}, \bibinfo {author} {\bibfnamefont {D.}~\bibnamefont {Cadamuro}},
  \bibinfo {author} {\bibfnamefont {M.}~\bibnamefont {Goodsell}}, \bibinfo
  {author} {\bibfnamefont {J.}~\bibnamefont {Jaeckel}}, \bibinfo {author}
  {\bibfnamefont {J.}~\bibnamefont {Redondo}},\ and\ \bibinfo {author}
  {\bibfnamefont {A.}~\bibnamefont {Ringwald}},\ }\bibfield  {title} {\bibinfo
  {title} {{WISPy Cold Dark Matter}},\ }\href
  {https://doi.org/10.1088/1475-7516/2012/06/013} {\bibfield  {journal}
  {\bibinfo  {journal} {JCAP}\ }\textbf {\bibinfo {volume} {06}},\ \bibinfo
  {pages} {013}},\ \Eprint {https://arxiv.org/abs/1201.5902} {arXiv:1201.5902
  [hep-ph]} \BibitemShut {NoStop}%
\bibitem [{\citenamefont {Braaten}\ and\ \citenamefont
  {Zhang}(2019)}]{braaten2019colloquium}%
  \BibitemOpen
  \bibfield  {author} {\bibinfo {author} {\bibfnamefont {E.}~\bibnamefont
  {Braaten}}\ and\ \bibinfo {author} {\bibfnamefont {H.}~\bibnamefont
  {Zhang}},\ }\bibfield  {title} {\bibinfo {title} {{Colloquium: The physics of
  axion stars}},\ }\href {https://doi.org/10.1103/RevModPhys.91.041002}
  {\bibfield  {journal} {\bibinfo  {journal} {Rev. Mod. Phys.}\ }\textbf
  {\bibinfo {volume} {91}},\ \bibinfo {pages} {041002} (\bibinfo {year}
  {2019})}\BibitemShut {NoStop}%
\bibitem [{\citenamefont {Freese}\ \emph {et~al.}(2013)\citenamefont {Freese},
  \citenamefont {Lisanti},\ and\ \citenamefont
  {Savage}}]{freese2013colloquium}%
  \BibitemOpen
  \bibfield  {author} {\bibinfo {author} {\bibfnamefont {K.}~\bibnamefont
  {Freese}}, \bibinfo {author} {\bibfnamefont {M.}~\bibnamefont {Lisanti}},\
  and\ \bibinfo {author} {\bibfnamefont {C.}~\bibnamefont {Savage}},\
  }\bibfield  {title} {\bibinfo {title} {Colloquium: Annual modulation of dark
  matter},\ }\href {https://doi.org/10.1103/RevModPhys.85.1561} {\bibfield
  {journal} {\bibinfo  {journal} {Rev. Mod. Phys.}\ }\textbf {\bibinfo {volume}
  {85}},\ \bibinfo {pages} {1561} (\bibinfo {year} {2013})}\BibitemShut
  {NoStop}%
\bibitem [{\citenamefont {Pillepich}\ \emph {et~al.}(2014)\citenamefont
  {Pillepich}, \citenamefont {Kuhlen}, \citenamefont {Guedes},\ and\
  \citenamefont {Madau}}]{pillepich2014distribution}%
  \BibitemOpen
  \bibfield  {author} {\bibinfo {author} {\bibfnamefont {A.}~\bibnamefont
  {Pillepich}}, \bibinfo {author} {\bibfnamefont {M.}~\bibnamefont {Kuhlen}},
  \bibinfo {author} {\bibfnamefont {J.}~\bibnamefont {Guedes}},\ and\ \bibinfo
  {author} {\bibfnamefont {P.}~\bibnamefont {Madau}},\ }\bibfield  {title}
  {\bibinfo {title} {{The distribution of dark matter in the Milky Way’s
  disk}},\ }\href {https://doi.org/10.1088/0004-637X/784/2/161} {\bibfield
  {journal} {\bibinfo  {journal} {Astrophys. J.}\ }\textbf {\bibinfo {volume}
  {784}},\ \bibinfo {pages} {161} (\bibinfo {year} {2014})}\BibitemShut
  {NoStop}%
\bibitem [{\citenamefont {Evans}\ \emph {et~al.}(2019)\citenamefont {Evans},
  \citenamefont {O’Hare},\ and\ \citenamefont
  {McCabe}}]{evans2019refinement}%
  \BibitemOpen
  \bibfield  {author} {\bibinfo {author} {\bibfnamefont {N.~W.}\ \bibnamefont
  {Evans}}, \bibinfo {author} {\bibfnamefont {C.~A.}\ \bibnamefont
  {O’Hare}},\ and\ \bibinfo {author} {\bibfnamefont {C.}~\bibnamefont
  {McCabe}},\ }\bibfield  {title} {\bibinfo {title} {{Refinement of the
  standard halo model for dark matter searches in light of the Gaia Sausage}},\
  }\href {https://doi.org/10.1103/PhysRevD.99.023012} {\bibfield  {journal}
  {\bibinfo  {journal} {Phys. Rev. D}\ }\textbf {\bibinfo {volume} {99}},\
  \bibinfo {pages} {023012} (\bibinfo {year} {2019})}\BibitemShut {NoStop}%
\bibitem [{\citenamefont {Hui}\ \emph {et~al.}(2017)\citenamefont {Hui},
  \citenamefont {Ostriker}, \citenamefont {Tremaine},\ and\ \citenamefont
  {Witten}}]{hui2017ultralight}%
  \BibitemOpen
  \bibfield  {author} {\bibinfo {author} {\bibfnamefont {L.}~\bibnamefont
  {Hui}}, \bibinfo {author} {\bibfnamefont {J.~P.}\ \bibnamefont {Ostriker}},
  \bibinfo {author} {\bibfnamefont {S.}~\bibnamefont {Tremaine}},\ and\
  \bibinfo {author} {\bibfnamefont {E.}~\bibnamefont {Witten}},\ }\bibfield
  {title} {\bibinfo {title} {Ultralight scalars as cosmological dark matter},\
  }\href {https://doi.org/10.1103/PhysRevD.95.043541} {\bibfield  {journal}
  {\bibinfo  {journal} {Phys. Rev. D}\ }\textbf {\bibinfo {volume} {95}},\
  \bibinfo {pages} {043541} (\bibinfo {year} {2017})}\BibitemShut {NoStop}%
\bibitem [{\citenamefont {Foster}\ \emph {et~al.}(2018)\citenamefont {Foster},
  \citenamefont {Rodd},\ and\ \citenamefont {Safdi}}]{foster2018revealing}%
  \BibitemOpen
  \bibfield  {author} {\bibinfo {author} {\bibfnamefont {J.~W.}\ \bibnamefont
  {Foster}}, \bibinfo {author} {\bibfnamefont {N.~L.}\ \bibnamefont {Rodd}},\
  and\ \bibinfo {author} {\bibfnamefont {B.~R.}\ \bibnamefont {Safdi}},\
  }\bibfield  {title} {\bibinfo {title} {Revealing the dark matter halo with
  axion direct detection},\ }\href {https://doi.org/10.1103/PhysRevD.97.123006}
  {\bibfield  {journal} {\bibinfo  {journal} {Phys. Rev. D}\ }\textbf {\bibinfo
  {volume} {97}},\ \bibinfo {pages} {123006} (\bibinfo {year}
  {2018})}\BibitemShut {NoStop}%
\bibitem [{\citenamefont {Lin}\ \emph {et~al.}(2018)\citenamefont {Lin},
  \citenamefont {Schive}, \citenamefont {Wong},\ and\ \citenamefont
  {Chiueh}}]{lin2018self}%
  \BibitemOpen
  \bibfield  {author} {\bibinfo {author} {\bibfnamefont {S.-C.}\ \bibnamefont
  {Lin}}, \bibinfo {author} {\bibfnamefont {H.-Y.}\ \bibnamefont {Schive}},
  \bibinfo {author} {\bibfnamefont {S.-K.}\ \bibnamefont {Wong}},\ and\
  \bibinfo {author} {\bibfnamefont {T.}~\bibnamefont {Chiueh}},\ }\bibfield
  {title} {\bibinfo {title} {Self-consistent construction of virialized wave
  dark matter halos},\ }\href {https://doi.org/10.1103/PhysRevD.97.103523}
  {\bibfield  {journal} {\bibinfo  {journal} {Phys. Rev. D}\ }\textbf {\bibinfo
  {volume} {97}},\ \bibinfo {pages} {103523} (\bibinfo {year}
  {2018})}\BibitemShut {NoStop}%
\bibitem [{\citenamefont {Centers}\ \emph {et~al.}(2021)\citenamefont
  {Centers}, \citenamefont {Blanchard}, \citenamefont {Conrad}, \citenamefont
  {Figueroa}, \citenamefont {Garcon}, \citenamefont {Gramolin}, \citenamefont
  {Jackson~Kimball}, \citenamefont {Lawson}, \citenamefont {Pelssers},
  \citenamefont {Smiga} \emph {et~al.}}]{centers2021stochastic}%
  \BibitemOpen
  \bibfield  {author} {\bibinfo {author} {\bibfnamefont {G.~P.}\ \bibnamefont
  {Centers}}, \bibinfo {author} {\bibfnamefont {J.~W.}\ \bibnamefont
  {Blanchard}}, \bibinfo {author} {\bibfnamefont {J.}~\bibnamefont {Conrad}},
  \bibinfo {author} {\bibfnamefont {N.~L.}\ \bibnamefont {Figueroa}}, \bibinfo
  {author} {\bibfnamefont {A.}~\bibnamefont {Garcon}}, \bibinfo {author}
  {\bibfnamefont {A.~V.}\ \bibnamefont {Gramolin}}, \bibinfo {author}
  {\bibfnamefont {D.~F.}\ \bibnamefont {Jackson~Kimball}}, \bibinfo {author}
  {\bibfnamefont {M.}~\bibnamefont {Lawson}}, \bibinfo {author} {\bibfnamefont
  {B.}~\bibnamefont {Pelssers}}, \bibinfo {author} {\bibfnamefont {J.~A.}\
  \bibnamefont {Smiga}}, \emph {et~al.},\ }\bibfield  {title} {\bibinfo {title}
  {Stochastic fluctuations of bosonic dark matter},\ }\href
  {https://doi.org/10.1038/s41467-021-27632-7} {\bibfield  {journal} {\bibinfo
  {journal} {Nature Comm.}\ }\textbf {\bibinfo {volume} {12}},\ \bibinfo
  {pages} {7321} (\bibinfo {year} {2021})}\BibitemShut {NoStop}%
\bibitem [{\citenamefont {Lisanti}\ \emph {et~al.}(2021)\citenamefont
  {Lisanti}, \citenamefont {Moschella},\ and\ \citenamefont
  {Terrano}}]{lisanti2021stochastic}%
  \BibitemOpen
  \bibfield  {author} {\bibinfo {author} {\bibfnamefont {M.}~\bibnamefont
  {Lisanti}}, \bibinfo {author} {\bibfnamefont {M.}~\bibnamefont {Moschella}},\
  and\ \bibinfo {author} {\bibfnamefont {W.}~\bibnamefont {Terrano}},\
  }\bibfield  {title} {\bibinfo {title} {Stochastic properties of ultralight
  scalar field gradients},\ }\href
  {https://doi.org/10.1103/PhysRevD.104.055037} {\bibfield  {journal} {\bibinfo
   {journal} {Phys. Rev. D}\ }\textbf {\bibinfo {volume} {104}},\ \bibinfo
  {pages} {055037} (\bibinfo {year} {2021})}\BibitemShut {NoStop}%
\bibitem [{\citenamefont {Graham}\ \emph
  {et~al.}(2016{\natexlab{a}})\citenamefont {Graham}, \citenamefont {Kaplan},
  \citenamefont {Mardon}, \citenamefont {Rajendran},\ and\ \citenamefont
  {Terrano}}]{graham2016dark}%
  \BibitemOpen
  \bibfield  {author} {\bibinfo {author} {\bibfnamefont {P.~W.}\ \bibnamefont
  {Graham}}, \bibinfo {author} {\bibfnamefont {D.~E.}\ \bibnamefont {Kaplan}},
  \bibinfo {author} {\bibfnamefont {J.}~\bibnamefont {Mardon}}, \bibinfo
  {author} {\bibfnamefont {S.}~\bibnamefont {Rajendran}},\ and\ \bibinfo
  {author} {\bibfnamefont {W.~A.}\ \bibnamefont {Terrano}},\ }\bibfield
  {title} {\bibinfo {title} {Dark matter direct detection with
  accelerometers},\ }\href {https://doi.org/10.1103/PhysRevD.93.075029}
  {\bibfield  {journal} {\bibinfo  {journal} {Phys. Rev. D}\ }\textbf {\bibinfo
  {volume} {93}},\ \bibinfo {pages} {075029} (\bibinfo {year}
  {2016}{\natexlab{a}})}\BibitemShut {NoStop}%
\bibitem [{\citenamefont {Safronova}\ \emph {et~al.}(2018)\citenamefont
  {Safronova}, \citenamefont {Budker}, \citenamefont {DeMille}, \citenamefont
  {Jackson~Kimball}, \citenamefont {Derevianko},\ and\ \citenamefont
  {Clark}}]{safronova2018search}%
  \BibitemOpen
  \bibfield  {author} {\bibinfo {author} {\bibfnamefont {M.}~\bibnamefont
  {Safronova}}, \bibinfo {author} {\bibfnamefont {D.}~\bibnamefont {Budker}},
  \bibinfo {author} {\bibfnamefont {D.}~\bibnamefont {DeMille}}, \bibinfo
  {author} {\bibfnamefont {D.~F.}\ \bibnamefont {Jackson~Kimball}}, \bibinfo
  {author} {\bibfnamefont {A.}~\bibnamefont {Derevianko}},\ and\ \bibinfo
  {author} {\bibfnamefont {C.~W.}\ \bibnamefont {Clark}},\ }\bibfield  {title}
  {\bibinfo {title} {Search for new physics with atoms and molecules},\ }\href
  {https://doi.org/10.1103/RevModPhys.90.025008} {\bibfield  {journal}
  {\bibinfo  {journal} {Rev. Mod. Phys.}\ }\textbf {\bibinfo {volume} {90}},\
  \bibinfo {pages} {025008} (\bibinfo {year} {2018})}\BibitemShut {NoStop}%
\bibitem [{\citenamefont {Sikivie}(1983)}]{sikivie1983experimental}%
  \BibitemOpen
  \bibfield  {author} {\bibinfo {author} {\bibfnamefont {P.}~\bibnamefont
  {Sikivie}},\ }\bibfield  {title} {\bibinfo {title} {{Experimental tests of
  the ``invisible'' axion}},\ }\href
  {https://doi.org/10.1103/PhysRevLett.51.1415} {\bibfield  {journal} {\bibinfo
   {journal} {Phys. Rev. Lett.}\ }\textbf {\bibinfo {volume} {51}},\ \bibinfo
  {pages} {1415} (\bibinfo {year} {1983})}\BibitemShut {NoStop}%
\bibitem [{\citenamefont {Asztalos}\ \emph {et~al.}(2010)\citenamefont
  {Asztalos}, \citenamefont {Carosi}, \citenamefont {Hagmann}, \citenamefont
  {Kinion}, \citenamefont {Van~Bibber}, \citenamefont {Hotz}, \citenamefont
  {Rosenberg}, \citenamefont {Rybka}, \citenamefont {Hoskins}, \citenamefont
  {Hwang} \emph {et~al.}}]{asztalos2010squid}%
  \BibitemOpen
  \bibfield  {author} {\bibinfo {author} {\bibfnamefont {S.~J.}\ \bibnamefont
  {Asztalos}}, \bibinfo {author} {\bibfnamefont {G.}~\bibnamefont {Carosi}},
  \bibinfo {author} {\bibfnamefont {C.}~\bibnamefont {Hagmann}}, \bibinfo
  {author} {\bibfnamefont {D.}~\bibnamefont {Kinion}}, \bibinfo {author}
  {\bibfnamefont {K.}~\bibnamefont {Van~Bibber}}, \bibinfo {author}
  {\bibfnamefont {M.}~\bibnamefont {Hotz}}, \bibinfo {author} {\bibfnamefont
  {L.}~\bibnamefont {Rosenberg}}, \bibinfo {author} {\bibfnamefont
  {G.}~\bibnamefont {Rybka}}, \bibinfo {author} {\bibfnamefont
  {J.}~\bibnamefont {Hoskins}}, \bibinfo {author} {\bibfnamefont
  {J.}~\bibnamefont {Hwang}}, \emph {et~al.},\ }\bibfield  {title} {\bibinfo
  {title} {{SQUID-based microwave cavity search for dark-matter axions}},\
  }\href {https://doi.org/10.1103/PhysRevLett.104.041301} {\bibfield  {journal}
  {\bibinfo  {journal} {Phys. Rev. Lett.}\ }\textbf {\bibinfo {volume} {104}},\
  \bibinfo {pages} {041301} (\bibinfo {year} {2010})}\BibitemShut {NoStop}%
\bibitem [{\citenamefont {Braine}\ \emph {et~al.}(2020)\citenamefont {Braine},
  \citenamefont {Cervantes}, \citenamefont {Crisosto}, \citenamefont {Du},
  \citenamefont {Kimes}, \citenamefont {Rosenberg}, \citenamefont {Rybka},
  \citenamefont {Yang}, \citenamefont {Bowring}, \citenamefont {Chou} \emph
  {et~al.}}]{braine2020extended}%
  \BibitemOpen
  \bibfield  {author} {\bibinfo {author} {\bibfnamefont {T.}~\bibnamefont
  {Braine}}, \bibinfo {author} {\bibfnamefont {R.}~\bibnamefont {Cervantes}},
  \bibinfo {author} {\bibfnamefont {N.}~\bibnamefont {Crisosto}}, \bibinfo
  {author} {\bibfnamefont {N.}~\bibnamefont {Du}}, \bibinfo {author}
  {\bibfnamefont {S.}~\bibnamefont {Kimes}}, \bibinfo {author} {\bibfnamefont
  {L.}~\bibnamefont {Rosenberg}}, \bibinfo {author} {\bibfnamefont
  {G.}~\bibnamefont {Rybka}}, \bibinfo {author} {\bibfnamefont
  {J.}~\bibnamefont {Yang}}, \bibinfo {author} {\bibfnamefont {D.}~\bibnamefont
  {Bowring}}, \bibinfo {author} {\bibfnamefont {A.}~\bibnamefont {Chou}}, \emph
  {et~al.},\ }\bibfield  {title} {\bibinfo {title} {Extended search for the
  invisible axion with the axion dark matter experiment},\ }\href
  {https://doi.org/10.1103/PhysRevLett.124.101303} {\bibfield  {journal}
  {\bibinfo  {journal} {Phys. Rev. Lett.}\ }\textbf {\bibinfo {volume} {124}},\
  \bibinfo {pages} {101303} (\bibinfo {year} {2020})}\BibitemShut {NoStop}%
\bibitem [{\citenamefont {Zhong}\ \emph {et~al.}(2018)\citenamefont {Zhong},
  \citenamefont {Al~Kenany}, \citenamefont {Backes}, \citenamefont {Brubaker},
  \citenamefont {Cahn}, \citenamefont {Carosi}, \citenamefont {Gurevich},
  \citenamefont {Kindel}, \citenamefont {Lamoreaux}, \citenamefont {Lehnert}
  \emph {et~al.}}]{zhong2018results}%
  \BibitemOpen
  \bibfield  {author} {\bibinfo {author} {\bibfnamefont {L.}~\bibnamefont
  {Zhong}}, \bibinfo {author} {\bibfnamefont {S.}~\bibnamefont {Al~Kenany}},
  \bibinfo {author} {\bibfnamefont {K.}~\bibnamefont {Backes}}, \bibinfo
  {author} {\bibfnamefont {B.}~\bibnamefont {Brubaker}}, \bibinfo {author}
  {\bibfnamefont {S.}~\bibnamefont {Cahn}}, \bibinfo {author} {\bibfnamefont
  {G.}~\bibnamefont {Carosi}}, \bibinfo {author} {\bibfnamefont
  {Y.}~\bibnamefont {Gurevich}}, \bibinfo {author} {\bibfnamefont
  {W.}~\bibnamefont {Kindel}}, \bibinfo {author} {\bibfnamefont
  {S.}~\bibnamefont {Lamoreaux}}, \bibinfo {author} {\bibfnamefont
  {K.}~\bibnamefont {Lehnert}}, \emph {et~al.},\ }\bibfield  {title} {\bibinfo
  {title} {{Results from phase 1 of the HAYSTAC microwave cavity axion
  experiment}},\ }\href {https://doi.org/10.1103/PhysRevD.97.092001} {\bibfield
   {journal} {\bibinfo  {journal} {Phys. Rev. D}\ }\textbf {\bibinfo {volume}
  {97}},\ \bibinfo {pages} {092001} (\bibinfo {year} {2018})}\BibitemShut
  {NoStop}%
\bibitem [{\citenamefont {Backes}\ \emph {et~al.}(2021)\citenamefont {Backes},
  \citenamefont {Palken}, \citenamefont {Kenany}, \citenamefont {Brubaker},
  \citenamefont {Cahn}, \citenamefont {Droster}, \citenamefont {Hilton},
  \citenamefont {Ghosh}, \citenamefont {Jackson}, \citenamefont {Lamoreaux}
  \emph {et~al.}}]{backes2021quantum}%
  \BibitemOpen
  \bibfield  {author} {\bibinfo {author} {\bibfnamefont {K.~M.}\ \bibnamefont
  {Backes}}, \bibinfo {author} {\bibfnamefont {D.~A.}\ \bibnamefont {Palken}},
  \bibinfo {author} {\bibfnamefont {S.~A.}\ \bibnamefont {Kenany}}, \bibinfo
  {author} {\bibfnamefont {B.~M.}\ \bibnamefont {Brubaker}}, \bibinfo {author}
  {\bibfnamefont {S.}~\bibnamefont {Cahn}}, \bibinfo {author} {\bibfnamefont
  {A.}~\bibnamefont {Droster}}, \bibinfo {author} {\bibfnamefont {G.~C.}\
  \bibnamefont {Hilton}}, \bibinfo {author} {\bibfnamefont {S.}~\bibnamefont
  {Ghosh}}, \bibinfo {author} {\bibfnamefont {H.}~\bibnamefont {Jackson}},
  \bibinfo {author} {\bibfnamefont {S.~K.}\ \bibnamefont {Lamoreaux}}, \emph
  {et~al.},\ }\bibfield  {title} {\bibinfo {title} {A quantum enhanced search
  for dark matter axions},\ }\href {https://doi.org/10.1038/s41586-021-03226-7}
  {\bibfield  {journal} {\bibinfo  {journal} {Nature}\ }\textbf {\bibinfo
  {volume} {590}},\ \bibinfo {pages} {238} (\bibinfo {year}
  {2021})}\BibitemShut {NoStop}%
\bibitem [{\citenamefont {Salemi}\ \emph {et~al.}(2021)\citenamefont {Salemi},
  \citenamefont {Foster}, \citenamefont {Ouellet}, \citenamefont {Gavin},
  \citenamefont {Pappas}, \citenamefont {Cheng}, \citenamefont {Richardson},
  \citenamefont {Henning}, \citenamefont {Kahn}, \citenamefont {Nguyen} \emph
  {et~al.}}]{salemi2021search}%
  \BibitemOpen
  \bibfield  {author} {\bibinfo {author} {\bibfnamefont {C.~P.}\ \bibnamefont
  {Salemi}}, \bibinfo {author} {\bibfnamefont {J.~W.}\ \bibnamefont {Foster}},
  \bibinfo {author} {\bibfnamefont {J.~L.}\ \bibnamefont {Ouellet}}, \bibinfo
  {author} {\bibfnamefont {A.}~\bibnamefont {Gavin}}, \bibinfo {author}
  {\bibfnamefont {K.~M.}\ \bibnamefont {Pappas}}, \bibinfo {author}
  {\bibfnamefont {S.}~\bibnamefont {Cheng}}, \bibinfo {author} {\bibfnamefont
  {K.~A.}\ \bibnamefont {Richardson}}, \bibinfo {author} {\bibfnamefont
  {R.}~\bibnamefont {Henning}}, \bibinfo {author} {\bibfnamefont
  {Y.}~\bibnamefont {Kahn}}, \bibinfo {author} {\bibfnamefont {R.}~\bibnamefont
  {Nguyen}}, \emph {et~al.},\ }\bibfield  {title} {\bibinfo {title} {{Search
  for low-mass axion dark matter with ABRACADABRA-10 cm}},\ }\href
  {https://doi.org/10.1103/PhysRevLett.127.081801} {\bibfield  {journal}
  {\bibinfo  {journal} {Phys. Rev. Lett.}\ }\textbf {\bibinfo {volume} {127}},\
  \bibinfo {pages} {081801} (\bibinfo {year} {2021})}\BibitemShut {NoStop}%
\bibitem [{\citenamefont {Gramolin}\ \emph {et~al.}(2021)\citenamefont
  {Gramolin}, \citenamefont {Aybas}, \citenamefont {Johnson}, \citenamefont
  {Adam},\ and\ \citenamefont {Sushkov}}]{gramolin2021search}%
  \BibitemOpen
  \bibfield  {author} {\bibinfo {author} {\bibfnamefont {A.~V.}\ \bibnamefont
  {Gramolin}}, \bibinfo {author} {\bibfnamefont {D.}~\bibnamefont {Aybas}},
  \bibinfo {author} {\bibfnamefont {D.}~\bibnamefont {Johnson}}, \bibinfo
  {author} {\bibfnamefont {J.}~\bibnamefont {Adam}},\ and\ \bibinfo {author}
  {\bibfnamefont {A.~O.}\ \bibnamefont {Sushkov}},\ }\bibfield  {title}
  {\bibinfo {title} {Search for axion-like dark matter with ferromagnets},\
  }\href {https://doi.org/10.1038/s41567-020-1006-6} {\bibfield  {journal}
  {\bibinfo  {journal} {Nature Physics}\ }\textbf {\bibinfo {volume} {17}},\
  \bibinfo {pages} {79} (\bibinfo {year} {2021})}\BibitemShut {NoStop}%
\bibitem [{\citenamefont {Andrew}\ \emph {et~al.}(2023)\citenamefont {Andrew},
  \citenamefont {Ahn}, \citenamefont {Kutlu}, \citenamefont {Kim},
  \citenamefont {Ko}, \citenamefont {Ivanov}, \citenamefont {Byun},
  \citenamefont {van Loo}, \citenamefont {Park}, \citenamefont {Jeong} \emph
  {et~al.}}]{andrew2023axion}%
  \BibitemOpen
  \bibfield  {author} {\bibinfo {author} {\bibfnamefont {K.~Y.}\ \bibnamefont
  {Andrew}}, \bibinfo {author} {\bibfnamefont {S.}~\bibnamefont {Ahn}},
  \bibinfo {author} {\bibfnamefont {{\c{C}}.}~\bibnamefont {Kutlu}}, \bibinfo
  {author} {\bibfnamefont {J.}~\bibnamefont {Kim}}, \bibinfo {author}
  {\bibfnamefont {B.~R.}\ \bibnamefont {Ko}}, \bibinfo {author} {\bibfnamefont
  {B.~I.}\ \bibnamefont {Ivanov}}, \bibinfo {author} {\bibfnamefont
  {H.}~\bibnamefont {Byun}}, \bibinfo {author} {\bibfnamefont {A.~F.}\
  \bibnamefont {van Loo}}, \bibinfo {author} {\bibfnamefont {S.}~\bibnamefont
  {Park}}, \bibinfo {author} {\bibfnamefont {J.}~\bibnamefont {Jeong}}, \emph
  {et~al.},\ }\bibfield  {title} {\bibinfo {title} {Axion dark matter search
  around 4.55 $\mu$ ev with dine-fischler-srednicki-zhitnitskii sensitivity},\
  }\href {https://doi.org/10.1103/PhysRevLett.130.071002} {\bibfield  {journal}
  {\bibinfo  {journal} {Phys. Rev. Lett.}\ }\textbf {\bibinfo {volume} {130}},\
  \bibinfo {pages} {071002} (\bibinfo {year} {2023})}\BibitemShut {NoStop}%
\bibitem [{\citenamefont {Wagner}\ \emph {et~al.}(2010)\citenamefont {Wagner},
  \citenamefont {Rybka}, \citenamefont {Hotz}, \citenamefont {Rosenberg},
  \citenamefont {Asztalos}, \citenamefont {Carosi}, \citenamefont {Hagmann},
  \citenamefont {Kinion}, \citenamefont {Van~Bibber}, \citenamefont {Hoskins}
  \emph {et~al.}}]{wagner2010search}%
  \BibitemOpen
  \bibfield  {author} {\bibinfo {author} {\bibfnamefont {A.}~\bibnamefont
  {Wagner}}, \bibinfo {author} {\bibfnamefont {G.}~\bibnamefont {Rybka}},
  \bibinfo {author} {\bibfnamefont {M.}~\bibnamefont {Hotz}}, \bibinfo {author}
  {\bibfnamefont {L.}~\bibnamefont {Rosenberg}}, \bibinfo {author}
  {\bibfnamefont {S.}~\bibnamefont {Asztalos}}, \bibinfo {author}
  {\bibfnamefont {G.}~\bibnamefont {Carosi}}, \bibinfo {author} {\bibfnamefont
  {C.}~\bibnamefont {Hagmann}}, \bibinfo {author} {\bibfnamefont
  {D.}~\bibnamefont {Kinion}}, \bibinfo {author} {\bibfnamefont
  {K.}~\bibnamefont {Van~Bibber}}, \bibinfo {author} {\bibfnamefont
  {J.}~\bibnamefont {Hoskins}}, \emph {et~al.},\ }\bibfield  {title} {\bibinfo
  {title} {{Search for hidden sector photons with the ADMX detector}},\ }\href
  {https://doi.org/10.1103/PhysRevLett.105.171801} {\bibfield  {journal}
  {\bibinfo  {journal} {Phys. Rev. Lett.}\ }\textbf {\bibinfo {volume} {105}},\
  \bibinfo {pages} {171801} (\bibinfo {year} {2010})}\BibitemShut {NoStop}%
\bibitem [{\citenamefont {Chaudhuri}\ \emph {et~al.}(2015)\citenamefont
  {Chaudhuri}, \citenamefont {Graham}, \citenamefont {Irwin}, \citenamefont
  {Mardon}, \citenamefont {Rajendran},\ and\ \citenamefont
  {Zhao}}]{chaudhuri2015radio}%
  \BibitemOpen
  \bibfield  {author} {\bibinfo {author} {\bibfnamefont {S.}~\bibnamefont
  {Chaudhuri}}, \bibinfo {author} {\bibfnamefont {P.~W.}\ \bibnamefont
  {Graham}}, \bibinfo {author} {\bibfnamefont {K.}~\bibnamefont {Irwin}},
  \bibinfo {author} {\bibfnamefont {J.}~\bibnamefont {Mardon}}, \bibinfo
  {author} {\bibfnamefont {S.}~\bibnamefont {Rajendran}},\ and\ \bibinfo
  {author} {\bibfnamefont {Y.}~\bibnamefont {Zhao}},\ }\bibfield  {title}
  {\bibinfo {title} {Radio for hidden-photon dark matter detection},\ }\href
  {https://doi.org/10.1103/PhysRevD.92.075012} {\bibfield  {journal} {\bibinfo
  {journal} {Phys. Rev. D}\ }\textbf {\bibinfo {volume} {92}},\ \bibinfo
  {pages} {075012} (\bibinfo {year} {2015})}\BibitemShut {NoStop}%
\bibitem [{\citenamefont {Phipps}\ \emph {et~al.}(2020)\citenamefont {Phipps},
  \citenamefont {Kuenstner}, \citenamefont {Chaudhuri}, \citenamefont {Dawson},
  \citenamefont {Young}, \citenamefont {FitzGerald}, \citenamefont {Froland},
  \citenamefont {Wells}, \citenamefont {Li}, \citenamefont {Cho} \emph
  {et~al.}}]{phipps2020exclusion}%
  \BibitemOpen
  \bibfield  {author} {\bibinfo {author} {\bibfnamefont {A.}~\bibnamefont
  {Phipps}}, \bibinfo {author} {\bibfnamefont {S.~E.}\ \bibnamefont
  {Kuenstner}}, \bibinfo {author} {\bibfnamefont {S.}~\bibnamefont
  {Chaudhuri}}, \bibinfo {author} {\bibfnamefont {C.~S.}\ \bibnamefont
  {Dawson}}, \bibinfo {author} {\bibfnamefont {B.~A.}\ \bibnamefont {Young}},
  \bibinfo {author} {\bibfnamefont {C.~T.}\ \bibnamefont {FitzGerald}},
  \bibinfo {author} {\bibfnamefont {H.}~\bibnamefont {Froland}}, \bibinfo
  {author} {\bibfnamefont {K.}~\bibnamefont {Wells}}, \bibinfo {author}
  {\bibfnamefont {D.}~\bibnamefont {Li}}, \bibinfo {author} {\bibfnamefont
  {H.}~\bibnamefont {Cho}}, \emph {et~al.},\ }\bibfield  {title} {\bibinfo
  {title} {{Exclusion limits on hidden-photon dark matter near 2 neV from a
  fixed-frequency superconducting lumped-element resonator}},\ }in\ \href
  {https://doi.org/10.1007/978-3-030-43761-9_16} {\emph {\bibinfo {booktitle}
  {Microwave Cavities and Detectors for Axion Research: Proceedings of the 3rd
  International Workshop}}}\ (\bibinfo {organization} {Springer},\ \bibinfo
  {year} {2020})\ p.\ \bibinfo {pages} {139}\BibitemShut {NoStop}%
\bibitem [{\citenamefont {Fedderke}\ \emph
  {et~al.}(2021{\natexlab{a}})\citenamefont {Fedderke}, \citenamefont {Graham},
  \citenamefont {Jackson~Kimball},\ and\ \citenamefont
  {Kalia}}]{Fedd21concept}%
  \BibitemOpen
  \bibfield  {author} {\bibinfo {author} {\bibfnamefont {M.~A.}\ \bibnamefont
  {Fedderke}}, \bibinfo {author} {\bibfnamefont {P.~W.}\ \bibnamefont
  {Graham}}, \bibinfo {author} {\bibfnamefont {D.~F.}\ \bibnamefont
  {Jackson~Kimball}},\ and\ \bibinfo {author} {\bibfnamefont {S.}~\bibnamefont
  {Kalia}},\ }\bibfield  {title} {\bibinfo {title} {Earth as a transducer for
  dark-photon dark-matter detection},\ }\href
  {https://doi.org/10.1103/PhysRevD.104.075023} {\bibfield  {journal} {\bibinfo
   {journal} {Phys. Rev. D}\ }\textbf {\bibinfo {volume} {104}},\ \bibinfo
  {pages} {075023} (\bibinfo {year} {2021}{\natexlab{a}})}\BibitemShut
  {NoStop}%
\bibitem [{\citenamefont {Fedderke}\ \emph
  {et~al.}(2021{\natexlab{b}})\citenamefont {Fedderke}, \citenamefont {Graham},
  \citenamefont {Jackson~Kimball},\ and\ \citenamefont {Kalia}}]{Fed21search}%
  \BibitemOpen
  \bibfield  {author} {\bibinfo {author} {\bibfnamefont {M.~A.}\ \bibnamefont
  {Fedderke}}, \bibinfo {author} {\bibfnamefont {P.~W.}\ \bibnamefont
  {Graham}}, \bibinfo {author} {\bibfnamefont {D.~F.}\ \bibnamefont
  {Jackson~Kimball}},\ and\ \bibinfo {author} {\bibfnamefont {S.}~\bibnamefont
  {Kalia}},\ }\bibfield  {title} {\bibinfo {title} {Search for dark-photon dark
  matter in the supermag geomagnetic field dataset},\ }\href
  {https://doi.org/10.1103/PhysRevD.104.095032} {\bibfield  {journal} {\bibinfo
   {journal} {Phys. Rev. D}\ }\textbf {\bibinfo {volume} {104}},\ \bibinfo
  {pages} {095032} (\bibinfo {year} {2021}{\natexlab{b}})}\BibitemShut
  {NoStop}%
\bibitem [{\citenamefont {Arza}\ \emph {et~al.}(2022)\citenamefont {Arza},
  \citenamefont {Fedderke}, \citenamefont {Graham}, \citenamefont
  {Jackson~Kimball},\ and\ \citenamefont {Kalia}}]{Arza22}%
  \BibitemOpen
  \bibfield  {author} {\bibinfo {author} {\bibfnamefont {A.}~\bibnamefont
  {Arza}}, \bibinfo {author} {\bibfnamefont {M.~A.}\ \bibnamefont {Fedderke}},
  \bibinfo {author} {\bibfnamefont {P.~W.}\ \bibnamefont {Graham}}, \bibinfo
  {author} {\bibfnamefont {D.~F.}\ \bibnamefont {Jackson~Kimball}},\ and\
  \bibinfo {author} {\bibfnamefont {S.}~\bibnamefont {Kalia}},\ }\bibfield
  {title} {\bibinfo {title} {Earth as a transducer for axion dark-matter
  detection},\ }\href {https://doi.org/10.1103/PhysRevD.105.095007} {\bibfield
  {journal} {\bibinfo  {journal} {Phys. Rev. D}\ }\textbf {\bibinfo {volume}
  {105}},\ \bibinfo {pages} {095007} (\bibinfo {year} {2022})}\BibitemShut
  {NoStop}%
\bibitem [{\citenamefont {Primakoff}(1951)}]{primakoff1951photo}%
  \BibitemOpen
  \bibfield  {author} {\bibinfo {author} {\bibfnamefont {H.}~\bibnamefont
  {Primakoff}},\ }\bibfield  {title} {\bibinfo {title} {Photo-production of
  neutral mesons in nuclear electric fields and the mean life of the neutral
  meson},\ }\href {https://doi.org/10.1103/PhysRev.81.899} {\bibfield
  {journal} {\bibinfo  {journal} {Phys. Rev.}\ }\textbf {\bibinfo {volume}
  {81}},\ \bibinfo {pages} {899} (\bibinfo {year} {1951})}\BibitemShut
  {NoStop}%
\bibitem [{\citenamefont {Raffelt}\ and\ \citenamefont
  {Seckel}(1988)}]{raffelt1988bounds}%
  \BibitemOpen
  \bibfield  {author} {\bibinfo {author} {\bibfnamefont {G.}~\bibnamefont
  {Raffelt}}\ and\ \bibinfo {author} {\bibfnamefont {D.}~\bibnamefont
  {Seckel}},\ }\bibfield  {title} {\bibinfo {title} {{Bounds on exotic-particle
  interactions from SN1987A}},\ }\href
  {https://doi.org/10.1103/PhysRevLett.60.1793} {\bibfield  {journal} {\bibinfo
   {journal} {Phys. Rev. Lett.}\ }\textbf {\bibinfo {volume} {60}},\ \bibinfo
  {pages} {1793} (\bibinfo {year} {1988})}\BibitemShut {NoStop}%
\bibitem [{SNI()}]{SNIPEHuntStory}%
  \BibitemOpen
  \href@noop {} {}\bibinfo {howpublished}
  {\url{https://en.wikipedia.org/wiki/Snipe_hunt}.}\BibitemShut {Stop}%
\bibitem [{\citenamefont {Bowen}\ \emph {et~al.}(2019)\citenamefont {Bowen},
  \citenamefont {Zhivun}, \citenamefont {Wickenbrock}, \citenamefont {Dumont},
  \citenamefont {Bale}, \citenamefont {Pankow}, \citenamefont {Dobler},
  \citenamefont {Wurtele},\ and\ \citenamefont {Budker}}]{Bowen2019Geo}%
  \BibitemOpen
  \bibfield  {author} {\bibinfo {author} {\bibfnamefont {T.~A.}\ \bibnamefont
  {Bowen}}, \bibinfo {author} {\bibfnamefont {E.}~\bibnamefont {Zhivun}},
  \bibinfo {author} {\bibfnamefont {A.}~\bibnamefont {Wickenbrock}}, \bibinfo
  {author} {\bibfnamefont {V.}~\bibnamefont {Dumont}}, \bibinfo {author}
  {\bibfnamefont {S.~D.}\ \bibnamefont {Bale}}, \bibinfo {author}
  {\bibfnamefont {C.}~\bibnamefont {Pankow}}, \bibinfo {author} {\bibfnamefont
  {G.}~\bibnamefont {Dobler}}, \bibinfo {author} {\bibfnamefont {J.~S.}\
  \bibnamefont {Wurtele}},\ and\ \bibinfo {author} {\bibfnamefont
  {D.}~\bibnamefont {Budker}},\ }\bibfield  {title} {\bibinfo {title} {A
  network of magnetometers for multi-scale urban science and informatics},\
  }\href {https://doi.org/10.5194/gi-8-129-2019} {\bibfield  {journal}
  {\bibinfo  {journal} {Geosci. Instrum. Methods Data Syst.}\ }\textbf
  {\bibinfo {volume} {8}},\ \bibinfo {pages} {129} (\bibinfo {year}
  {2019})}\BibitemShut {NoStop}%
\bibitem [{\citenamefont {Dumont}\ \emph {et~al.}(2022)\citenamefont {Dumont},
  \citenamefont {Bowen}, \citenamefont {Roglans}, \citenamefont {Dobler},
  \citenamefont {Sharma}, \citenamefont {Karpf}, \citenamefont {Bale},
  \citenamefont {Wickenbrock}, \citenamefont {Zhivun}, \citenamefont {Kornack},
  \citenamefont {Wurtele},\ and\ \citenamefont {Budker}}]{Dumont2022JAP}%
  \BibitemOpen
  \bibfield  {author} {\bibinfo {author} {\bibfnamefont {V.}~\bibnamefont
  {Dumont}}, \bibinfo {author} {\bibfnamefont {T.~A.}\ \bibnamefont {Bowen}},
  \bibinfo {author} {\bibfnamefont {R.}~\bibnamefont {Roglans}}, \bibinfo
  {author} {\bibfnamefont {G.}~\bibnamefont {Dobler}}, \bibinfo {author}
  {\bibfnamefont {M.~S.}\ \bibnamefont {Sharma}}, \bibinfo {author}
  {\bibfnamefont {A.}~\bibnamefont {Karpf}}, \bibinfo {author} {\bibfnamefont
  {S.~D.}\ \bibnamefont {Bale}}, \bibinfo {author} {\bibfnamefont
  {A.}~\bibnamefont {Wickenbrock}}, \bibinfo {author} {\bibfnamefont
  {E.}~\bibnamefont {Zhivun}}, \bibinfo {author} {\bibfnamefont
  {T.}~\bibnamefont {Kornack}}, \bibinfo {author} {\bibfnamefont {J.~S.}\
  \bibnamefont {Wurtele}},\ and\ \bibinfo {author} {\bibfnamefont
  {D.}~\bibnamefont {Budker}},\ }\bibfield  {title} {\bibinfo {title} {Do
  cities have a unique magnetic pulse?},\ }\href
  {https://doi.org/10.1063/5.0088264} {\bibfield  {journal} {\bibinfo
  {journal} {J. of Appl. Phys.}\ }\textbf {\bibinfo {volume} {131}},\ \bibinfo
  {pages} {204902} (\bibinfo {year} {2022})}\BibitemShut {NoStop}%
\bibitem [{Sup()}]{SuperMAGwebsite}%
  \BibitemOpen
  \href@noop {} {}\bibinfo {howpublished}
  {\url{http://supermag.jhuapl.edu}.}\BibitemShut {Stop}%
\bibitem [{\citenamefont {Gjerloev}(2009)}]{Gjerloev:2009wsd}%
  \BibitemOpen
  \bibfield  {author} {\bibinfo {author} {\bibfnamefont {J.~W.}\ \bibnamefont
  {Gjerloev}},\ }\bibfield  {title} {\bibinfo {title} {{A Global Ground-Based
  Magnetometer Initiative}},\ }\href {https://doi.org/10.1029/2009EO270002}
  {\bibfield  {journal} {\bibinfo  {journal} {Eos}\ }\textbf {\bibinfo {volume}
  {90}},\ \bibinfo {pages} {230} (\bibinfo {year} {2009})}\BibitemShut
  {NoStop}%
\bibitem [{\citenamefont {Gjerloev}(2012)}]{Gjerloev:2012sdg}%
  \BibitemOpen
  \bibfield  {author} {\bibinfo {author} {\bibfnamefont {J.~W.}\ \bibnamefont
  {Gjerloev}},\ }\bibfield  {title} {\bibinfo {title} {The {SuperMAG} data
  processing technique},\ }\href {https://doi.org/10.1029/2012JA017683}
  {\bibfield  {journal} {\bibinfo  {journal} {J. Geophys. Res. Space Phys.}\
  }\textbf {\bibinfo {volume} {117}},\ \bibinfo {pages} {A09213} (\bibinfo
  {year} {2012})}\BibitemShut {NoStop}%
\bibitem [{\citenamefont {Constable}\ and\ \citenamefont
  {Constable}(2004)}]{Constable}%
  \BibitemOpen
  \bibfield  {author} {\bibinfo {author} {\bibfnamefont {C.~G.}\ \bibnamefont
  {Constable}}\ and\ \bibinfo {author} {\bibfnamefont {S.~C.}\ \bibnamefont
  {Constable}},\ }\bibinfo {title} {Satellite magnetic field measurements:
  Applications in studying the deep earth},\ in\ \href
  {https://doi.org/https://doi.org/10.1029/150GM13} {\emph {\bibinfo
  {booktitle} {The State of the Planet: Frontiers and Challenges in
  Geophysics}}}\ (\bibinfo  {publisher} {American Geophysical Union, AGU},\
  \bibinfo {year} {2004})\ pp.\ \bibinfo {pages} {147--159}\BibitemShut
  {NoStop}%
\bibitem [{\citenamefont {Cvetic}\ and\ \citenamefont
  {Langacker}(1996)}]{Cve96}%
  \BibitemOpen
  \bibfield  {author} {\bibinfo {author} {\bibfnamefont {M.}~\bibnamefont
  {Cvetic}}\ and\ \bibinfo {author} {\bibfnamefont {P.}~\bibnamefont
  {Langacker}},\ }\bibfield  {title} {\bibinfo {title} {{Implications of
  Abelian extended gauge structures from string models}},\ }\href
  {https://doi.org/10.1103/PhysRevD.54.3570} {\bibfield  {journal} {\bibinfo
  {journal} {Phys. Rev. D}\ }\textbf {\bibinfo {volume} {54}},\ \bibinfo
  {pages} {3570} (\bibinfo {year} {1996})}\BibitemShut {NoStop}%
\bibitem [{\citenamefont {Holdom}(1986)}]{Hol86}%
  \BibitemOpen
  \bibfield  {author} {\bibinfo {author} {\bibfnamefont {B.}~\bibnamefont
  {Holdom}},\ }\bibfield  {title} {\bibinfo {title} {{Two U(1)'s and $\epsilon$
  charge shifts}},\ }\href {https://doi.org/10.1016/0370-2693(86)91377-8}
  {\bibfield  {journal} {\bibinfo  {journal} {Phys. Lett. B}\ }\textbf
  {\bibinfo {volume} {166}},\ \bibinfo {pages} {196} (\bibinfo {year}
  {1986})}\BibitemShut {NoStop}%
\bibitem [{\citenamefont {Graham}\ \emph {et~al.}(2014)\citenamefont {Graham},
  \citenamefont {Mardon}, \citenamefont {Rajendran},\ and\ \citenamefont
  {Zhao}}]{graham2014parametrically}%
  \BibitemOpen
  \bibfield  {author} {\bibinfo {author} {\bibfnamefont {P.~W.}\ \bibnamefont
  {Graham}}, \bibinfo {author} {\bibfnamefont {J.}~\bibnamefont {Mardon}},
  \bibinfo {author} {\bibfnamefont {S.}~\bibnamefont {Rajendran}},\ and\
  \bibinfo {author} {\bibfnamefont {Y.}~\bibnamefont {Zhao}},\ }\bibfield
  {title} {\bibinfo {title} {Parametrically enhanced hidden photon search},\
  }\href {https://doi.org/10.1103/PhysRevD.90.075017} {\bibfield  {journal}
  {\bibinfo  {journal} {Phys. Rev. D}\ }\textbf {\bibinfo {volume} {90}},\
  \bibinfo {pages} {075017} (\bibinfo {year} {2014})}\BibitemShut {NoStop}%
\bibitem [{\citenamefont {Graham}\ \emph
  {et~al.}(2016{\natexlab{b}})\citenamefont {Graham}, \citenamefont {Mardon},\
  and\ \citenamefont {Rajendran}}]{Graham:2015rva}%
  \BibitemOpen
  \bibfield  {author} {\bibinfo {author} {\bibfnamefont {P.~W.}\ \bibnamefont
  {Graham}}, \bibinfo {author} {\bibfnamefont {J.}~\bibnamefont {Mardon}},\
  and\ \bibinfo {author} {\bibfnamefont {S.}~\bibnamefont {Rajendran}},\
  }\bibfield  {title} {\bibinfo {title} {{Vector Dark Matter from Inflationary
  Fluctuations}},\ }\href {https://doi.org/10.1103/PhysRevD.93.103520}
  {\bibfield  {journal} {\bibinfo  {journal} {Phys. Rev. D}\ }\textbf {\bibinfo
  {volume} {93}},\ \bibinfo {pages} {103520} (\bibinfo {year}
  {2016}{\natexlab{b}})}\BibitemShut {NoStop}%
\bibitem [{\citenamefont {Ahmed}\ \emph {et~al.}(2020)\citenamefont {Ahmed},
  \citenamefont {Grzadkowski},\ and\ \citenamefont
  {Socha}}]{ahmed2020gravitational}%
  \BibitemOpen
  \bibfield  {author} {\bibinfo {author} {\bibfnamefont {A.}~\bibnamefont
  {Ahmed}}, \bibinfo {author} {\bibfnamefont {B.}~\bibnamefont {Grzadkowski}},\
  and\ \bibinfo {author} {\bibfnamefont {A.}~\bibnamefont {Socha}},\ }\bibfield
   {title} {\bibinfo {title} {Gravitational production of vector dark matter},\
  }\href {https://doi.org/10.1007/JHEP08(2020)059} {\bibfield  {journal}
  {\bibinfo  {journal} {J. High Energy Phys.}\ }\textbf {\bibinfo {volume}
  {2020}}\bibinfo  {number} { (8)},\ \bibinfo {pages} {59}}\BibitemShut
  {NoStop}%
\bibitem [{\citenamefont {Kolb}\ and\ \citenamefont
  {Long}(2021)}]{kolb2021completely}%
  \BibitemOpen
\bibfield  {number} {  }\bibfield  {author} {\bibinfo {author} {\bibfnamefont
  {E.~W.}\ \bibnamefont {Kolb}}\ and\ \bibinfo {author} {\bibfnamefont {A.~J.}\
  \bibnamefont {Long}},\ }\bibfield  {title} {\bibinfo {title} {Completely dark
  photons from gravitational particle production during the inflationary era},\
  }\href {https://doi.org/10.1007/JHEP03(2021)283} {\bibfield  {journal}
  {\bibinfo  {journal} {J. High Energy Phys.}\ }\textbf {\bibinfo {volume}
  {2021}}\bibinfo  {number} { (3)},\ \bibinfo {pages} {283}}\BibitemShut
  {NoStop}%
\bibitem [{\citenamefont {Adshead}\ \emph {et~al.}(2023)\citenamefont
  {Adshead}, \citenamefont {Lozanov},\ and\ \citenamefont
  {Weiner}}]{Adshead:2023qiw}%
  \BibitemOpen
\bibfield  {number} {  }\bibfield  {author} {\bibinfo {author} {\bibfnamefont
  {P.}~\bibnamefont {Adshead}}, \bibinfo {author} {\bibfnamefont {K.~D.}\
  \bibnamefont {Lozanov}},\ and\ \bibinfo {author} {\bibfnamefont {Z.~J.}\
  \bibnamefont {Weiner}},\ }\bibfield  {title} {\bibinfo {title} {{Dark photon
  dark matter from an oscillating dilaton}},\ }\href@noop {} {\  (\bibinfo
  {year} {2023})},\ \Eprint {https://arxiv.org/abs/2301.07718}
  {arXiv:2301.07718 [hep-ph]} \BibitemShut {NoStop}%
\bibitem [{\citenamefont {Nelson}\ and\ \citenamefont
  {Scholtz}(2011)}]{nelson2011dark}%
  \BibitemOpen
  \bibfield  {author} {\bibinfo {author} {\bibfnamefont {A.~E.}\ \bibnamefont
  {Nelson}}\ and\ \bibinfo {author} {\bibfnamefont {J.}~\bibnamefont
  {Scholtz}},\ }\bibfield  {title} {\bibinfo {title} {Dark light, dark matter,
  and the misalignment mechanism},\ }\href
  {https://doi.org/10.1103/PhysRevD.84.103501} {\bibfield  {journal} {\bibinfo
  {journal} {Phys. Rev. D}\ }\textbf {\bibinfo {volume} {84}},\ \bibinfo
  {pages} {103501} (\bibinfo {year} {2011})}\BibitemShut {NoStop}%
\bibitem [{\citenamefont {Arias}\ \emph
  {et~al.}(2012{\natexlab{b}})\citenamefont {Arias}, \citenamefont {Cadamuro},
  \citenamefont {Goodsell}, \citenamefont {Jaeckel}, \citenamefont {Redondo},\
  and\ \citenamefont {Ringwald}}]{Paola_Arias:2012jcap}%
  \BibitemOpen
  \bibfield  {author} {\bibinfo {author} {\bibfnamefont {P.}~\bibnamefont
  {Arias}}, \bibinfo {author} {\bibfnamefont {D.}~\bibnamefont {Cadamuro}},
  \bibinfo {author} {\bibfnamefont {M.}~\bibnamefont {Goodsell}}, \bibinfo
  {author} {\bibfnamefont {J.}~\bibnamefont {Jaeckel}}, \bibinfo {author}
  {\bibfnamefont {J.}~\bibnamefont {Redondo}},\ and\ \bibinfo {author}
  {\bibfnamefont {A.}~\bibnamefont {Ringwald}},\ }\bibfield  {title} {\bibinfo
  {title} {Wispy cold dark matter},\ }\href
  {https://doi.org/10.1088/1475-7516/2012/06/013} {\bibfield  {journal}
  {\bibinfo  {journal} {J. Cosmol. Astropart. Phys.}\ }\textbf {\bibinfo
  {volume} {06}}\bibinfo  {number} { (06)},\ \bibinfo {pages}
  {013}}\BibitemShut {NoStop}%
\bibitem [{\citenamefont {Caputo}\ \emph {et~al.}(2021)\citenamefont {Caputo},
  \citenamefont {O'Hare}, \citenamefont {Millar},\ and\ \citenamefont
  {Vitagliano}}]{Caputo:2021eaa}%
  \BibitemOpen
\bibfield  {number} {  }\bibfield  {author} {\bibinfo {author} {\bibfnamefont
  {A.}~\bibnamefont {Caputo}}, \bibinfo {author} {\bibfnamefont {C.~A.~J.}\
  \bibnamefont {O'Hare}}, \bibinfo {author} {\bibfnamefont {A.~J.}\
  \bibnamefont {Millar}},\ and\ \bibinfo {author} {\bibfnamefont
  {E.}~\bibnamefont {Vitagliano}},\ }\bibfield  {title} {\bibinfo {title}
  {{Dark photon limits: a cookbook}},\ }\href@noop {} {\  (\bibinfo {year}
  {2021})},\ \Eprint {https://arxiv.org/abs/2105.04565} {arXiv:2105.04565
  [hep-ph]} \BibitemShut {NoStop}%
\bibitem [{\citenamefont {Read}(2014)}]{read2014local}%
  \BibitemOpen
  \bibfield  {author} {\bibinfo {author} {\bibfnamefont {J.~I.}\ \bibnamefont
  {Read}},\ }\bibfield  {title} {\bibinfo {title} {The local dark matter
  density},\ }\href {https://doi.org/10.1088/0954-3899/41/6/063101} {\bibfield
  {journal} {\bibinfo  {journal} {J. Phys. G: Nucl. Part. Phys.}\ }\textbf
  {\bibinfo {volume} {41}},\ \bibinfo {pages} {063101} (\bibinfo {year}
  {2014})}\BibitemShut {NoStop}%
\bibitem [{\citenamefont {Bland-Hawthorn}\ and\ \citenamefont
  {Gerhard}(2016)}]{bland2016galaxy}%
  \BibitemOpen
  \bibfield  {author} {\bibinfo {author} {\bibfnamefont {J.}~\bibnamefont
  {Bland-Hawthorn}}\ and\ \bibinfo {author} {\bibfnamefont {O.}~\bibnamefont
  {Gerhard}},\ }\bibfield  {title} {\bibinfo {title} {The galaxy in context:
  structural, kinematic, and integrated properties},\ }\href
  {https://doi.org/10.1146/annurev-astro-081915-023441} {\bibfield  {journal}
  {\bibinfo  {journal} {Annu. Rev. Astron. Astrophys.}\ }\textbf {\bibinfo
  {volume} {54}},\ \bibinfo {pages} {529} (\bibinfo {year} {2016})}\BibitemShut
  {NoStop}%
\bibitem [{\citenamefont {Peccei}\ and\ \citenamefont
  {Quinn}(1977)}]{Peccei:1977hh}%
  \BibitemOpen
  \bibfield  {author} {\bibinfo {author} {\bibfnamefont {R.}~\bibnamefont
  {Peccei}}\ and\ \bibinfo {author} {\bibfnamefont {H.~R.}\ \bibnamefont
  {Quinn}},\ }\bibfield  {title} {\bibinfo {title} {{CP Conservation in the
  Presence of Instantons}},\ }\href
  {https://doi.org/10.1103/PhysRevLett.38.1440} {\bibfield  {journal} {\bibinfo
   {journal} {Phys. Rev. Lett.}\ }\textbf {\bibinfo {volume} {38}},\ \bibinfo
  {pages} {1440} (\bibinfo {year} {1977})}\BibitemShut {NoStop}%
\bibitem [{\citenamefont {Weinberg}(1978)}]{Weinberg:1977ma}%
  \BibitemOpen
  \bibfield  {author} {\bibinfo {author} {\bibfnamefont {S.}~\bibnamefont
  {Weinberg}},\ }\bibfield  {title} {\bibinfo {title} {{A New Light Boson?}},\
  }\href {https://doi.org/10.1103/PhysRevLett.40.223} {\bibfield  {journal}
  {\bibinfo  {journal} {Phys. Rev. Lett.}\ }\textbf {\bibinfo {volume} {40}},\
  \bibinfo {pages} {223} (\bibinfo {year} {1978})}\BibitemShut {NoStop}%
\bibitem [{\citenamefont {Wilczek}(1978)}]{Wilczek:1977pj}%
  \BibitemOpen
  \bibfield  {author} {\bibinfo {author} {\bibfnamefont {F.}~\bibnamefont
  {Wilczek}},\ }\bibfield  {title} {\bibinfo {title} {{Problem of Strong $P$
  and $T$ Invariance in the Presence of Instantons}},\ }\href
  {https://doi.org/10.1103/PhysRevLett.40.279} {\bibfield  {journal} {\bibinfo
  {journal} {Phys. Rev. Lett.}\ }\textbf {\bibinfo {volume} {40}},\ \bibinfo
  {pages} {279} (\bibinfo {year} {1978})}\BibitemShut {NoStop}%
\bibitem [{\citenamefont {Preskill}\ \emph {et~al.}(1983)\citenamefont
  {Preskill}, \citenamefont {Wise},\ and\ \citenamefont
  {Wilczek}}]{preskill1983cosmology}%
  \BibitemOpen
  \bibfield  {author} {\bibinfo {author} {\bibfnamefont {J.}~\bibnamefont
  {Preskill}}, \bibinfo {author} {\bibfnamefont {M.~B.}\ \bibnamefont {Wise}},\
  and\ \bibinfo {author} {\bibfnamefont {F.}~\bibnamefont {Wilczek}},\
  }\bibfield  {title} {\bibinfo {title} {Cosmology of the invisible axion},\
  }\href {https://doi.org/10.1016/0370-2693(83)90637-8} {\bibfield  {journal}
  {\bibinfo  {journal} {Phys. Lett. B}\ }\textbf {\bibinfo {volume} {120}},\
  \bibinfo {pages} {127} (\bibinfo {year} {1983})}\BibitemShut {NoStop}%
\bibitem [{\citenamefont {Abbott}\ and\ \citenamefont
  {Sikivie}(1983)}]{Abbott:1982af}%
  \BibitemOpen
  \bibfield  {author} {\bibinfo {author} {\bibfnamefont {L.~F.}\ \bibnamefont
  {Abbott}}\ and\ \bibinfo {author} {\bibfnamefont {P.}~\bibnamefont
  {Sikivie}},\ }\bibfield  {title} {\bibinfo {title} {{A Cosmological Bound on
  the Invisible Axion}},\ }\href {https://doi.org/10.1016/0370-2693(83)90638-X}
  {\bibfield  {journal} {\bibinfo  {journal} {Phys. Lett. B}\ }\textbf
  {\bibinfo {volume} {120}},\ \bibinfo {pages} {133} (\bibinfo {year}
  {1983})}\BibitemShut {NoStop}%
\bibitem [{\citenamefont {Dine}\ and\ \citenamefont
  {Fischler}(1983)}]{Dine:1982ah}%
  \BibitemOpen
  \bibfield  {author} {\bibinfo {author} {\bibfnamefont {M.}~\bibnamefont
  {Dine}}\ and\ \bibinfo {author} {\bibfnamefont {W.}~\bibnamefont
  {Fischler}},\ }\bibfield  {title} {\bibinfo {title} {{The Not So Harmless
  Axion}},\ }\href {https://doi.org/10.1016/0370-2693(83)90639-1} {\bibfield
  {journal} {\bibinfo  {journal} {Phys. Lett. B}\ }\textbf {\bibinfo {volume}
  {120}},\ \bibinfo {pages} {137} (\bibinfo {year} {1983})}\BibitemShut
  {NoStop}%
\bibitem [{\citenamefont {Graham}\ and\ \citenamefont
  {Scherlis}(2018)}]{graham2018stochastic}%
  \BibitemOpen
  \bibfield  {author} {\bibinfo {author} {\bibfnamefont {P.~W.}\ \bibnamefont
  {Graham}}\ and\ \bibinfo {author} {\bibfnamefont {A.}~\bibnamefont
  {Scherlis}},\ }\bibfield  {title} {\bibinfo {title} {Stochastic axion
  scenario},\ }\href {https://doi.org/10.1103/PhysRevD.98.035017} {\bibfield
  {journal} {\bibinfo  {journal} {Phys. Rev. D}\ }\textbf {\bibinfo {volume}
  {98}},\ \bibinfo {pages} {035017} (\bibinfo {year} {2018})}\BibitemShut
  {NoStop}%
\bibitem [{\citenamefont {Svrcek}\ and\ \citenamefont
  {Witten}(2006)}]{svrcek2006axions}%
  \BibitemOpen
  \bibfield  {author} {\bibinfo {author} {\bibfnamefont {P.}~\bibnamefont
  {Svrcek}}\ and\ \bibinfo {author} {\bibfnamefont {E.}~\bibnamefont
  {Witten}},\ }\bibfield  {title} {\bibinfo {title} {Axions in string theory},\
  }\href {https://doi.org/10.1088/1126-6708/2006/06/051} {\bibfield  {journal}
  {\bibinfo  {journal} {J. High Energy Phys.}\ }\textbf {\bibinfo {volume}
  {2006}}\bibinfo  {number} { (06)},\ \bibinfo {pages} {051}}\BibitemShut
  {NoStop}%
\bibitem [{\citenamefont {Arvanitaki}\ \emph {et~al.}(2010)\citenamefont
  {Arvanitaki}, \citenamefont {Dimopoulos}, \citenamefont {Dubovsky},
  \citenamefont {Kaloper},\ and\ \citenamefont
  {March-Russell}}]{arvanitaki2010string}%
  \BibitemOpen
\bibfield  {number} {  }\bibfield  {author} {\bibinfo {author} {\bibfnamefont
  {A.}~\bibnamefont {Arvanitaki}}, \bibinfo {author} {\bibfnamefont
  {S.}~\bibnamefont {Dimopoulos}}, \bibinfo {author} {\bibfnamefont
  {S.}~\bibnamefont {Dubovsky}}, \bibinfo {author} {\bibfnamefont
  {N.}~\bibnamefont {Kaloper}},\ and\ \bibinfo {author} {\bibfnamefont
  {J.}~\bibnamefont {March-Russell}},\ }\bibfield  {title} {\bibinfo {title}
  {String axiverse},\ }\href {https://doi.org/10.1103/PhysRevD.81.123530}
  {\bibfield  {journal} {\bibinfo  {journal} {Phys. Rev. D}\ }\textbf {\bibinfo
  {volume} {81}},\ \bibinfo {pages} {123530} (\bibinfo {year}
  {2010})}\BibitemShut {NoStop}%
\bibitem [{\citenamefont {Graham}\ \emph
  {et~al.}(2015{\natexlab{b}})\citenamefont {Graham}, \citenamefont {Kaplan},\
  and\ \citenamefont {Rajendran}}]{graham2015cosmological}%
  \BibitemOpen
  \bibfield  {author} {\bibinfo {author} {\bibfnamefont {P.~W.}\ \bibnamefont
  {Graham}}, \bibinfo {author} {\bibfnamefont {D.~E.}\ \bibnamefont {Kaplan}},\
  and\ \bibinfo {author} {\bibfnamefont {S.}~\bibnamefont {Rajendran}},\
  }\bibfield  {title} {\bibinfo {title} {Cosmological relaxation of the
  electroweak scale},\ }\href {https://doi.org/10.1103/PhysRevLett.115.221801}
  {\bibfield  {journal} {\bibinfo  {journal} {Phys. Rev. Lett.}\ }\textbf
  {\bibinfo {volume} {115}},\ \bibinfo {pages} {221801} (\bibinfo {year}
  {2015}{\natexlab{b}})}\BibitemShut {NoStop}%
\bibitem [{\citenamefont {Alexander}\ \emph {et~al.}(2023)\citenamefont
  {Alexander}, \citenamefont {Gilmer}, \citenamefont {Manton},\ and\
  \citenamefont {McDonough}}]{alexander2023piaxion}%
  \BibitemOpen
  \bibfield  {author} {\bibinfo {author} {\bibfnamefont {S.}~\bibnamefont
  {Alexander}}, \bibinfo {author} {\bibfnamefont {H.}~\bibnamefont {Gilmer}},
  \bibinfo {author} {\bibfnamefont {T.}~\bibnamefont {Manton}},\ and\ \bibinfo
  {author} {\bibfnamefont {E.}~\bibnamefont {McDonough}},\ }\href@noop {}
  {\bibinfo {title} {The $\pi$-axion and $\pi$-axiverse of dark qcd}} (\bibinfo
  {year} {2023}),\ \Eprint {https://arxiv.org/abs/2304.11176} {arXiv:2304.11176
  [hep-ph]} \BibitemShut {NoStop}%
\bibitem [{\citenamefont {Sentman}(2017)}]{sentman2017schumann}%
  \BibitemOpen
  \bibfield  {author} {\bibinfo {author} {\bibfnamefont {D.~D.}\ \bibnamefont
  {Sentman}},\ }\bibfield  {title} {\bibinfo {title} {Schumann resonances},\
  }in\ \href {https://doi.org/10.1201/9780203719503} {\emph {\bibinfo
  {booktitle} {Handbook of atmospheric electrodynamics}}}\ (\bibinfo
  {publisher} {CRC Press},\ \bibinfo {year} {2017})\ pp.\ \bibinfo {pages}
  {267--295}\BibitemShut {NoStop}%
\bibitem [{\citenamefont {Rodríguez-Camacho}\ \emph
  {et~al.}(2022)\citenamefont {Rodríguez-Camacho}, \citenamefont {Salinas},
  \citenamefont {Carrión}, \citenamefont {Portí}, \citenamefont
  {Fornieles-Callejón},\ and\ \citenamefont
  {Toledo-Redondo}}]{Rodriguez-Camacho}%
  \BibitemOpen
  \bibfield  {author} {\bibinfo {author} {\bibfnamefont {J.}~\bibnamefont
  {Rodríguez-Camacho}}, \bibinfo {author} {\bibfnamefont {A.}~\bibnamefont
  {Salinas}}, \bibinfo {author} {\bibfnamefont {M.~C.}\ \bibnamefont
  {Carrión}}, \bibinfo {author} {\bibfnamefont {J.}~\bibnamefont {Portí}},
  \bibinfo {author} {\bibfnamefont {J.}~\bibnamefont {Fornieles-Callejón}},\
  and\ \bibinfo {author} {\bibfnamefont {S.}~\bibnamefont {Toledo-Redondo}},\
  }\bibfield  {title} {\bibinfo {title} {Four year study of the schumann
  resonance regular variations using the sierra nevada station ground-based
  magnetometers},\ }\href
  {https://doi.org/https://doi.org/10.1029/2021JD036051} {\bibfield  {journal}
  {\bibinfo  {journal} {Journal of Geophysical Research: Atmospheres}\ }\textbf
  {\bibinfo {volume} {127}},\ \bibinfo {pages} {e2021JD036051} (\bibinfo {year}
  {2022})}\BibitemShut {NoStop}%
\bibitem [{\citenamefont {Alken}\ \emph {et~al.}(2021)\citenamefont {Alken},
  \citenamefont {Th{\'e}bault}, \citenamefont {Beggan}, \citenamefont {Amit},
  \citenamefont {Aubert}, \citenamefont {Baerenzung}, \citenamefont {Bondar},
  \citenamefont {Brown}, \citenamefont {Califf}, \citenamefont {Chambodut},
  \citenamefont {Chulliat}, \citenamefont {Cox}, \citenamefont {Finlay},
  \citenamefont {Fournier}, \citenamefont {Gillet}, \citenamefont {Grayver},
  \citenamefont {Hammer}, \citenamefont {Holschneider}, \citenamefont {Huder},
  \citenamefont {Hulot}, \citenamefont {Jager}, \citenamefont {Kloss},
  \citenamefont {Korte}, \citenamefont {Kuang}, \citenamefont {Kuvshinov},
  \citenamefont {Langlais}, \citenamefont {L{\'e}ger}, \citenamefont {Lesur},
  \citenamefont {Livermore}, \citenamefont {Lowes}, \citenamefont {Macmillan},
  \citenamefont {Magnes}, \citenamefont {Mandea}, \citenamefont {Marsal},
  \citenamefont {Matzka}, \citenamefont {Metman}, \citenamefont {Minami},
  \citenamefont {Morschhauser}, \citenamefont {Mound}, \citenamefont {Nair},
  \citenamefont {Nakano}, \citenamefont {Olsen}, \citenamefont
  {Pav{\'o}n-Carrasco}, \citenamefont {Petrov}, \citenamefont {Ropp},
  \citenamefont {Rother}, \citenamefont {Sabaka}, \citenamefont {Sanchez},
  \citenamefont {Saturnino}, \citenamefont {Schnepf}, \citenamefont {Shen},
  \citenamefont {Stolle}, \citenamefont {Tangborn}, \citenamefont
  {T{\o}ffner-Clausen}, \citenamefont {Toh}, \citenamefont {Torta},
  \citenamefont {Varner}, \citenamefont {Vervelidou}, \citenamefont {Vigneron},
  \citenamefont {Wardinski}, \citenamefont {Wicht}, \citenamefont {Woods},
  \citenamefont {Yang}, \citenamefont {Zeren},\ and\ \citenamefont
  {Zhou}}]{IGRF}%
  \BibitemOpen
  \bibfield  {author} {\bibinfo {author} {\bibfnamefont {P.}~\bibnamefont
  {Alken}}, \bibinfo {author} {\bibfnamefont {E.}~\bibnamefont {Th{\'e}bault}},
  \bibinfo {author} {\bibfnamefont {C.~D.}\ \bibnamefont {Beggan}}, \bibinfo
  {author} {\bibfnamefont {H.}~\bibnamefont {Amit}}, \bibinfo {author}
  {\bibfnamefont {J.}~\bibnamefont {Aubert}}, \bibinfo {author} {\bibfnamefont
  {J.}~\bibnamefont {Baerenzung}}, \bibinfo {author} {\bibfnamefont {T.~N.}\
  \bibnamefont {Bondar}}, \bibinfo {author} {\bibfnamefont {W.~J.}\
  \bibnamefont {Brown}}, \bibinfo {author} {\bibfnamefont {S.}~\bibnamefont
  {Califf}}, \bibinfo {author} {\bibfnamefont {A.}~\bibnamefont {Chambodut}},
  \bibinfo {author} {\bibfnamefont {A.}~\bibnamefont {Chulliat}}, \bibinfo
  {author} {\bibfnamefont {G.~A.}\ \bibnamefont {Cox}}, \bibinfo {author}
  {\bibfnamefont {C.~C.}\ \bibnamefont {Finlay}}, \bibinfo {author}
  {\bibfnamefont {A.}~\bibnamefont {Fournier}}, \bibinfo {author}
  {\bibfnamefont {N.}~\bibnamefont {Gillet}}, \bibinfo {author} {\bibfnamefont
  {A.}~\bibnamefont {Grayver}}, \bibinfo {author} {\bibfnamefont {M.~D.}\
  \bibnamefont {Hammer}}, \bibinfo {author} {\bibfnamefont {M.}~\bibnamefont
  {Holschneider}}, \bibinfo {author} {\bibfnamefont {L.}~\bibnamefont {Huder}},
  \bibinfo {author} {\bibfnamefont {G.}~\bibnamefont {Hulot}}, \bibinfo
  {author} {\bibfnamefont {T.}~\bibnamefont {Jager}}, \bibinfo {author}
  {\bibfnamefont {C.}~\bibnamefont {Kloss}}, \bibinfo {author} {\bibfnamefont
  {M.}~\bibnamefont {Korte}}, \bibinfo {author} {\bibfnamefont
  {W.}~\bibnamefont {Kuang}}, \bibinfo {author} {\bibfnamefont
  {A.}~\bibnamefont {Kuvshinov}}, \bibinfo {author} {\bibfnamefont
  {B.}~\bibnamefont {Langlais}}, \bibinfo {author} {\bibfnamefont {J.-M.}\
  \bibnamefont {L{\'e}ger}}, \bibinfo {author} {\bibfnamefont {V.}~\bibnamefont
  {Lesur}}, \bibinfo {author} {\bibfnamefont {P.~W.}\ \bibnamefont
  {Livermore}}, \bibinfo {author} {\bibfnamefont {F.~J.}\ \bibnamefont
  {Lowes}}, \bibinfo {author} {\bibfnamefont {S.}~\bibnamefont {Macmillan}},
  \bibinfo {author} {\bibfnamefont {W.}~\bibnamefont {Magnes}}, \bibinfo
  {author} {\bibfnamefont {M.}~\bibnamefont {Mandea}}, \bibinfo {author}
  {\bibfnamefont {S.}~\bibnamefont {Marsal}}, \bibinfo {author} {\bibfnamefont
  {J.}~\bibnamefont {Matzka}}, \bibinfo {author} {\bibfnamefont {M.~C.}\
  \bibnamefont {Metman}}, \bibinfo {author} {\bibfnamefont {T.}~\bibnamefont
  {Minami}}, \bibinfo {author} {\bibfnamefont {A.}~\bibnamefont
  {Morschhauser}}, \bibinfo {author} {\bibfnamefont {J.~E.}\ \bibnamefont
  {Mound}}, \bibinfo {author} {\bibfnamefont {M.}~\bibnamefont {Nair}},
  \bibinfo {author} {\bibfnamefont {S.}~\bibnamefont {Nakano}}, \bibinfo
  {author} {\bibfnamefont {N.}~\bibnamefont {Olsen}}, \bibinfo {author}
  {\bibfnamefont {F.~J.}\ \bibnamefont {Pav{\'o}n-Carrasco}}, \bibinfo {author}
  {\bibfnamefont {V.~G.}\ \bibnamefont {Petrov}}, \bibinfo {author}
  {\bibfnamefont {G.}~\bibnamefont {Ropp}}, \bibinfo {author} {\bibfnamefont
  {M.}~\bibnamefont {Rother}}, \bibinfo {author} {\bibfnamefont {T.~J.}\
  \bibnamefont {Sabaka}}, \bibinfo {author} {\bibfnamefont {S.}~\bibnamefont
  {Sanchez}}, \bibinfo {author} {\bibfnamefont {D.}~\bibnamefont {Saturnino}},
  \bibinfo {author} {\bibfnamefont {N.~R.}\ \bibnamefont {Schnepf}}, \bibinfo
  {author} {\bibfnamefont {X.}~\bibnamefont {Shen}}, \bibinfo {author}
  {\bibfnamefont {C.}~\bibnamefont {Stolle}}, \bibinfo {author} {\bibfnamefont
  {A.}~\bibnamefont {Tangborn}}, \bibinfo {author} {\bibfnamefont
  {L.}~\bibnamefont {T{\o}ffner-Clausen}}, \bibinfo {author} {\bibfnamefont
  {H.}~\bibnamefont {Toh}}, \bibinfo {author} {\bibfnamefont {J.~M.}\
  \bibnamefont {Torta}}, \bibinfo {author} {\bibfnamefont {J.}~\bibnamefont
  {Varner}}, \bibinfo {author} {\bibfnamefont {F.}~\bibnamefont {Vervelidou}},
  \bibinfo {author} {\bibfnamefont {P.}~\bibnamefont {Vigneron}}, \bibinfo
  {author} {\bibfnamefont {I.}~\bibnamefont {Wardinski}}, \bibinfo {author}
  {\bibfnamefont {J.}~\bibnamefont {Wicht}}, \bibinfo {author} {\bibfnamefont
  {A.}~\bibnamefont {Woods}}, \bibinfo {author} {\bibfnamefont
  {Y.}~\bibnamefont {Yang}}, \bibinfo {author} {\bibfnamefont {Z.}~\bibnamefont
  {Zeren}},\ and\ \bibinfo {author} {\bibfnamefont {B.}~\bibnamefont {Zhou}},\
  }\bibfield  {title} {\bibinfo {title} {{International Geomagnetic Reference
  Field: the thirteenth generation}},\ }\href
  {https://doi.org/https://doi.org/10.1186/s40623-020-01288-x} {\bibfield
  {journal} {\bibinfo  {journal} {Earth Planets Space}\ }\textbf {\bibinfo
  {volume} {73}},\ \bibinfo {pages} {49} (\bibinfo {year} {2021})}\BibitemShut
  {NoStop}%
\bibitem [{\citenamefont {Budker}\ \emph {et~al.}(2002)\citenamefont {Budker},
  \citenamefont {Kimball}, \citenamefont {Yashchuk},\ and\ \citenamefont
  {Zolotorev}}]{budker2002nonlinear}%
  \BibitemOpen
  \bibfield  {author} {\bibinfo {author} {\bibfnamefont {D.}~\bibnamefont
  {Budker}}, \bibinfo {author} {\bibfnamefont {D.}~\bibnamefont {Kimball}},
  \bibinfo {author} {\bibfnamefont {V.}~\bibnamefont {Yashchuk}},\ and\
  \bibinfo {author} {\bibfnamefont {M.}~\bibnamefont {Zolotorev}},\ }\bibfield
  {title} {\bibinfo {title} {Nonlinear magneto-optical rotation with
  frequency-modulated light},\ }\href
  {https://doi.org/10.1103/PhysRevA.65.055403} {\bibfield  {journal} {\bibinfo
  {journal} {Phys. Rev. A}\ }\textbf {\bibinfo {volume} {65}},\ \bibinfo
  {pages} {055403} (\bibinfo {year} {2002})}\BibitemShut {NoStop}%
\bibitem [{\citenamefont {Gawlik}\ \emph {et~al.}(2006)\citenamefont {Gawlik},
  \citenamefont {Krzemie{\'n}}, \citenamefont {Pustelny}, \citenamefont
  {Sangla}, \citenamefont {Zachorowski}, \citenamefont {Graf}, \citenamefont
  {Sushkov},\ and\ \citenamefont {Budker}}]{gawlik2006nonlinear}%
  \BibitemOpen
  \bibfield  {author} {\bibinfo {author} {\bibfnamefont {W.}~\bibnamefont
  {Gawlik}}, \bibinfo {author} {\bibfnamefont {L.}~\bibnamefont
  {Krzemie{\'n}}}, \bibinfo {author} {\bibfnamefont {S.}~\bibnamefont
  {Pustelny}}, \bibinfo {author} {\bibfnamefont {D.}~\bibnamefont {Sangla}},
  \bibinfo {author} {\bibfnamefont {J.}~\bibnamefont {Zachorowski}}, \bibinfo
  {author} {\bibfnamefont {M.}~\bibnamefont {Graf}}, \bibinfo {author}
  {\bibfnamefont {A.}~\bibnamefont {Sushkov}},\ and\ \bibinfo {author}
  {\bibfnamefont {D.}~\bibnamefont {Budker}},\ }\bibfield  {title} {\bibinfo
  {title} {Nonlinear magneto-optical rotation with amplitude modulated light},\
  }\bibfield  {journal} {\bibinfo  {journal} {Appl. Phys. Lett.}\ }\textbf
  {\bibinfo {volume} {88}},\ \href {https://doi.org/10.1063/1.2190457}
  {10.1063/1.2190457} (\bibinfo {year} {2006})\BibitemShut {NoStop}%
\bibitem [{\citenamefont {Jackson~Kimball}\ \emph {et~al.}(2009)\citenamefont
  {Jackson~Kimball}, \citenamefont {Jacome}, \citenamefont {Guttikonda},
  \citenamefont {Bahr},\ and\ \citenamefont {Chan}}]{kimball2009magnetometric}%
  \BibitemOpen
  \bibfield  {author} {\bibinfo {author} {\bibfnamefont {D.}~\bibnamefont
  {Jackson~Kimball}}, \bibinfo {author} {\bibfnamefont {L.~R.}\ \bibnamefont
  {Jacome}}, \bibinfo {author} {\bibfnamefont {S.}~\bibnamefont {Guttikonda}},
  \bibinfo {author} {\bibfnamefont {E.~J.}\ \bibnamefont {Bahr}},\ and\
  \bibinfo {author} {\bibfnamefont {L.~F.}\ \bibnamefont {Chan}},\ }\bibfield
  {title} {\bibinfo {title} {Magnetometric sensitivity optimization for
  nonlinear optical rotation with frequency-modulated light: Rubidium d2
  line},\ }\bibfield  {journal} {\bibinfo  {journal} {J. Appl. Phys.}\ }\textbf
  {\bibinfo {volume} {106}},\ \href {https://doi.org/10.1063/1.3225917}
  {10.1063/1.3225917} (\bibinfo {year} {2009})\BibitemShut {NoStop}%
\bibitem [{\citenamefont {Survey}()}]{USGS}%
  \BibitemOpen
  \bibfield  {author} {\bibinfo {author} {\bibfnamefont {U.~G.}\ \bibnamefont
  {Survey}},\ }\href@noop {} {\bibinfo {title} {Usgs geomagnetic survey}},\
  \bibinfo {howpublished} {\url{https://geomag.usgs.gov/ws/docs}},\ \bibinfo
  {note} {accessed: 2023-03-10}\BibitemShut {NoStop}%
\bibitem [{Spa()}]{SpaceDotCom}%
  \BibitemOpen
  \href@noop {} {\bibinfo {title} {Solar storm forecast}},\ \bibinfo
  {howpublished}
  {\url{https://www.space.com/moderate-solar-storm-forecast-july-23}},\
  \bibinfo {note} {accessed: 2023-03-10}\BibitemShut {NoStop}%
\bibitem [{atl(2015)}]{atlas}%
  \BibitemOpen
  \href@noop {} {\bibinfo {title} {{World Atlas of Ground Conductivities,
  \href{https://www.itu.int/rec/R-REC-P.832-4-201507-I/en}{Rec.~ITU-R
  P.832-4}}}},\ \bibinfo {howpublished} {{International Telecommunication
  Union}} (\bibinfo {year} {2015})\BibitemShut {NoStop}%
\bibitem [{\citenamefont {Takeda}\ and\ \citenamefont
  {Araki}(1985)}]{Takeda:1985hcf}%
  \BibitemOpen
  \bibfield  {author} {\bibinfo {author} {\bibfnamefont {M.}~\bibnamefont
  {Takeda}}\ and\ \bibinfo {author} {\bibfnamefont {T.}~\bibnamefont {Araki}},\
  }\bibfield  {title} {\bibinfo {title} {{Electric conductivity of the
  ionosphere and nocturnal currents}},\ }\href
  {https://doi.org/10.1016/0021-9169(85)90043-1} {\bibfield  {journal}
  {\bibinfo  {journal} {J. Atmos. Terr. Phys.}\ }\textbf {\bibinfo {volume}
  {47}},\ \bibinfo {pages} {601} (\bibinfo {year} {1985})}\BibitemShut
  {NoStop}%
\bibitem [{\citenamefont {Richmond}\ and\ \citenamefont
  {Thayer}(2000)}]{GM118}%
  \BibitemOpen
  \bibfield  {author} {\bibinfo {author} {\bibfnamefont {A.}~\bibnamefont
  {Richmond}}\ and\ \bibinfo {author} {\bibfnamefont {J.}~\bibnamefont
  {Thayer}},\ }\bibfield  {title} {\bibinfo {title}
  {\href{https://agupubs.onlinelibrary.wiley.com/doi/10.1029/GM118p0131}{{Ionospheric
  Electrodynamics: A Tutorial}}},\ }in\ \href@noop {} {\emph {\bibinfo
  {booktitle} {{Magnetospheric Current Systems (Geophysical Monograph
  118)}}}},\ \bibinfo {editor} {edited by\ \bibinfo {editor} {\bibfnamefont
  {S.}~\bibnamefont {Ohtani}}, \bibinfo {editor} {\bibfnamefont
  {R.}~\bibnamefont {Fujii}}, \bibinfo {editor} {\bibfnamefont
  {M.}~\bibnamefont {Hesse}},\ and\ \bibinfo {editor} {\bibfnamefont {R.~L.}\
  \bibnamefont {Lysak}}}\ (\bibinfo  {publisher} {American Geophysical Union},\
  \bibinfo {address} {Washington, DC},\ \bibinfo {year} {2000})\ pp.\ \bibinfo
  {pages} {131--146}\BibitemShut {NoStop}%
\bibitem [{\citenamefont {Bloch}\ and\ \citenamefont
  {Kalia}(2023)}]{bloch2023curl}%
  \BibitemOpen
  \bibfield  {author} {\bibinfo {author} {\bibfnamefont {I.~M.}\ \bibnamefont
  {Bloch}}\ and\ \bibinfo {author} {\bibfnamefont {S.}~\bibnamefont {Kalia}},\
  }\href@noop {} {\bibinfo {title} {Curl up with a good $\mathbf b$: Detecting
  ultralight dark matter with differential magnetometry}} (\bibinfo {year}
  {2023}),\ \Eprint {https://arxiv.org/abs/2308.10931} {arXiv:2308.10931
  [hep-ph]} \BibitemShut {NoStop}%
\bibitem [{coh()}]{cohare_git}%
  \BibitemOpen
  \href@noop {} {}\bibinfo {howpublished}
  {\url{https://cajohare.github.io/AxionLimits/}.}\BibitemShut {Stop}%
\bibitem [{\citenamefont {Jiang}\ \emph {et~al.}(2023)\citenamefont {Jiang},
  \citenamefont {Hong}, \citenamefont {Hu}, \citenamefont {Chen}, \citenamefont
  {Yang}, \citenamefont {Hu}, \citenamefont {Yang}, \citenamefont {Shu},
  \citenamefont {Zhao},\ and\ \citenamefont {Peng}}]{jiang2023search}%
  \BibitemOpen
  \bibfield  {author} {\bibinfo {author} {\bibfnamefont {M.}~\bibnamefont
  {Jiang}}, \bibinfo {author} {\bibfnamefont {T.}~\bibnamefont {Hong}},
  \bibinfo {author} {\bibfnamefont {D.}~\bibnamefont {Hu}}, \bibinfo {author}
  {\bibfnamefont {Y.}~\bibnamefont {Chen}}, \bibinfo {author} {\bibfnamefont
  {F.}~\bibnamefont {Yang}}, \bibinfo {author} {\bibfnamefont {T.}~\bibnamefont
  {Hu}}, \bibinfo {author} {\bibfnamefont {X.}~\bibnamefont {Yang}}, \bibinfo
  {author} {\bibfnamefont {J.}~\bibnamefont {Shu}}, \bibinfo {author}
  {\bibfnamefont {Y.}~\bibnamefont {Zhao}},\ and\ \bibinfo {author}
  {\bibfnamefont {X.}~\bibnamefont {Peng}},\ }\bibfield  {title} {\bibinfo
  {title} {Search for dark photons with synchronized quantum sensor network},\
  }\bibfield  {journal} {\bibinfo  {journal} {arXiv:2305.00890}\ }\href
  {https://doi.org/10.48550/arXiv.2305.00890} {10.48550/arXiv.2305.00890}
  (\bibinfo {year} {2023})\BibitemShut {NoStop}%
\bibitem [{\citenamefont {Fischbach}\ \emph {et~al.}(1994)\citenamefont
  {Fischbach}, \citenamefont {Kloor}, \citenamefont {Langel}, \citenamefont
  {Lui},\ and\ \citenamefont {Peredo}}]{DarkPhoton_Earth:1994}%
  \BibitemOpen
  \bibfield  {author} {\bibinfo {author} {\bibfnamefont {E.}~\bibnamefont
  {Fischbach}}, \bibinfo {author} {\bibfnamefont {H.}~\bibnamefont {Kloor}},
  \bibinfo {author} {\bibfnamefont {R.~A.}\ \bibnamefont {Langel}}, \bibinfo
  {author} {\bibfnamefont {A.~T.~Y.}\ \bibnamefont {Lui}},\ and\ \bibinfo
  {author} {\bibfnamefont {M.}~\bibnamefont {Peredo}},\ }\bibfield  {title}
  {\bibinfo {title} {New geomagnetic limits on the photon mass and on
  long-range forces coexisting with electromagnetism},\ }\href
  {https://doi.org/10.1103/PhysRevLett.73.514} {\bibfield  {journal} {\bibinfo
  {journal} {Phys. Rev. Lett.}\ }\textbf {\bibinfo {volume} {73}},\ \bibinfo
  {pages} {514} (\bibinfo {year} {1994})}\BibitemShut {NoStop}%
\bibitem [{\citenamefont {Marocco}(2021)}]{dph_Jupiter:2021}%
  \BibitemOpen
  \bibfield  {author} {\bibinfo {author} {\bibfnamefont {G.}~\bibnamefont
  {Marocco}},\ }\bibfield  {title} {\bibinfo {title} {{Dark photon limits from
  magnetic fields and astrophysical plasmas}},\ }\href@noop {} {\  (\bibinfo
  {year} {2021})},\ \Eprint {https://arxiv.org/abs/2110.02875}
  {arXiv:2110.02875 [hep-ph]} \BibitemShut {NoStop}%
\bibitem [{\citenamefont {Bhoonah}\ \emph {et~al.}(2019)\citenamefont
  {Bhoonah}, \citenamefont {Bramante}, \citenamefont {Elahi},\ and\
  \citenamefont {Schon}}]{dark_photon_gas_cloud:2019prd}%
  \BibitemOpen
  \bibfield  {author} {\bibinfo {author} {\bibfnamefont {A.}~\bibnamefont
  {Bhoonah}}, \bibinfo {author} {\bibfnamefont {J.}~\bibnamefont {Bramante}},
  \bibinfo {author} {\bibfnamefont {F.}~\bibnamefont {Elahi}},\ and\ \bibinfo
  {author} {\bibfnamefont {S.}~\bibnamefont {Schon}},\ }\bibfield  {title}
  {\bibinfo {title} {Galactic center gas clouds and novel bounds on ultralight
  dark photon, vector portal, strongly interacting, composite, and super-heavy
  dark matter},\ }\href {https://doi.org/10.1103/PhysRevD.100.023001}
  {\bibfield  {journal} {\bibinfo  {journal} {Phys. Rev. D}\ }\textbf {\bibinfo
  {volume} {100}},\ \bibinfo {pages} {023001} (\bibinfo {year}
  {2019})}\BibitemShut {NoStop}%
\bibitem [{\citenamefont {Dubovsky}\ and\ \citenamefont
  {Hernández-Chifflet}(2015)}]{dph_IGM_Dubovsky:2015jcap}%
  \BibitemOpen
  \bibfield  {author} {\bibinfo {author} {\bibfnamefont {S.}~\bibnamefont
  {Dubovsky}}\ and\ \bibinfo {author} {\bibfnamefont {G.}~\bibnamefont
  {Hernández-Chifflet}},\ }\bibfield  {title} {\bibinfo {title} {Heating up
  the galaxy with hidden photons},\ }\href
  {https://doi.org/10.1088/1475-7516/2015/12/054} {\bibfield  {journal}
  {\bibinfo  {journal} {J. Cosmol. Astropart. Phys.}\ }\textbf {\bibinfo
  {volume} {2015}}\bibinfo  {number} { (12)},\ \bibinfo {pages}
  {054}}\BibitemShut {NoStop}%
\bibitem [{\citenamefont {Wadekar}\ and\ \citenamefont
  {Farrar}(2021)}]{dph_LeoT:2021prd}%
  \BibitemOpen
\bibfield  {number} {  }\bibfield  {author} {\bibinfo {author} {\bibfnamefont
  {D.}~\bibnamefont {Wadekar}}\ and\ \bibinfo {author} {\bibfnamefont {G.~R.}\
  \bibnamefont {Farrar}},\ }\bibfield  {title} {\bibinfo {title} {Gas-rich
  dwarf galaxies as a new probe of dark matter interactions with ordinary
  matter},\ }\href {https://doi.org/10.1103/PhysRevD.103.123028} {\bibfield
  {journal} {\bibinfo  {journal} {Phys. Rev. D}\ }\textbf {\bibinfo {volume}
  {103}},\ \bibinfo {pages} {123028} (\bibinfo {year} {2021})}\BibitemShut
  {NoStop}%
\bibitem [{\citenamefont {Caputo}\ \emph
  {et~al.}(2020{\natexlab{a}})\citenamefont {Caputo}, \citenamefont {Liu},
  \citenamefont {Mishra-Sharma},\ and\ \citenamefont
  {Ruderman}}]{dph_cobe_firas_caputo:2021prl}%
  \BibitemOpen
  \bibfield  {author} {\bibinfo {author} {\bibfnamefont {A.}~\bibnamefont
  {Caputo}}, \bibinfo {author} {\bibfnamefont {H.}~\bibnamefont {Liu}},
  \bibinfo {author} {\bibfnamefont {S.}~\bibnamefont {Mishra-Sharma}},\ and\
  \bibinfo {author} {\bibfnamefont {J.~T.}\ \bibnamefont {Ruderman}},\
  }\bibfield  {title} {\bibinfo {title} {Dark photon oscillations in our
  inhomogeneous universe},\ }\href
  {https://doi.org/10.1103/PhysRevLett.125.221303} {\bibfield  {journal}
  {\bibinfo  {journal} {Phys. Rev. Lett.}\ }\textbf {\bibinfo {volume} {125}},\
  \bibinfo {pages} {221303} (\bibinfo {year} {2020}{\natexlab{a}})}\BibitemShut
  {NoStop}%
\bibitem [{\citenamefont {McDermott}\ and\ \citenamefont
  {Witte}(2020)}]{dph_cobe_firas_witte:2020prg}%
  \BibitemOpen
  \bibfield  {author} {\bibinfo {author} {\bibfnamefont {S.~D.}\ \bibnamefont
  {McDermott}}\ and\ \bibinfo {author} {\bibfnamefont {S.~J.}\ \bibnamefont
  {Witte}},\ }\bibfield  {title} {\bibinfo {title} {Cosmological evolution of
  light dark photon dark matter},\ }\href
  {https://doi.org/10.1103/PhysRevD.101.063030} {\bibfield  {journal} {\bibinfo
   {journal} {Phys. Rev. D}\ }\textbf {\bibinfo {volume} {101}},\ \bibinfo
  {pages} {063030} (\bibinfo {year} {2020})}\BibitemShut {NoStop}%
\bibitem [{\citenamefont {Caputo}\ \emph
  {et~al.}(2020{\natexlab{b}})\citenamefont {Caputo}, \citenamefont {Liu},
  \citenamefont {Mishra-Sharma},\ and\ \citenamefont
  {Ruderman}}]{dph_HeII:2020prl}%
  \BibitemOpen
  \bibfield  {author} {\bibinfo {author} {\bibfnamefont {A.}~\bibnamefont
  {Caputo}}, \bibinfo {author} {\bibfnamefont {H.}~\bibnamefont {Liu}},
  \bibinfo {author} {\bibfnamefont {S.}~\bibnamefont {Mishra-Sharma}},\ and\
  \bibinfo {author} {\bibfnamefont {J.~T.}\ \bibnamefont {Ruderman}},\
  }\bibfield  {title} {\bibinfo {title} {Dark photon oscillations in our
  inhomogeneous universe},\ }\href
  {https://doi.org/10.1103/PhysRevLett.125.221303} {\bibfield  {journal}
  {\bibinfo  {journal} {Phys. Rev. Lett.}\ }\textbf {\bibinfo {volume} {125}},\
  \bibinfo {pages} {221303} (\bibinfo {year} {2020}{\natexlab{b}})}\BibitemShut
  {NoStop}%
\bibitem [{\citenamefont {{CAST Collaboration}}\ \emph
  {et~al.}(2017)\citenamefont {{CAST Collaboration}}, \citenamefont
  {Anastassopoulos}, \citenamefont {Aune}, \citenamefont {Barth}, \citenamefont
  {Belov}, \citenamefont {Bräuninger}, \citenamefont {Cantatore},
  \citenamefont {Carmona}, \citenamefont {Castel}, \citenamefont {Cetin},
  \citenamefont {Christensen}, \citenamefont {Collar}, \citenamefont {Dafni},
  \citenamefont {Davenport}, \citenamefont {Decker}, \citenamefont {Dermenev},
  \citenamefont {Desch}, \citenamefont {Eleftheriadis}, \citenamefont
  {Fanourakis}, \citenamefont {Ferrer-Ribas}, \citenamefont {Fischer},
  \citenamefont {García}, \citenamefont {Gardikiotis}, \citenamefont {Garza},
  \citenamefont {Gazis}, \citenamefont {Geralis}, \citenamefont {Giomataris},
  \citenamefont {Gninenko}, \citenamefont {Hailey}, \citenamefont {Hasinoff},
  \citenamefont {Hoffmann}, \citenamefont {Iguaz}, \citenamefont {Irastorza},
  \citenamefont {Jakobsen}, \citenamefont {Jacoby}, \citenamefont {Jakovčić},
  \citenamefont {Kaminski}, \citenamefont {Karuza}, \citenamefont {Kralj},
  \citenamefont {Krčmar}, \citenamefont {Kostoglou}, \citenamefont {Krieger},
  \citenamefont {Lakić}, \citenamefont {Laurent}, \citenamefont {Liolios},
  \citenamefont {Ljubičić}, \citenamefont {Luzón}, \citenamefont {Maroudas},
  \citenamefont {Miceli}, \citenamefont {Neff}, \citenamefont {Ortega},
  \citenamefont {Papaevangelou}, \citenamefont {Paraschou}, \citenamefont
  {Pivovaroff}, \citenamefont {Raffelt}, \citenamefont {Rosu}, \citenamefont
  {Ruz}, \citenamefont {Chóliz}, \citenamefont {Savvidis}, \citenamefont
  {Schmidt}, \citenamefont {Semertzidis}, \citenamefont {Solanki},
  \citenamefont {Stewart}, \citenamefont {Vafeiadis}, \citenamefont {Vogel},
  \citenamefont {Yildiz},\ and\ \citenamefont {Zioutas}}]{CAST:2017}%
  \BibitemOpen
  \bibfield  {author} {\bibinfo {author} {\bibnamefont {{CAST Collaboration}}},
  \bibinfo {author} {\bibfnamefont {V.}~\bibnamefont {Anastassopoulos}},
  \bibinfo {author} {\bibfnamefont {S.}~\bibnamefont {Aune}}, \bibinfo {author}
  {\bibfnamefont {K.}~\bibnamefont {Barth}}, \bibinfo {author} {\bibfnamefont
  {A.}~\bibnamefont {Belov}}, \bibinfo {author} {\bibfnamefont
  {H.}~\bibnamefont {Bräuninger}}, \bibinfo {author} {\bibfnamefont
  {G.}~\bibnamefont {Cantatore}}, \bibinfo {author} {\bibfnamefont {J.~M.}\
  \bibnamefont {Carmona}}, \bibinfo {author} {\bibfnamefont {J.~F.}\
  \bibnamefont {Castel}}, \bibinfo {author} {\bibfnamefont {S.~A.}\
  \bibnamefont {Cetin}}, \bibinfo {author} {\bibfnamefont {F.}~\bibnamefont
  {Christensen}}, \bibinfo {author} {\bibfnamefont {J.~I.}\ \bibnamefont
  {Collar}}, \bibinfo {author} {\bibfnamefont {T.}~\bibnamefont {Dafni}},
  \bibinfo {author} {\bibfnamefont {M.}~\bibnamefont {Davenport}}, \bibinfo
  {author} {\bibfnamefont {T.~A.}\ \bibnamefont {Decker}}, \bibinfo {author}
  {\bibfnamefont {A.}~\bibnamefont {Dermenev}}, \bibinfo {author}
  {\bibfnamefont {K.}~\bibnamefont {Desch}}, \bibinfo {author} {\bibfnamefont
  {C.}~\bibnamefont {Eleftheriadis}}, \bibinfo {author} {\bibfnamefont
  {G.}~\bibnamefont {Fanourakis}}, \bibinfo {author} {\bibfnamefont
  {E.}~\bibnamefont {Ferrer-Ribas}}, \bibinfo {author} {\bibfnamefont
  {H.}~\bibnamefont {Fischer}}, \bibinfo {author} {\bibfnamefont {J.~A.}\
  \bibnamefont {García}}, \bibinfo {author} {\bibfnamefont {A.}~\bibnamefont
  {Gardikiotis}}, \bibinfo {author} {\bibfnamefont {J.~G.}\ \bibnamefont
  {Garza}}, \bibinfo {author} {\bibfnamefont {E.~N.}\ \bibnamefont {Gazis}},
  \bibinfo {author} {\bibfnamefont {T.}~\bibnamefont {Geralis}}, \bibinfo
  {author} {\bibfnamefont {I.}~\bibnamefont {Giomataris}}, \bibinfo {author}
  {\bibfnamefont {S.}~\bibnamefont {Gninenko}}, \bibinfo {author}
  {\bibfnamefont {C.~J.}\ \bibnamefont {Hailey}}, \bibinfo {author}
  {\bibfnamefont {M.~D.}\ \bibnamefont {Hasinoff}}, \bibinfo {author}
  {\bibfnamefont {D.~H.~H.}\ \bibnamefont {Hoffmann}}, \bibinfo {author}
  {\bibfnamefont {F.~J.}\ \bibnamefont {Iguaz}}, \bibinfo {author}
  {\bibfnamefont {I.~G.}\ \bibnamefont {Irastorza}}, \bibinfo {author}
  {\bibfnamefont {A.}~\bibnamefont {Jakobsen}}, \bibinfo {author}
  {\bibfnamefont {J.}~\bibnamefont {Jacoby}}, \bibinfo {author} {\bibfnamefont
  {K.}~\bibnamefont {Jakovčić}}, \bibinfo {author} {\bibfnamefont
  {J.}~\bibnamefont {Kaminski}}, \bibinfo {author} {\bibfnamefont
  {M.}~\bibnamefont {Karuza}}, \bibinfo {author} {\bibfnamefont
  {N.}~\bibnamefont {Kralj}}, \bibinfo {author} {\bibfnamefont
  {M.}~\bibnamefont {Krčmar}}, \bibinfo {author} {\bibfnamefont
  {S.}~\bibnamefont {Kostoglou}}, \bibinfo {author} {\bibfnamefont
  {C.}~\bibnamefont {Krieger}}, \bibinfo {author} {\bibfnamefont
  {B.}~\bibnamefont {Lakić}}, \bibinfo {author} {\bibfnamefont {J.~M.}\
  \bibnamefont {Laurent}}, \bibinfo {author} {\bibfnamefont {A.}~\bibnamefont
  {Liolios}}, \bibinfo {author} {\bibfnamefont {A.}~\bibnamefont {Ljubičić}},
  \bibinfo {author} {\bibfnamefont {G.}~\bibnamefont {Luzón}}, \bibinfo
  {author} {\bibfnamefont {M.}~\bibnamefont {Maroudas}}, \bibinfo {author}
  {\bibfnamefont {L.}~\bibnamefont {Miceli}}, \bibinfo {author} {\bibfnamefont
  {S.}~\bibnamefont {Neff}}, \bibinfo {author} {\bibfnamefont {I.}~\bibnamefont
  {Ortega}}, \bibinfo {author} {\bibfnamefont {T.}~\bibnamefont
  {Papaevangelou}}, \bibinfo {author} {\bibfnamefont {K.}~\bibnamefont
  {Paraschou}}, \bibinfo {author} {\bibfnamefont {M.~J.}\ \bibnamefont
  {Pivovaroff}}, \bibinfo {author} {\bibfnamefont {G.}~\bibnamefont {Raffelt}},
  \bibinfo {author} {\bibfnamefont {M.}~\bibnamefont {Rosu}}, \bibinfo {author}
  {\bibfnamefont {J.}~\bibnamefont {Ruz}}, \bibinfo {author} {\bibfnamefont
  {E.~R.}\ \bibnamefont {Chóliz}}, \bibinfo {author} {\bibfnamefont
  {I.}~\bibnamefont {Savvidis}}, \bibinfo {author} {\bibfnamefont
  {S.}~\bibnamefont {Schmidt}}, \bibinfo {author} {\bibfnamefont {Y.~K.}\
  \bibnamefont {Semertzidis}}, \bibinfo {author} {\bibfnamefont {S.~K.}\
  \bibnamefont {Solanki}}, \bibinfo {author} {\bibfnamefont {L.}~\bibnamefont
  {Stewart}}, \bibinfo {author} {\bibfnamefont {T.}~\bibnamefont {Vafeiadis}},
  \bibinfo {author} {\bibfnamefont {J.~K.}\ \bibnamefont {Vogel}}, \bibinfo
  {author} {\bibfnamefont {S.~C.}\ \bibnamefont {Yildiz}},\ and\ \bibinfo
  {author} {\bibfnamefont {K.}~\bibnamefont {Zioutas}},\ }\bibfield  {title}
  {\bibinfo {title} {{New CAST limit on the axion–photon interaction}},\
  }\href {https://doi.org/https://doi.org/10.1038/nphys4109} {\bibfield
  {journal} {\bibinfo  {journal} {Nature Physics}\ }\textbf {\bibinfo {volume}
  {13}},\ \bibinfo {pages} {584} (\bibinfo {year} {2017})}\BibitemShut
  {NoStop}%
\bibitem [{\citenamefont {Caputo}\ \emph {et~al.}(2022)\citenamefont {Caputo},
  \citenamefont {Janka}, \citenamefont {Raffelt},\ and\ \citenamefont
  {Vitagliano}}]{SNe:2022prl}%
  \BibitemOpen
  \bibfield  {author} {\bibinfo {author} {\bibfnamefont {A.}~\bibnamefont
  {Caputo}}, \bibinfo {author} {\bibfnamefont {H.-T.}\ \bibnamefont {Janka}},
  \bibinfo {author} {\bibfnamefont {G.}~\bibnamefont {Raffelt}},\ and\ \bibinfo
  {author} {\bibfnamefont {E.}~\bibnamefont {Vitagliano}},\ }\bibfield  {title}
  {\bibinfo {title} {Low-energy supernovae severely constrain radiative
  particle decays},\ }\href {https://doi.org/10.1103/PhysRevLett.128.221103}
  {\bibfield  {journal} {\bibinfo  {journal} {Phys. Rev. Lett.}\ }\textbf
  {\bibinfo {volume} {128}},\ \bibinfo {pages} {221103} (\bibinfo {year}
  {2022})}\BibitemShut {NoStop}%
\bibitem [{\citenamefont {Wouters}\ and\ \citenamefont
  {Brun}(2013)}]{hydra_a_Wouters_2013}%
  \BibitemOpen
  \bibfield  {author} {\bibinfo {author} {\bibfnamefont {D.}~\bibnamefont
  {Wouters}}\ and\ \bibinfo {author} {\bibfnamefont {P.}~\bibnamefont {Brun}},\
  }\bibfield  {title} {\bibinfo {title} {Constraints on axion-like particles
  from x-ray observations of the hydra galaxy cluster},\ }\href
  {https://doi.org/10.1088/0004-637X/772/1/44} {\bibfield  {journal} {\bibinfo
  {journal} {Astrophys. J.}\ }\textbf {\bibinfo {volume} {772}},\ \bibinfo
  {pages} {44} (\bibinfo {year} {2013})}\BibitemShut {NoStop}%
\bibitem [{\citenamefont {Dessert}\ \emph {et~al.}(2020)\citenamefont
  {Dessert}, \citenamefont {Foster},\ and\ \citenamefont
  {Safdi}}]{Superstar_cluster:2020prl}%
  \BibitemOpen
  \bibfield  {author} {\bibinfo {author} {\bibfnamefont {C.}~\bibnamefont
  {Dessert}}, \bibinfo {author} {\bibfnamefont {J.~W.}\ \bibnamefont
  {Foster}},\ and\ \bibinfo {author} {\bibfnamefont {B.~R.}\ \bibnamefont
  {Safdi}},\ }\bibfield  {title} {\bibinfo {title} {X-ray searches for axions
  from super star clusters},\ }\href
  {https://doi.org/10.1103/PhysRevLett.125.261102} {\bibfield  {journal}
  {\bibinfo  {journal} {Phys. Rev. Lett.}\ }\textbf {\bibinfo {volume} {125}},\
  \bibinfo {pages} {261102} (\bibinfo {year} {2020})}\BibitemShut {NoStop}%
\bibitem [{\citenamefont {Marsh}\ \emph {et~al.}(2017)\citenamefont {Marsh},
  \citenamefont {Russell}, \citenamefont {Fabian}, \citenamefont {McNamara},
  \citenamefont {Nulsen},\ and\ \citenamefont {Reynolds}}]{M87_Marsh_2017}%
  \BibitemOpen
  \bibfield  {author} {\bibinfo {author} {\bibfnamefont {M.~D.}\ \bibnamefont
  {Marsh}}, \bibinfo {author} {\bibfnamefont {H.~R.}\ \bibnamefont {Russell}},
  \bibinfo {author} {\bibfnamefont {A.~C.}\ \bibnamefont {Fabian}}, \bibinfo
  {author} {\bibfnamefont {B.~R.}\ \bibnamefont {McNamara}}, \bibinfo {author}
  {\bibfnamefont {P.}~\bibnamefont {Nulsen}},\ and\ \bibinfo {author}
  {\bibfnamefont {C.~S.}\ \bibnamefont {Reynolds}},\ }\bibfield  {title}
  {\bibinfo {title} {A new bound on axion-like particles},\ }\href
  {https://doi.org/10.1088/1475-7516/2017/12/036} {\bibfield  {journal}
  {\bibinfo  {journal} {J. Cosmol. Astropart. Phys.}\ }\textbf {\bibinfo
  {volume} {2017}}\bibinfo  {number} { (12)},\ \bibinfo {pages}
  {036}}\BibitemShut {NoStop}%
\bibitem [{\citenamefont {{Sisk Reyn{\'e}s}}\ \emph {et~al.}(2021)\citenamefont
  {{Sisk Reyn{\'e}s}}, \citenamefont {Matthews}, \citenamefont {Reynolds},
  \citenamefont {Russell}, \citenamefont {Smith},\ and\ \citenamefont
  {Marsh}}]{H1821_643:2021mnra}%
  \BibitemOpen
\bibfield  {number} {  }\bibfield  {author} {\bibinfo {author} {\bibfnamefont
  {J.}~\bibnamefont {{Sisk Reyn{\'e}s}}}, \bibinfo {author} {\bibfnamefont
  {J.~H.}\ \bibnamefont {Matthews}}, \bibinfo {author} {\bibfnamefont {C.~S.}\
  \bibnamefont {Reynolds}}, \bibinfo {author} {\bibfnamefont {H.~R.}\
  \bibnamefont {Russell}}, \bibinfo {author} {\bibfnamefont {R.~N.}\
  \bibnamefont {Smith}},\ and\ \bibinfo {author} {\bibfnamefont {M.~C.~D.}\
  \bibnamefont {Marsh}},\ }\bibfield  {title} {\bibinfo {title} {{New
  constraints on light axion-like particles using \emph{Chandra} transmission
  grating spectroscopy of the powerful cluster-hosted quasar H1821+643}},\
  }\href {https://doi.org/10.1093/mnras/stab3464} {\bibfield  {journal}
  {\bibinfo  {journal} {Mon. Not. R. Astron. Soc.}\ }\textbf {\bibinfo {volume}
  {510}},\ \bibinfo {pages} {1264} (\bibinfo {year} {2021})}\BibitemShut
  {NoStop}%
\bibitem [{\citenamefont {Votis}\ \emph {et~al.}(2018)\citenamefont {Votis},
  \citenamefont {Tatsis}, \citenamefont {Christofilakis}, \citenamefont
  {Chronopoulos}, \citenamefont {Kostarakis}, \citenamefont {Tritakis},\ and\
  \citenamefont {Repapis}}]{votis2018new}%
  \BibitemOpen
  \bibfield  {author} {\bibinfo {author} {\bibfnamefont {C.~I.}\ \bibnamefont
  {Votis}}, \bibinfo {author} {\bibfnamefont {G.}~\bibnamefont {Tatsis}},
  \bibinfo {author} {\bibfnamefont {V.}~\bibnamefont {Christofilakis}},
  \bibinfo {author} {\bibfnamefont {S.~K.}\ \bibnamefont {Chronopoulos}},
  \bibinfo {author} {\bibfnamefont {P.}~\bibnamefont {Kostarakis}}, \bibinfo
  {author} {\bibfnamefont {V.}~\bibnamefont {Tritakis}},\ and\ \bibinfo
  {author} {\bibfnamefont {C.}~\bibnamefont {Repapis}},\ }\bibfield  {title}
  {\bibinfo {title} {{A new portable ELF Schumann resonance receiver: Design
  and detailed analysis of the antenna and the analog front-end}},\ }\href
  {https://doi.org/10.1186/s13638-018-1157-7} {\bibfield  {journal} {\bibinfo
  {journal} {EURASIP Journal on Wireless Communications and Networking}\
  }\textbf {\bibinfo {volume} {2018}},\ \bibinfo {pages} {155} (\bibinfo {year}
  {2018})}\BibitemShut {NoStop}%
\bibitem [{\citenamefont {Poliakov}\ \emph {et~al.}(2017)\citenamefont
  {Poliakov}, \citenamefont {Reznikov}, \citenamefont {Shchennikov},
  \citenamefont {Kopytenko},\ and\ \citenamefont
  {Samsonov}}]{poliakov2017range}%
  \BibitemOpen
  \bibfield  {author} {\bibinfo {author} {\bibfnamefont {S.}~\bibnamefont
  {Poliakov}}, \bibinfo {author} {\bibfnamefont {B.}~\bibnamefont {Reznikov}},
  \bibinfo {author} {\bibfnamefont {A.}~\bibnamefont {Shchennikov}}, \bibinfo
  {author} {\bibfnamefont {E.}~\bibnamefont {Kopytenko}},\ and\ \bibinfo
  {author} {\bibfnamefont {B.}~\bibnamefont {Samsonov}},\ }\bibfield  {title}
  {\bibinfo {title} {The range of induction-coil magnetic field sensors for
  geophysical explorations},\ }\href
  {https://doi.org/10.3103/S0747923917010078} {\bibfield  {journal} {\bibinfo
  {journal} {Seismic instruments}\ }\textbf {\bibinfo {volume} {53}},\ \bibinfo
  {pages} {1} (\bibinfo {year} {2017})}\BibitemShut {NoStop}%
\bibitem [{\citenamefont {Hospodarsky}(2016)}]{hospodarsky2016spaced}%
  \BibitemOpen
  \bibfield  {author} {\bibinfo {author} {\bibfnamefont {G.~B.}\ \bibnamefont
  {Hospodarsky}},\ }\bibfield  {title} {\bibinfo {title} {Spaced-based search
  coil magnetometers},\ }\href {https://doi.org/10.1002/2016JA022565}
  {\bibfield  {journal} {\bibinfo  {journal} {Journal of Geophysical Research:
  Space Physics}\ }\textbf {\bibinfo {volume} {121}},\ \bibinfo {pages} {12068}
  (\bibinfo {year} {2016})}\BibitemShut {NoStop}%
\end{thebibliography}%

\end{document}